\documentclass[12pt,preprint]{aastex}
\usepackage{graphicx}
\usepackage{rotating}
\usepackage{longtable}
\usepackage{lscape}
\usepackage{geometry}
\usepackage{epsfig}
\usepackage[maxfloats=43]{morefloats}

\begin{document}

\title{A Statistical Study of GRB X-ray Flares: Evidence of Ubiquitous Bulk Acceleration in the Emission Region}

\author{Lan-Wei Jia\altaffilmark{1,2}, Z. Lucas Uhm\altaffilmark{2}, Bing Zhang\altaffilmark{2,3,4}}

\altaffiltext{1}{School of Physics and Electronic Engineering, Guangzhou University, Guangzhou 510006, China; lwjia@gzhu.edu.cn}
\altaffiltext{2}{Department of Physics and Astronomy, University of Nevada, Las Vegas, NV 89154, USA; uhm@physics.unlv.edu, zhang@physics.unlv.edu}
\altaffiltext{3}{Department of Astronomy, School of Physics, Peking University, Beijing 100871, China}
\altaffiltext{4}{Kavli Institute for Astronomy and Astrophysics, Peking University, Beijing, 100871, China}
\begin{abstract}
When emission in a conical relativistic jet ceases abruptly (or decays sharply), the observed decay light curve is controlled by the high-latitude ``curvature effect". Recently, Uhm \& Zhang found that the decay slopes of three GRB X-ray flares are steeper than what the standard model predicts. This requires bulk acceleration of the emission region, which is consistent with a Poynting-flux-dominated outflow. In this paper, we systematically analyze a sample of 85 bright X-ray flares detected in 63 {\em Swift} GRBs, and investigate the relationship between the temporal decay index $\alpha$ and spectral index $\beta$ during the steep decay phase of these flares. The $\alpha$ value depends on the choice of the zero time point $t_0$. We adopt two methods. ``Method I" takes $t_0^{\rm I}$ as the first rising data point of each flare, and is the most conservative approach. We find that at 99.9$\%$ condifence level 56/85 flares have decay slopes steeper than the simplest curvature effect prediction, and therefore, are in the acceleration regime. ``Method II" extrapolates the rising light curve of each flare backwards until the flux density is three orders of magnitude lower than the peak flux density, and defines the corresponding time as the time zero point ($t_{0}^{\rm II}$). We find that 74/85 flares fall into the acceleration regime at 99.9$\%$ condifence level. This suggests that bulk acceleration is common, may be even ubiquitous among X-ray flares, pointing towards a Poynting-flux-dominated jet composition for these events.
\end{abstract}
\keywords{gamma-ray burst: general--methods: statistical}
\section{Introduction}

Gamma-ray bursts (GRBs) are the brightest electromagnetic explosions in the Universe. Despite decades of investigations, the physical mechanism to produce the observed emission is not identified. One fundamental question is: what is the composition of the relativistic jets (e.g., Zhang 2011; Kumar \& Zhang 2015; Pe'er 2015)?

One widely discussed model for GRBs is the so called ``fireball shock'' model (Goodman 1986; Pacz\'ynski 1986). Within this model, a matter dominated outflow (with negligible fraction of magnetic energy) initially in the form of a hot fireball undergoes a rapid acceleration by converting its thermal energy to kinetic energy (Shemi \& Piran 1990; M\'esz\'aros et al. 1993; Piran et al. 1993; Kobayashi et al. 1999), and the outflow later dissipates its kinetic energy in internal shocks (Rees \& M\'esz\'aros 1994; Kobayashi et al. 1997) or the external shocks (Rees \& M\'esz\'aros 1992; M\'esz\'aros \& Rees 1993) to power the observed GRB emission. 

Alternatively, the outflow can be Poynting-flux-dominated (i.e., $\sigma \gg 1$, where $\sigma$ is the magnetization parameter defined as the ratio between the Poynting flux and the plasma matter flux). Within this scenario, the Poynting flux energy may be also converted to the kinetic energy, but in a much slower pace (e.g., Drenkhahn \& Spruit 2002; Komissarov et al. 2009; Granot et al. 2011). Magnetic dissipation likely happens through reconnection or current instabilities, which convert the Poynting flux energy directly to particle energy and radiation. There are two types of such models. The first type of models, which may be relevant for a striped wind magnetic field configuration, invokes rapid dissipation at small radii (below the photosphere), so that such dissipations would enhance the photosphere emission (e.g., Thompson 1994; Drenkhahn 2002; Giannios 2008), and the outflow already reaches $\sigma \leq 1$ at the photosphere. The other type of models, which may be more relevant for helical magnetic configurations, envisages that direct magnetic dissipation is prohibited at small radii, but is triggered abruptly at a large emission radius, by e.g., internal collisions (i.e., the ICMART model, Zhang \& Yan 2011). Such a scenario is supported by the lack of, or the weak photosphere emission component observed in the majority of GRBs (Zhang \& Pe'er 2009; Guiriec et al. 2011, 2015; Axelsson et al. 2012; Burgess et al. 2014; Gao \& Zhang 2015). 

Great efforts have been made to reproduce the observed temporal (e.g., Hasc\"oet et al. 2012; Zhang \& Zhang 2014) and spectral (e.g., Pe'er et al. 2006; Daigne et al. 2011; Vurm et al. 2011; Lundman et al. 2013; Deng \& Zhang 2014; Uhm \& Zhang 2014; Yu et al. 2015; Zhang et al. 2015) properties of GRBs using models with distinct jet compositions. However, definitive conclusions could not be drawn due to the lack of a ``smoking-gun" signature in light curves or spectra that uniquely belongs to one type of model. 

Such a ``smoking-gun" feature is available from the dynamical evolution of the jet. As discussed above, a fireball has a rapid acceleration early on, and can only reduce kinetic energy (due to internal shock dissipation) at large radii from the central engine. A Poynting-flux-dominated jet with $\sigma > 1$ in the emission region, on the other hand, is still undergoing gradual acceleration when emission happens (Gao \& Zhang 2015). More importantly, if the emission is powered by an abrupt dissipation of the magnetic energy in the flow, part of the dissipated energy would be used to directly accelerate the flow (Drenkhahn \& Spruit 2002; Zhang \& Zhang 2014). As a result, {\em bulk acceleration during the emission phase is expected, which could be a ``smoking-gun" signature of a Poynting-flux-dominated outflow}.

Recently, Uhm \& Zhang (2015b) discovered the evidence of rapid bulk acceleration in X-ray flares following GRBs, based on an analysis of the ``curvature effect" of flares. The curvature effect is a well-studied effect to quantify the decay light curve of relativistic jet emission. When emission in a conical relativistic jet ceases abruptly (or decays sharply), the observed flux is controlled by emission from high-latitudes from the line of sight, which has a progressively lower Doppler factor, and hence, a lower flux. If the emission region moves with a constant Lorentz factor, there is a simple relationship between the decay index $\alpha$ and the spectral index $\beta$ (in the convention of $F_{\rm \nu}(t) \propto t^{-\alpha} \nu^{-\beta}$), which reads (Kumar \& Panaitescu 2000, see also Dermer 2004; Dyks et al. 2005; Uhm \& Zhang 2015a)
\begin{equation}
\alpha=2+\beta.
\label{curvature}
\end{equation}

This effect has been applied to interpret the decay segment of early X-ray afterglow light curve (Zhang et al. 2006), X-ray flares (Liang et al. 2006; Mu et al. 2015), or the decay tail of the GRB prompt emission (Qin et al. 2006; Jia 2008; Shenoy et al. 2013). Since the decay index $\alpha$ depends on the choice of the zero time point $t_0$, this curvature effect can be tested only if the $t_0$ effect is properly taken into account (Zhang et al. 2006). Uhm \& Zhang (2015a) recently found that the well-known relation in equation (1) is valid only if the relativistic spherical shell moves with a constant Lorentz factor $\Gamma$. The decay slope is steeper (shallower) than the value predicted by this formula if the emission region is undergoing acceleration (deceleration) when the emission ceases. Using the most conservative method of defining $t_0$, Uhm \& Zhang (2015b) showed that the decay slopes of three example X-ray flares (in GRBs 140108A, 110820A, and 090621A) are too steep to be accounted for by Eq.(1). They drew the conclusion that the emission regions of these three flares must be undergoing significant bulk acceleration when emission ceases. A detailed theoretical modeling of GRB 140108A flare confirmed that an accelerating emission region can simultaneously account for the light curve and spectral evolution of the flare. They then drew the conclusion that the jet composition of these three flares is Poynting-flux-dominated.

X-ray flares (Burrows et al. 2005) are observed in nearly half GRBs. They occur after the GRB prompt emission phase as late as $\sim 10^5$ seconds, show rapid rise and fall, and are consistent with the extension of prompt emission into the weaker and softer regime (Burrows et al. 2005; Chincarini et al. 2007, 2010; Margutti et al. 2010; Wang \& Dai 2013; Guidorzi et al. 2015; Troja et al. 2015; Yi et al. 2015).  Physically they are likely produced with the similar mechanism as the prompt emission, likely due to internal energy dissipation in a low-power wind at later epochs (e.g., Burrows et al. 2005; Fan \& Wei 2005; Zhang et al. 2006; Lazzati \& Perna 2007; Maxham \& Zhang 2009; Zhang et al. 2014).  Based on an energetic argument, Fan et al. (2005) suggested that X-ray flares in short GRBs have to be powered by a magnetized central engine.

Thanks to more than 10-year of observations of GRB early afterglow with the {\em Swift} satellite, plentiful of X-ray flare data have been collected. Some statistical studies of X-ray flares have been published in the past. Liang et al. (2006) first pointed out that the required $t_0$'s to interpret the decay phase of X-ray flares using the curvature effect model are associated with the flares, suggesting that X-ray flares most likely are triggered by late central engine activities. Chincarini et al. (2007) and Falcone et al. (2007) performed a detailed analysis of the statistical properties of X-ray flare light curves and spectra, respectively. Lazzati et al. (2008) investigated the global decay of X-ray flare luminosity as a function of time and made a connection to the accretion history of the black hole central engine. The possible connection between X-ray flares and prompt emission properties have been extensively studied (Falcone et al. 2006; Chincarini et al. 2010; Margutti et al. 2010; Guidorzi et al. 2015). Some efforts to estimate the properties of the X-ray flare outflows (e.g., Lorentz factor) have been carried out by various authors (Jin et al. 2010; Mu et al. 2015; Troja et al. 2015; Yi et al. 2015). By comparing the X-ray flare properties with those of solar flares, Wang \& Dai (2013) reached the conclusion that the underlying physical mechanism may be attributed to a self-organized criticality system, which is consistent with magnetic reconnection processes.

In this paper, we perform a systematic study of X-ray flares that significantly differs from previous studies. Prompted by the finding of Uhm \& Zhang (2015b) that some X-ray flares demand bulk acceleration in the emission region, we systematically analyze the decay properties of all the X-ray flares, aiming at addressing what fraction of X-ray flares demand bulk acceleration, and hence, a Poynting-flux-dominated outflow. Our approach closely follows that of Uhm \& Zhang (2015b).

The paper is organized as follows: A description of the sample selection criteria and data analysis methodology are presented in \S 2. A statistical study of the properties of various measured and/or derived parameters as well as their possible correlations are presented in \S 3. The conclusions are drawn in \S4 with some discussion.  
\section{Sample and Data Analysis}

We present an extensive temporal and spectral analysis for the X-ray flares observed with the X-Ray Telescope (XRT) onboard {\em Swift} over a span of 10 years from 2005 to 2014. The data are directly taken from the UK Swift Science Data Center at the University of Leicester (Evans et al. 2007, 2009) from the website http://www.swift.ac.uk/burst\_analyser/. For the light curves, we adopt the flux density light curve at 10 keV that is available from the light curve. This is because Eq.(1) is valid for flux density $F_\nu$. The XRT-band photon index ($\Gamma$) curves for each GRB are also directly downloaded, which can be used to derive the spectral index $\beta = \Gamma -1$. 

Since we mainly focus on the decay segment of the flares to investigate the curvature effect, we adopt the following two criteria to select the samples. First, we require that the decay tail is long enough, i.e., $F_{\nu, \rm p}/F_{\nu, \rm e} > 10$, where $F_{\nu, \rm p}$ and $F_{\nu, \rm e}$ are the flux densities of the flare at the peak and at the end of the flare, respectively. Second, we visually inspect the light curves to assure that the decay segment of the flares is clean, i.e., without superposition of other flares or significant fluctuations. Based on these two criteria, we finally come up with a sample of 85 flares detected in 63 different {\em Swift} GRBs\footnote{Our flare sample is therefore biased towards strong single flares. The conclusion drawn in this paper applies to all the flares as long as the overlapping flares and dimmer flares are not physically different from the bright single flares.}. They are listed in Table 1. Flares in a same GRB are distinguished by a number based on the sequence they appear in the GRB.

An important factor to correctly delineate the curvature effect is the so-called ``$t_0$ effect" (Zhang et al. 2006). Since afterglow light curves are plotted in the log-log scale, different zero time points would lead to different decay power laws. The convention of plotting afterglow light curves is that the GRB trigger time ($t_{\rm trigger}$) is defined as the zero time point. X-ray flares, on the other hand, mark distinct emission episodes from the prompt emission phase, which is likely caused by restarts of the central engine. If an X-ray flare starts at a new time $t_0$, the relevant decay slope should be $-d \ln F_\nu/ d \ln (t-t_0)$ rather than $-d \ln F_\nu/ d \ln (t-t_{\rm trigger})$.
Properly shifting $t_0$ is therefore essential to derive the correct temporal decay index $\alpha$ in the light curve of X-ray flares.

With a given $t_0$, one can perform a temporal fit to the X-ray flare light curve \footnote{We note that an X-ray flare can be in principle a superposition of multiple sub-flares. In some cases, the sub-flares have similar decay slopes as the main flare, but have low amplitudes, so that they cannot be identified with our fitting algorithm without very careful visual inspections. Some others have significant sub-flares with distinct decay indices. We have eliminated those cases from our flare sample.} with a smooth broken power-law function 
\begin{equation}
F_{\nu}(t)=F_{\nu, 0}\left[\left(\frac{t+t_0}{t_{b}+t_0}\right)^{\alpha_{1}\omega}+\left(\frac{t+t_0}{t_{b}+t_0}\right)^{\alpha_{2}\omega}\right]^{-\frac{1}{\omega}},
\label{eq:curve}
\end{equation}
where $\alpha_{1}$ and $\alpha_{2}$ are the temporal slopes for the rising part and the decaying part of the flare, respectively, $t_{b}$ is the break (peak) time, and $\omega$ measures the sharpness of the peak. An IDL subroutine named ``MPFITFUN.PRO"\footnote{http://cow.physics.wisc.edu/$\sim$craigm/idl/down/mpfitfun.pro} is employed during our fitting to the light curves of the flares, and a Levenberg-Marquardt least-square fit to the data for a given model is performed to optimize the model parameters. Through the fitting, one gets the best-fit parameters of the observed peak time ($t_{\rm p}^{\rm obs}$), the peak flux density ($F_{\rm \nu, p}$), and the observed temporal slopes for the rising part ($\alpha_{\rm 1}^{\rm obs}$) and the decaying part ($\alpha_{\rm 2}^{\rm obs}$). They are listed in Table 1. For those GRBs with redshift measurements (37 altogether), we also calculated the peak luminosity of the flares at 10 keV ($L_{\rm p}$) based on
\begin{equation}\label{eq:eiso}
  L_{\rm p}={4\pi D_{\rm L}^2} F_{\rm \nu, p} \nu_{\rm 10 keV},
\end{equation}
where $\nu_{\rm 10 keV}$ is the corresponding frequency for 10 keV, and $D_{\rm L}$ is the luminosity distance. A concordance cosmology with parameters $H_0 = 70$ km s$^{-1}$ Mpc$^{-1}$, $\Omega_M=0.30$, and $\Omega_{\Lambda}=0.70$ is adopted. The $z$ and $L_p$ results are also presented in Table 1.

Practically, the true $t_0$ for each flare is very difficult to find out. One needs detailed theoretical modeling in order to get a rough estimate of the true $t_0$ (Uhm \& Zhang 2015b). In any case, an {\em upper limit} to $t_0$ is available, which is the first rising data point of the flare. This point is usually defined by the background afterglow (power law component) flux level, and therefore is not intrinsic. Nonetheless, if one takes it as $t_0$, the decay light curve would be the shallowest among the possible allowed values. Uhm \& Zhang (2015b) found that even for these cases, the three X-ray flares showed even steeper light curves than the predicted decay based on Eq.(1).  They then concluded, based on the most conservative argument, that the emission regions of the three flares are undergoing significant bulk acceleration. 

In this paper, we adopt two possible $t_0$ values to perform our analysis. These two approaches are
\begin{itemize}
\item Method I: Take the first data point of a flare as the zero point, which we denote as $t_{0}^{\rm I}$. This is the most conservative approach;
\item Method II: It is almost certain that an X-ray flare starts earlier than the first observed point, with a flux (much) lower than the peak flux. Subject to uncertainties, our second method assumes that X-ray flares all start from a flux level that is three orders of magnitudes lower than the peak flux. Practically, we extrapolate the best fit curve based on Eq.(\ref{eq:curve}) downwards until $F_\nu$ is three orders of magnitudes lower than $F_{\nu,p}$, and define the corresponding time as $t_{0}^{\rm II}$.
\end{itemize}

In order to eliminate the influence of $t_0$ to the slope of the decaying part in X-ray flare, for both methods we shift the original observed light curve (left panel of each plot , and the blue light curve in the middle and right panel of each plot in Figure 1) to the left by subtracting $t_{0}$ in the horizontal axis (middle panel for Method I and right panel for Method II). The photon index curves are also shifted correspondingly (lower panels of each plot). For each new light curve, we perform a temporal fit to the light curve with a smooth broken power-law function 
\begin{equation}\label{SBPL}
F_{\nu}(t)=F_{\nu, 0}\left[\left(\frac{t}{t_{b}}\right)^{\alpha_{1}\omega}+\left(\frac{t}{t_{b}}\right)^{\alpha_{2}\omega}\right]^{-\frac{1}{\omega}},
\end{equation}
where $\alpha_{1}$, $\alpha_{2}$, $t_{b}$, and $\omega$ have the same meanings as those in equation (2), but are $t_0$-corrected. All the best-fit parameters are collected in Table 2, with the superscript ``I" or ``II" denoting for the Method I and II, respectively.

Following Uhm \& Zhang (2015b), next we compare the observed decay light curve ($t_0$ corrected) with the predicted decay light curve based on Eq.(1). Similar to the light curves, the photon index $\Gamma$ curves are also shifted by a corresponding $t_0$. By fitting the temporal evolution of $\beta = \Gamma-1$ as a function of $(t-t_0)$ (green curves in the middle/right bottom panels of each plot in Figure 1), we derive the time dependent decay slope $\hat\alpha_2$ predicted by the curvature effect based on Eq.(1) (also listed in Table 2). The corresponding predicted decay light curves (with arbitrary normalization) are marked as green curves in the middle/right upper panels of each plot in Figure 1. If this green curve is shallower than the data, the X-ray flare should be in the ``acceleration regime" (Uhm \& Zhang 2015b).
\section{Results and statistical analyses}

The criterion to judge whether the emission region is undergoing acceleration is to compare the measured decay slope of the flare after correcting $t_0$ (denoted as $\alpha_2^{\rm I}$ and $\alpha_2^{\rm II}$ for Methods I and II, respectively) with the predicted decay slope based on the curvature effect (denoted as $\hat\alpha_2^{\rm I}$ and $\hat\alpha_2^{\rm II}$ for the two methods, respectively). Let us define
\begin{equation}
\Delta \alpha = \alpha_2 - \hat\alpha_2.
\end{equation}
The X-ray flare is in the acceleration regime if $\Delta\alpha > 0$. 

Using Method I, we find that $\Delta \alpha^{\rm I}  =  \alpha_{\rm 2}^{\rm I}-\hat{\alpha}_{\rm 2}^{\rm I} > 0$ is satisfied for 56 out of 85 flares at the 99.9$\%$ condifence level. Since Method I uses the first observed point as $t_0$, the $t_0$-corrected light curve has the shallowest decay slope, so that the results are most conservative. This suggests that at least 56/85 X-ray flares are in the acceleration regime. The rest of flares have $\Delta \alpha^{\rm I} \leq 0$\footnote{$\Delta \alpha^{\rm I} \leq 0$ is defined if its central value is negative. Some of them can be consistent with zero within errors.}. However, it does not mean that the X-ray flares are in constant Lorentz factor motion or even in deceleration. Indeed, for Method II when we correct for a possibly more realistic (but likely still not accurate) $t_0$, more flares are in the acceleration regime\footnote{The X-ray flares of GRB 050502B and GRB 060904B already have the first data point deeper than $10^{-3} F_{\nu,p}$, so that Method II are irrelevant for these two flares. In any case, they are already in the acceleration regime.}. At the 99.9$\%$ confidence level, we find 74 out of 85 X-ray flares are in the acceleration regime. We inspect the remaining four flares (GRB 060607A-1, GRB 090407-2, GRB 090812-2 and GRB 140817A) closely, and find that the $t_0$-shifted light curves display significant fluctuations during the decay phase, likely due to overlapping small flares (see Figure 1 for details). This suggests that the decay phase is not controlled by the curvature effect, so that the possibility that these flares are also in the acceleration regime is not ruled out. Therefore, our results suggest that bulk acceleration is {\em ubiquitous} among X-ray flares.

In order to better understand X-ray flare physics, we perform a series of statistical analysis among various parameters. These parameters include directly-observed ones (the peak time $t_{\rm p}^{\rm obs}$ and the temporal slopes for the rising phase $\alpha_{\rm 1}^{\rm obs}$ and the decaying phase $\alpha_{\rm 2}^{\rm obs}$) and the measured or predicted parameters for $t_0$-shifted light curves  (e.g., the peak time $t_{\rm p}^{\rm I}$, the temporal slopes for the rising phase $\alpha_{\rm 1}^{\rm I}$ and the decaying phase $\alpha_{\rm 2}^{\rm I}$, the predicted decay slope $\hat{\alpha}_{\rm 2}^{\rm I}$ based on the simple curvature effect, derived from Method I; and the corresponding parameters, $t_{\rm p}^{\rm II}$, $\alpha_{\rm 1}^{\rm II}$, $\alpha_{\rm 2}^{\rm II}$, and $\hat{\alpha}_{\rm 2}^{\rm II}$, derived from Method II). Beyond these, we also define the following parameters: $\Delta \alpha^{\rm I}$ ($=\alpha_{\rm 2}^{\rm I}-\hat{\alpha}_{\rm 2}^{\rm I}$), $\Delta t_{\rm 1}^{\rm I}$ ($=t_{\rm p}^{\rm obs}-t_{0}^{\rm I}$), $\Delta t_{\rm 2}^{\rm I}$ ($=t_{\rm e}^{\rm I}-t_{\rm p}^{\rm obs}$) and $\Delta \alpha^{\rm II}$ ($=\alpha_{\rm 2}^{\rm II}-\hat{\alpha}_{\rm 2}^{\rm II}$), $\Delta t_{\rm 1}^{\rm II}$ ($=t_{\rm p}^{\rm obs}-t_{0}^{\rm II}$), $\Delta t_{\rm 2}^{\rm II}$ ($=t_{\rm e}^{\rm II}-t_{\rm p}^{\rm obs}$), corresponding to Method I and Method II, respectively. Here $t_{\rm e}^{\rm I}$ (defined as the last observed data point in the flare) and $t_{\rm e}^{\rm II}$ (defined by the time in the decaying segment when the flux is three orders of magnitudes lower than the peak flux) represent the ending data point of each flare for Method I and Method II, respectively. All these parameters are reported in Table 2.

X-ray flares are known to have a decreasing amplitude as a function of time (e.g., Chincarini et al. 2007, 2010; Lazzati et al. 2008; Margutti et al. 2010). In Fig.2a, we show the flare peak flux density at 10 keV as a function of the flare peak time in the observer frame ($t_p^{\rm obs}$). A general negative dependence (with a slope $-1.29\pm 0.24$, but with a poor correlation index $r=-0.134$) is seen. In Fig.2b, we show more intrinsic 10 keV peak luminosity as a function of rest-frame peak time ($t_p^{\rm obs}/(1+z)$). Again a negative dependence with large scatter is seen ($r=-0.129$), with a steeper slope $-1.95\pm 0.25$. For comparison,  Lazzati et al. (2008), Margutti et al. (2010) and Chincarini et al. (2010) all found that the average flare luminosity declines as a power law in time, but with somewhat different slopes, i.e. $-1.5\pm 0.16$,  $-2.7\pm 0.5$, and $-1.9\pm 0.1$, respectively. 

Another general trend found in previous studies is that in logarithmic scales, the flare relative width ($\Delta t/t_p$) seems to be universally distributed around 0.1 (Burrows et al. 2005; Chincarini et al. 2007, 2010; Margutti et al. 2010). This suggests that flares are narrow, and the rising and decaying slopes may not significantly depend on the epochs when they occur. On the other hand, if X-ray flares indeed require restarts of the central engine, then late flares are expected to appear narrower due to the wrong choice of their $t_0$ as the GRB trigger time (Zhang et al. 2006). It is therefore interesting to look into the possible ``narrowing" effect of flares as a function of time. In our sample, there are 22 GRBs that have 2 flares. In Figures 3(a) and 3(b) we display the observed rising ($\alpha_1^{\rm obs}$) and decaying ($\alpha_2^{\rm obs}$) slopes as a function of time. Those flares within a same burst are marked with the same color and connected with a line. The black points denote those flares that are single for a GRB. One can see that in most cases, the later flares seem to have both steeper rising and decaying slopes than their earlier counterparts, suggesting a possible $t_0$ effect. However, there are still a small fraction of flares that show an opposite trend. This suggests that the intrinsic distributions of the rising and decaying slopes of flares are wide, which would counter-balance the $t_0$ effect. In Fig.3c, we show the scatter plot of $\alpha_1^{\rm obs}$ and $\alpha_2^{\rm obs}$ for all the flares. A positive correlation (slope $0.28 \pm 0.06$) with large scatter ($r=0.414$) is seen. This is consistent with the $t_0$ effect, but again suggests a large scatter in the intrinsic distributions of the rising and decaying slopes among flares.  

Figure 4 shows the scatter plots of various times ($t_0$, rising time $\Delta t_1 = t_p^{\rm obs} - t_0$, and decaying time $\Delta t_2 = t_e - t_p^{\rm obs}$ for both Methods I and II) as a function of the observed peak time $t_p^{\rm obs}$, as well as $\Delta t_1 - \Delta t_2$ plots for both methods. As expected, $t_p^{\rm obs}$ sets an upper limit to $t_0$ (Fig.4I(a) and Fig.4II(a), red dashed line indicates the equality $t_0 = t_p^{\rm obs}$). Both the rising time and decaying time are positively correlated with $t_p^{\rm obs}$. This is another manifestation that $\Delta t/t_p \sim$ constant, as found in previous works. As expected, $\Delta t_1$ and $\Delta t_2$ are generally correlated, suggesting that the shape of flares in logarithmic scale essentially does not vary with time.

Figure 5 displays the scatter plots of the flare properties after the $t_0$ effect is corrected: $\alpha_1$ vs. $\alpha_2$ (Fig.5I(a) and Fig.5II(a)), $\alpha_1$ vs. $t_p$ (Fig.5I(b) and Fig.5II(b)), $\alpha_2$ vs. $t_p$ (Fig.5I(c) and Fig.5II(c)). As expected, no obvious correlations are found, and the scatter plots show the intrinsic scatter of the X-ray flare properties. In particular, the lack of the positive correlation between $\alpha_1$ and $\alpha_2$ (in contrast to the weak correlation between $\alpha_1^{\rm obs}$ and $\alpha_2^{\rm obs}$) is a strong indication of the $t_0$ effect as discussed above.

Figure 6 shows the dependences of $\Delta \alpha$ values on $t_p^{\rm obs}$, $t_p$, $\alpha_1$ and $\alpha_2$ for both Methods I and II. No significant correlations are found in other scatter plots, except a strong correlation between $\Delta \alpha$ and $\alpha_2$ in linear scale. This is understandable directly from the definition $\Delta \alpha = \alpha_2 - \hat\alpha_2$.  Within individual GRBs with multiple X-ray flares, some show decreasing $\Delta \alpha$ at later epochs. However, opposite trend can be seen in other bursts. Physically, different flares reflect different epochs of central engine activities. The degree of acceleration depends on the magnetization parameter of the outflow at that particular epoch. No first principle prediction is available regarding the evolution of magnetization of the GRB central engine.
\section{Conclusions and Discussion}
In this paper, we presented a statistical study of 85 bright X-ray flares following 63 GRBs detected by {\em Swift}, aiming at testing whether the emission regions of the flares are under bulk acceleration, based on the decay properties of the flares. Using two methods to estimate the beginning time of the flares ($t_0^{\rm I}$ as the beginning of the observed flare, and $t_0^{\rm II}$ as the time when the extrapolated flare flux is three orders of magnitude lower than the peak flux), we confirm the claim of Uhm \& Zhang (2015b) that X-ray flares usually undergo significant bulk acceleration. Even using the most conservative Method I, at the 99.9$\%$ confidence level we found that 56/85 flares are in the acceleration regime. With Method II, at the 99.9$\%$ confidence level 74/85 flares are in the acceleration regime. This suggests that bulk acceleration is a very common, may be even a ubiquitous property of X-ray flares.

Our results suggest that the composition of X-ray flare outflows are in the Poynting-flux-dominated regime. The emission of X-ray photons involves abrupt dissipation of Poynting-flux energy with $\sigma > 1$ in the emission region, with part of the dissipated energy directly used to accelerate the jet. Such a feature is a natural consequence of the ICMART model (Zhang \& Yan 2011; Zhang \& Zhang 2014). Since X-ray flares share many properties of GRB prompt emission pulses (Burrows et al. 2005; Chincarini et al. 2007, 2010; Margutti et al. 2010), it is natural to expect that the prompt emission of at least some GRBs is produced by the similar mechanism. This is consistent with other observational and theoretical arguments, e.g., the lack of or the weak quasi-thermal photosphere emission observed in most GRBs (Zhang \& Pe'er 2009; Guiriec et al. 2011, 2015; Axelsson et al. 2012; Burgess et al. 2014; Gao \& Zhang 2015), and the ability of optically-thin fast-cooling synchrotron emission to account for the spectra of GRBs with typical Band function parameters (Uhm \& Zhang 2014; Zhang et al. 2015)\footnote{GRBs with narrow spectra are dominated by bright photosphere emission, and likely have a matter-dominated jet composition (e.g., Ryde et al. 2010; Pe'er et al. 2012; Gao \& Zhang 2015; Yu et al. 2015).}. 

A decay index $\alpha$ steeper than $2+\beta$ may also be achieved by invoking anisotropic emission in the jet co-moving frame (e.g. Beloborodov et al. 2011; Hasco\"et et al. 2015). However, strong spectral evolution observed during both the rising and decaying phases of X-ray flares needs to be interpreted as well within such a scenario. The observational data, including both temporal and spectral properties of flares, are shown to be successfully reproduced within a simple physical model invoking synchrotron radiation from a rapidly-accelerating emission region, which points towards a Poynting-flux-dominated jet composition in X-ray flares (Uhm \& Zhang 2015b).
\begin{acknowledgements}
This work is supported by the National Basic Research Program of China (973 Program, grant No. 2014CB845800), the National Natural Science Foundation of China (Grant No. 11203008), the Excellent Youth Foundation of Guangdong Province (Grant No. YQ2015128), and the Guangzhou Education Bureau (Grant No. 1201410593). LWJ acknowledges a scholarship from China Scholarship Council (201408440204) for support. This work made use of data supplied by the UK Swift Science Data Center at the University of Leicester. 
\end{acknowledgements}

\clearpage
\begin{deluxetable}{lclclll}
\tabletypesize{\tiny}
\tablewidth{0pt}
\tablecaption{Observed data of 85 X-ray flares in our sample.}
\tablehead{
\colhead{flare name\tablenotemark{a}}&
\colhead{$z$\tablenotemark{b}}&
\colhead{$F_{\rm \nu, p}$ ($\mu$Jy @ 10 keV)\tablenotemark{c}}&
\colhead{$L_{\rm p}$ ($10^{48}$ erg/s)\tablenotemark{d}}&
\colhead{$t_{\rm p}^{\rm obs}$ (s)\tablenotemark{e}}&
\colhead{$\alpha_{\rm 1}^{\rm obs}$$\,$\tablenotemark{f}}&
\colhead{$\alpha_{\rm 2}^{\rm obs}$$\,$\tablenotemark{g}} 
}
\startdata
GRB 050502B	     & 5.2	       &$35.52\pm1.96$       &  $245.07\pm13.52$      & $708.10\pm1.75$  & $9.52\pm0.13$    & $17.84\pm0.33$  \\
GRB 050713A	     &  ...	       &$506.56\pm24.29$   &  ......                              & $107.17\pm0.21$  & $97.02\pm1.29$  & $35.35\pm0.52$  \\
GRB 050822 -1	     & 1.434      &$3.01\pm0.10$         &  $0.93\pm0.03$            & $238.18\pm1.99$  & $13.23\pm0.32$   & $31.36\pm0.31$  \\
GRB 050822 -2	     & 1.434      &$1.60\pm0.11$         &  $0.49\pm0.03$            & $424.29\pm1.06$  & $57.38\pm0.85$   & $45.03\pm0.42$  \\
GRB 060111A	     & 2.32	      &$120.28\pm1.08$     &  $120.44\pm1.08$        & $278.61\pm0.44$  & $34.81\pm0.63$   & $11.08\pm0.12$  \\
GRB 060204B -1   & ...	      &$500.15\pm18.94$   &  ......                              & $118.75\pm0.24$  & $70.02\pm1.50$   & $47.51\pm0.99$  \\
GRB 060204B -2   & ...	      &$26.23\pm1.89$       &  ......                              & $313.20\pm1.40$  & $69.36\pm1.37$   & $40.41\pm0.50$  \\
GRB 060210 -1     & 3.91	      &$125.60\pm2.52$     &   $443.91\pm8.89$       & $194.62\pm0.57$  & $35.10\pm0.37$   & $26.78\pm0.46$  \\
\enddata
\tablenotetext{a}{Flares are listed in chronological order by GRB date;}
\tablenotetext{b}{The redshift got from {\em http://www.mpe.mpg.de/$\sim$jcg/grbgen.html};}
\tablenotetext{c}{The peak flux density at 10 keV, in unit of $\mu$Jy;}
\tablenotetext{d}{The peak luminosity, in unit of $10^{48}$ erg/s;}
\tablenotetext{e}{The peak time of the observed light curve, in unit of s;}
\tablenotetext{f}{The slope for the rising part of X-ray flare;}
\tablenotetext{g}{The slope for the decaying part of X-ray flare.}
\end{deluxetable}
\clearpage
\newgeometry{left=0.5cm,right=2.5cm,top=1cm,bottom=0.5cm}
\begin{landscape}
\begin{deluxetable}{llcccccllcccccccc}
\setlength{\tabcolsep}{2pt}
\tabletypesize{\tiny}
\tablewidth{0pt}
\tablecaption{Parameters of 85 X-ray flares in our sample derived with our two methods.}
\tablehead{
\colhead{flare name\tablenotemark{a}}&
\colhead{$t_{0}^{\rm I} (s)$\tablenotemark{b}}&
\colhead{$t_{p}^{\rm I} (s)$\tablenotemark{c}}&
\colhead{$\alpha_{1}^{\rm I}$$\,$\tablenotemark{d}}&
\colhead{$\alpha_{2}^{\rm I}$$\,$\tablenotemark{e}}&
\colhead{$\hat{\alpha}_{2}^{\rm I}$$\,$\tablenotemark{f}}&
\colhead{$\Delta$$\alpha^{\rm I}$$\,$\tablenotemark{g}}&
\colhead{$\Delta$$t_{1}^{\rm I} (s)$$\,$\tablenotemark{h}}&
\colhead{$\Delta$$t_{2}^{\rm I} (s)$$\,$\tablenotemark{i}}&
\colhead{$t_{0}^{\rm II} (s)$\tablenotemark{j}}&
\colhead{$t_{p}^{\rm II}$ (s)\tablenotemark{k}}&
\colhead{$\alpha_{1}^{\rm II}$$\,$\tablenotemark{l}}&
\colhead{$\alpha_{2}^{\rm II}$$\,$\tablenotemark{m}}&
\colhead{$\hat{\alpha}_{2}^{\rm II}$$\,$\tablenotemark{n}}&
\colhead{$\Delta$$\alpha^{\rm II}$$\,$\tablenotemark{o}}&
\colhead{$\Delta$$t_{1}^{\rm II} (s)$\tablenotemark{p}}&
\colhead{$\Delta$$t_{2}^{\rm II} (s)$\tablenotemark{q}}
}
\startdata
GRB 050502B       & $137.84_{-69.60}^{+141.79}$  & 560.21$\pm$0.88    & 7.87$\pm$0.26    & 12.14$\pm$0.25   & 3.31$\pm$0.01   & 8.83$\pm$0.25     & $ 570.26_{-69.62}^{+141.80}$   & $ 420.16_{-23.61}^{+22.53}$    &             ......                             &           ......                    &             ......             &            ......                &            ......            &             ......            &            ......              &          ......                  \\
GRB 050713A       & $95.91_{-1.81}^{+1.72}$          & 12.59$\pm$0.29      & 3.66$\pm$0.28     & 4.44$\pm$0.10     & 3.28$\pm$0.05     & 1.15$\pm$0.11    & $ 11.27_{-1.82}^{+1.73}$            & $ 44.83_{-1.38}^{+1.48}$             & 92.25$\pm$0.36                     & 15.90$\pm$0.28          & 5.69$\pm$0.42       & 4.94$\pm$0.09        & 3.32$\pm$0.05      & 1.62$\pm$0.10      & 14.93$\pm$0.42      & 45.03$\pm$1.23      \\
GRB 050822 -1     & $198.38_{-2.05}^{+2.31}$        & 44.11$\pm$1.37      & 0.39$\pm$0.01     & 5.24$\pm$0.18      & 3.68$\pm$0.03     & 1.57$\pm$0.19     & $ 39.80_{-2.86}^{+3.05}$          & $ 47.37_{-7.13}^{+5.48}$             & 56.91$\pm$0.14                     & 176.19$\pm$1.46        & 7.90$\pm$0.28       & 13.97$\pm$0.45      & 3.70$\pm$0.03      & 10.27$\pm$0.45    & 181.27$\pm$2.00     & 116.23$\pm$2.37    \\
GRB 050822 -2     & $374.65_{-20.08}^{+17.52}$    & 49.83$\pm$1.13      & 4.66$\pm$0.26     & 4.55$\pm$0.21      & 4.59$\pm$0.03     & -0.05$\pm$0.22     & $ 49.64_{-20.11}^{+17.56}$      & $ 63.95_{-1.82}^{+1.77}$             & 330.60$\pm$0.37                  & 93.05$\pm$1.21          & 11.13$\pm$0.64     & 6.89$\pm$0.24        & 4.60$\pm$0.03       & 2.29$\pm$0.24      & 93.69$\pm$1.12       & 136.85$\pm$1.80    \\
\enddata
\tablenotetext{a}{Flares are listed in chronological order by GRB date;}
\tablenotetext{b}{The trigger time used in the light curve fitting, with Method I;}
\tablenotetext{c}{The peak time of the shifted light curve, with Method I;}
\tablenotetext{d}{The slope for the rising part of X-ray flare in the rest frame, with Method I;}
\tablenotetext{e}{The slope for the decaying part of X-ray flare in the rest frame, with Method I;}
\tablenotetext{f}{The slope of the curvature effect curve in the rest frame, with Method I;}
\tablenotetext{g}{One of the derived quantities, $\Delta \alpha^{\rm I}=\alpha_{\rm 2}^{\rm I}-\hat{\alpha}_{\rm 2}^{\rm I}$;}
\tablenotetext{h}{One of the derived quantities, $\Delta t_{\rm 1}^{\rm I}=t_{\rm p}^{\rm obs}-t_{0}^{\rm I}$;}
\tablenotetext{i}{One of the derived quantities, $\Delta t_{\rm 2}^{\rm I}=t_{\rm e}^{\rm I}-t_{\rm p}^{\rm obs}$;}
\tablenotetext{j}{The trigger time used in the light curve fitting, with Method II;}
\tablenotetext{k}{The peak time of the shifted light curve, with Method II;}
\tablenotetext{l}{The slope for the rising part of X-ray flare in the rest frame, with Method II;}
\tablenotetext{m}{The slope for the decaying part of X-ray flare in the rest frame, with Method II;}
\tablenotetext{n}{The slope of the curvature effect curve in the rest frame, with Method II;}
\tablenotetext{o}{One of the derived quantities, $\Delta \alpha^{\rm II}=\alpha_{\rm 2}^{\rm II}-\hat{\alpha}_{\rm 2}^{\rm II}$;}
\tablenotetext{p}{One of the derived quantities, $\Delta t_{\rm 1}^{\rm II}=t_{\rm p}^{\rm obs}-t_{0}^{\rm II}$;}
\tablenotetext{q}{One of the derived quantities, $\Delta t_{\rm 2}^{\rm II}=t_{\rm e}^{\rm II}-t_{\rm p}^{\rm obs}$.}
\end{deluxetable}
\end{landscape}
\restoregeometry
\clearpage
\setlength{\voffset}{-18mm}
\begin{figure*}\centering
\includegraphics[angle=0,scale=0.99,width=0.99\textwidth,height=0.54\textheight]{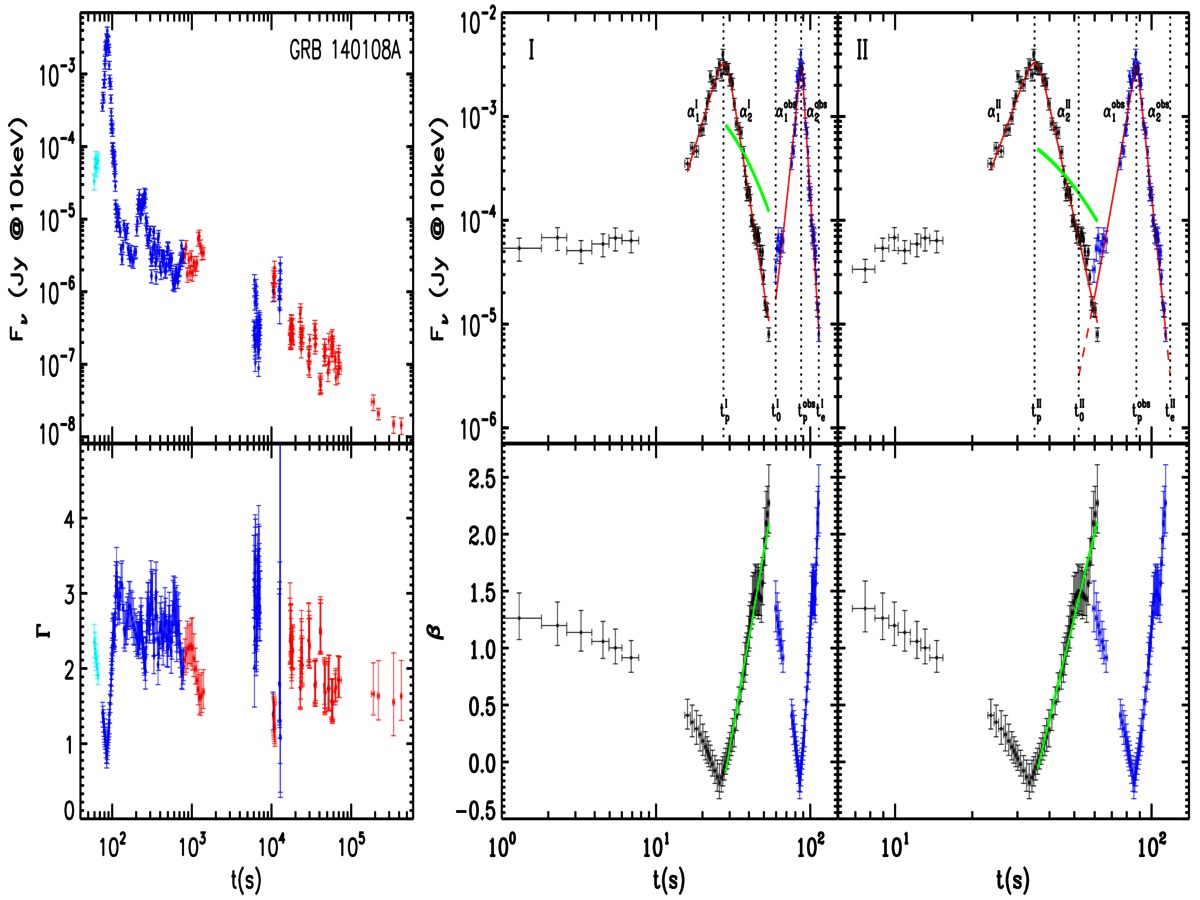}
\caption{\small Lightcurves of the original and $t_0$-corrected flares (upper panel of each plot) and their spectral evolution (lower panel of each plot, photon index $\Gamma$ for the lower-left panel and spectral index $\beta$ for the lower-middle and lower-right panels of each plot) for all the 85 flares studied in our paper. The original data are presented in the left panel of each plot, and are directly taken from the website of the UK Swift Science Data Center at the University of Leicester ({\em http://www.swift.ac.uk/burst\_analyser/docs.php\#usage}). The flare-only light curves and the spectral index $\beta$ evolution curves derived using Method I and II are presented in the middle (denoted as ``I") and right (denoted as ``II") panels of each plot, respectively. Method II for GRBs 050502B and 060904B is irrelevant, so that their panel ``II" is missing. For the light curves, the observed and the $t_0$-corrected light curves are represented by blue and black data points, respectively. The red solid curves are broken-power-law fits to the two light curves. Dashed red curves are the extension of the best fit curve in Method II. In the bottom panel, the $t_0$-shifted $\beta$ curve is fitted by a green curve, which is used to give predicted decaying light curve (with arbitrary normalization) (green curve in the upper panel). The vertical dotted lines mark various characteristic times: the peak time of the shifted light curve ($t_{\rm p}^{\rm I}$ and $t_{\rm p}^{\rm II}$ for Method I and II, respectively), the starting time ($t_{\rm 0}^{\rm I}$ and $t_{\rm 0}^{\rm II}$ for Method I and Method II, respectively), the observed peak time $t_{\rm p}^{\rm obs}$, and the ending time ($t_{\rm e}^{\rm I}$ and $t_{\rm e}^{\rm II}$ for Method I and II, respectively) of the light curves for each flare.}
\end{figure*}
\begin{figure*}\centering
\includegraphics[angle=0,scale=0.99,width=0.99\textwidth,height=0.75\textheight]{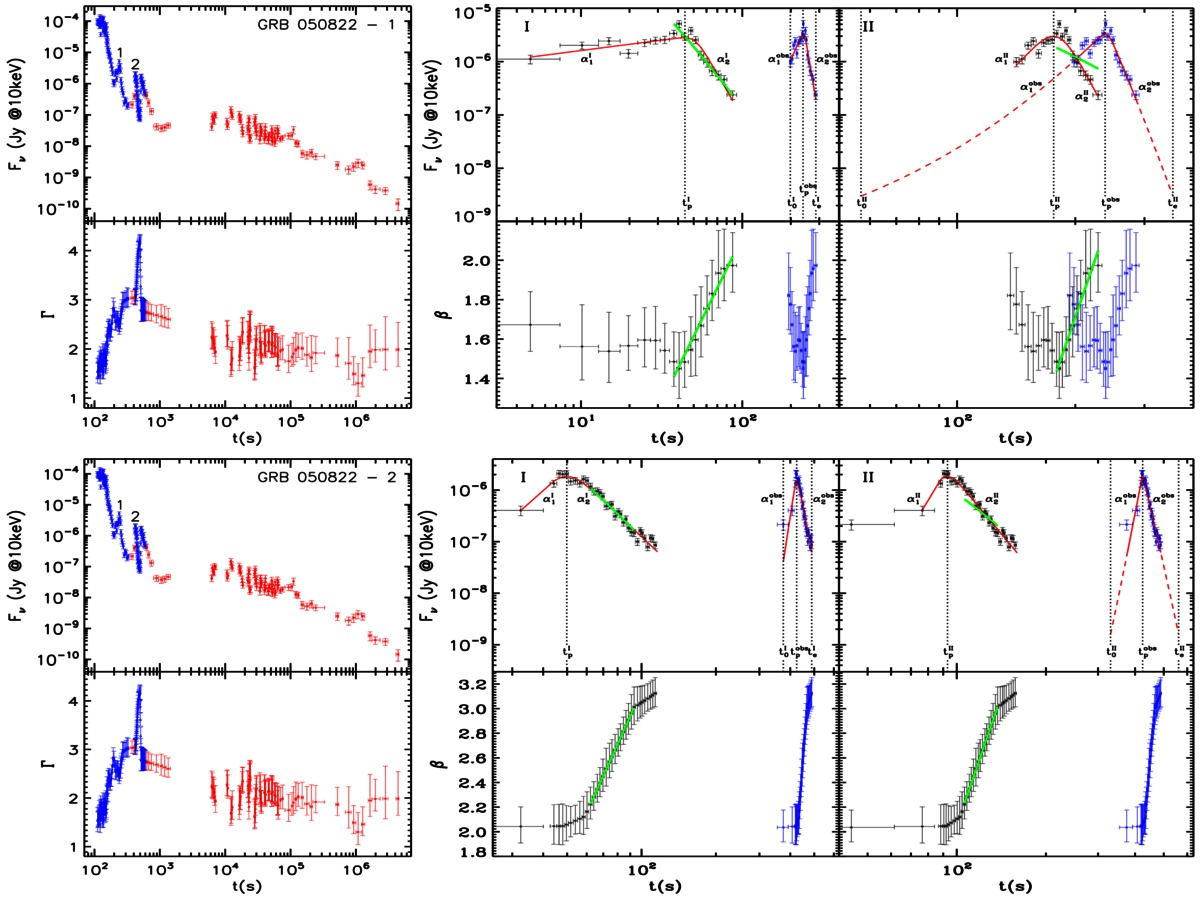}
\center{Fig. 1--- Continued}
\end{figure*}
\begin{figure*}\centering
\includegraphics[angle=0,scale=0.99,width=0.99\textwidth,height=0.75\textheight]{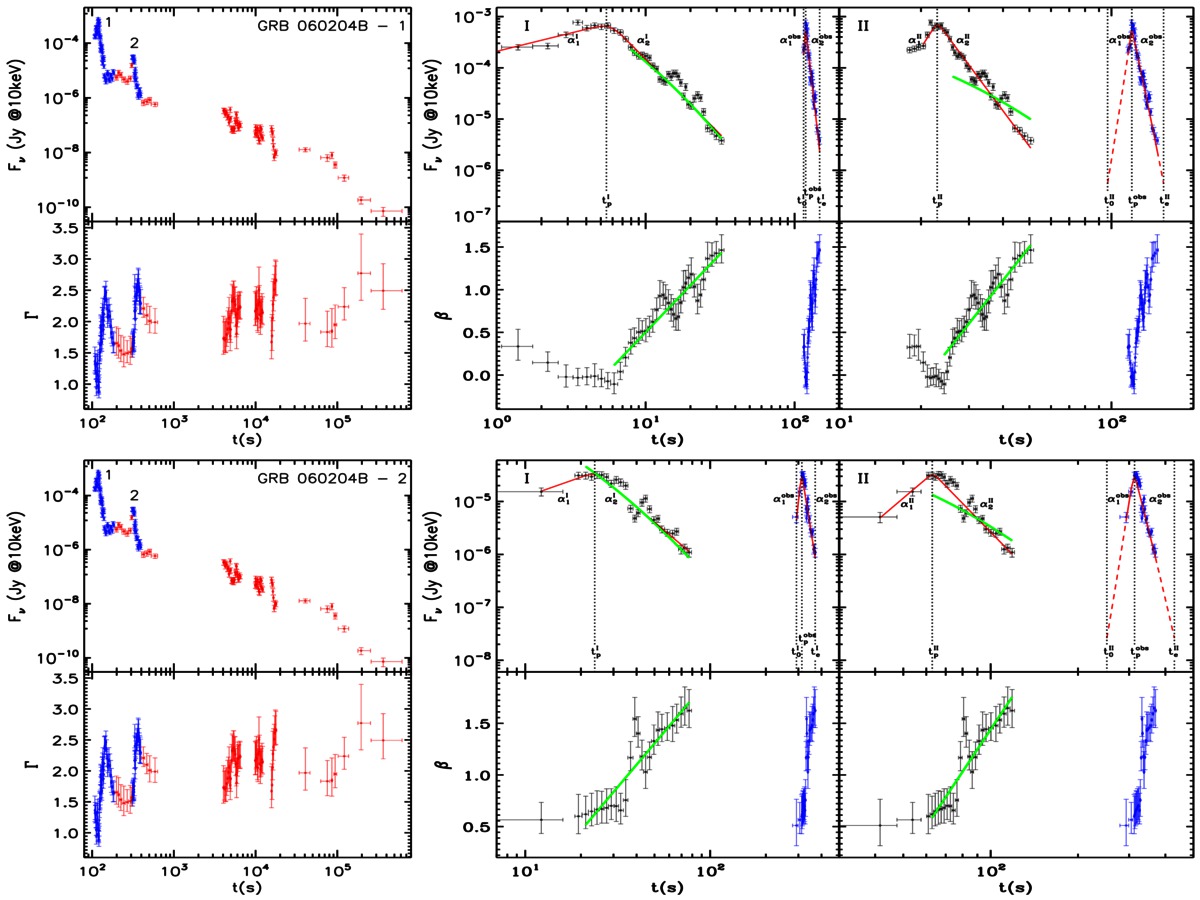}
\center{Fig. 1--- Continued}
\end{figure*}
\begin{figure*}\centering
\includegraphics[angle=0,scale=0.99,width=0.99\textwidth,height=0.75\textheight]{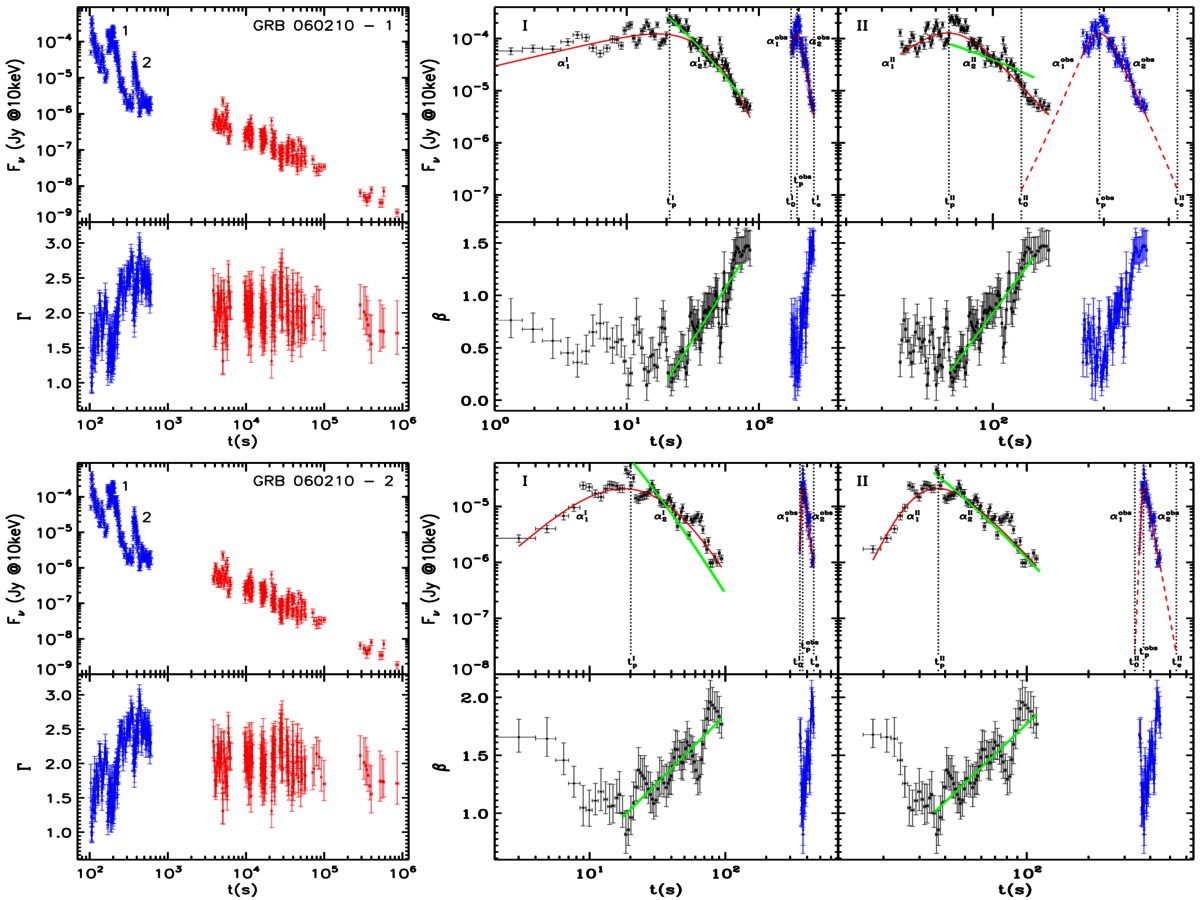}
\center{Fig. 1--- Continued}
\end{figure*}
\begin{figure*}\centering
\includegraphics[angle=0,scale=0.99,width=0.99\textwidth,height=0.75\textheight]{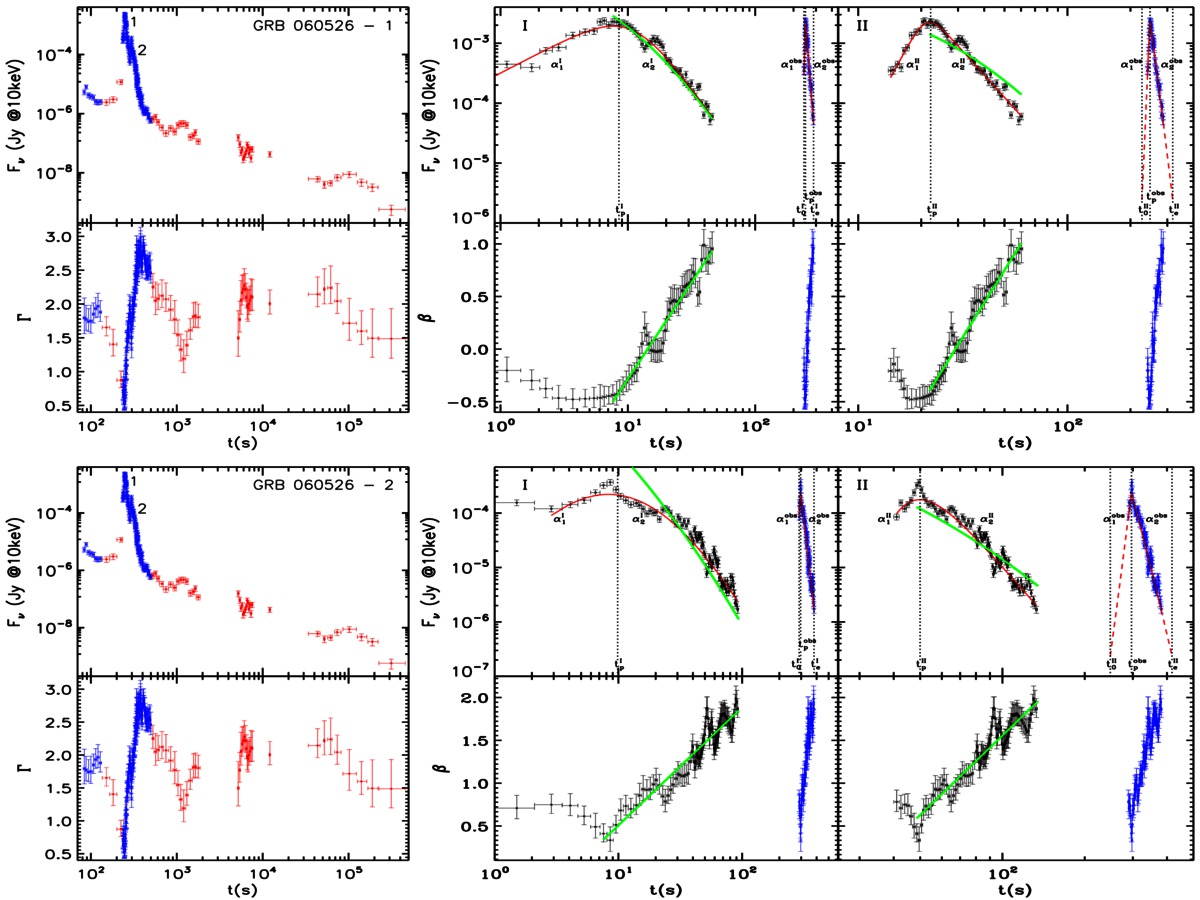}
\center{Fig. 1--- Continued}
\end{figure*}
\begin{figure*}\centering
\includegraphics[angle=0,scale=0.99,width=0.99\textwidth,height=0.75\textheight]{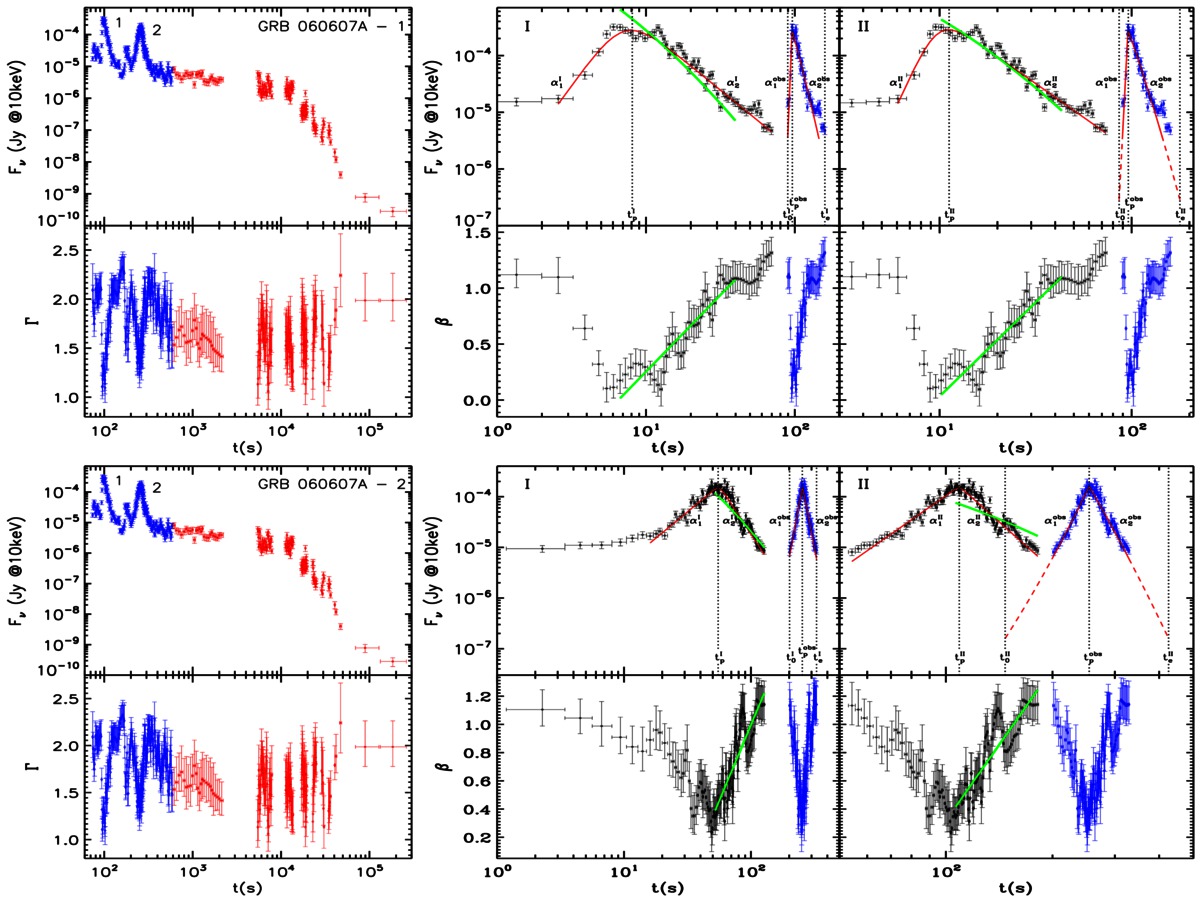}
\center{Fig. 1--- Continued}
\end{figure*}
\begin{figure*}\centering
\includegraphics[angle=0,scale=0.99,width=0.99\textwidth,height=0.75\textheight]{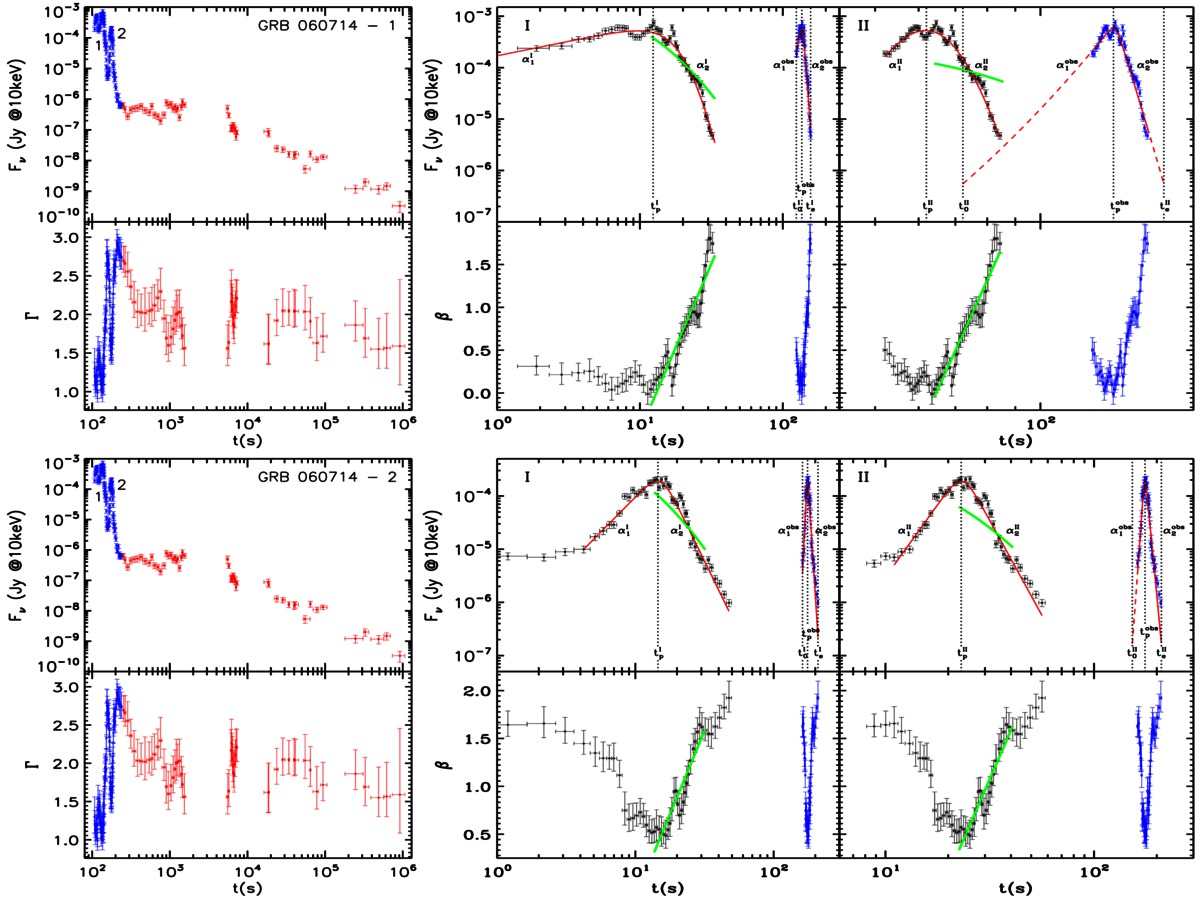} 
\center{Fig. 1--- Continued}
\end{figure*}
\begin{figure*}\centering
\includegraphics[angle=0,scale=0.99,width=0.99\textwidth,height=0.75\textheight]{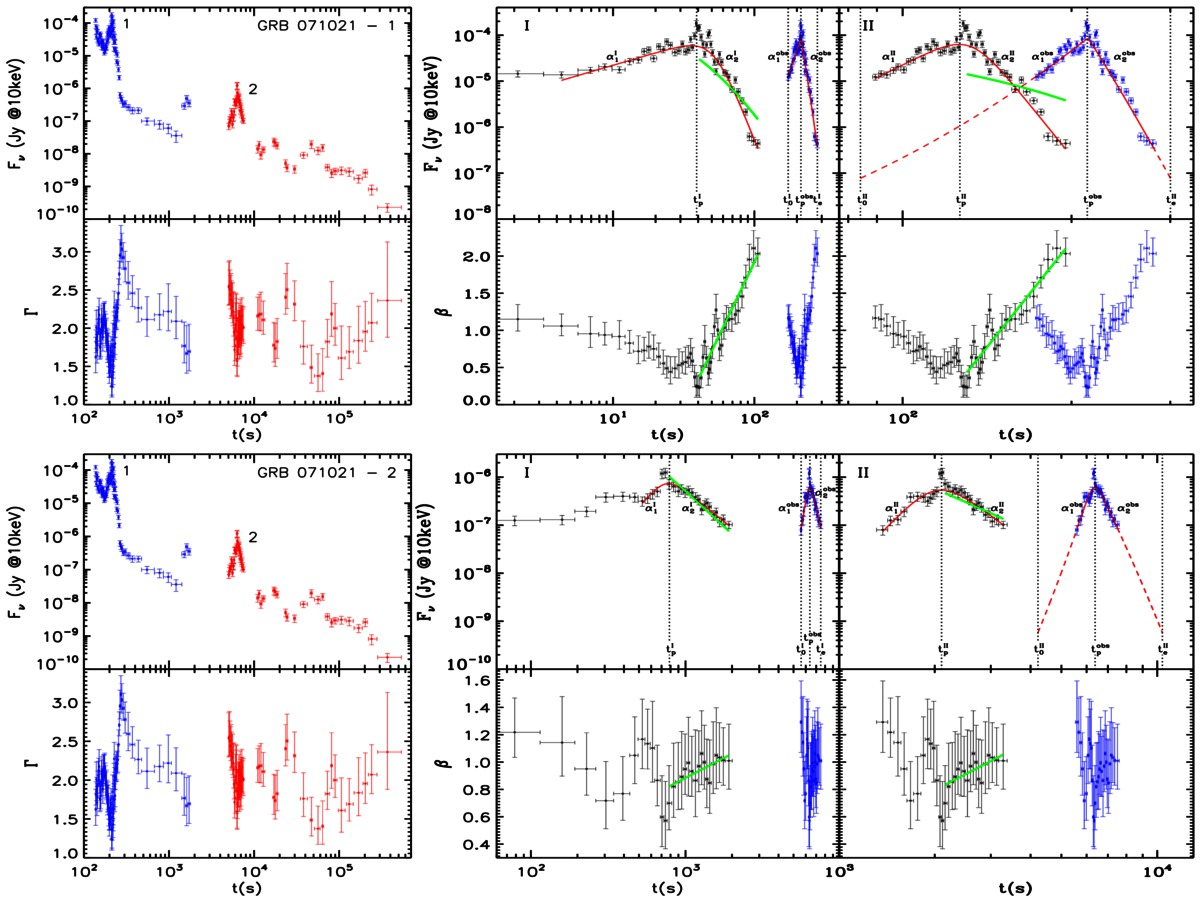}
\center{Fig. 1--- Continued}
\end{figure*}
\begin{figure*}\centering
\includegraphics[angle=0,scale=0.99,width=0.99\textwidth,height=0.75\textheight]{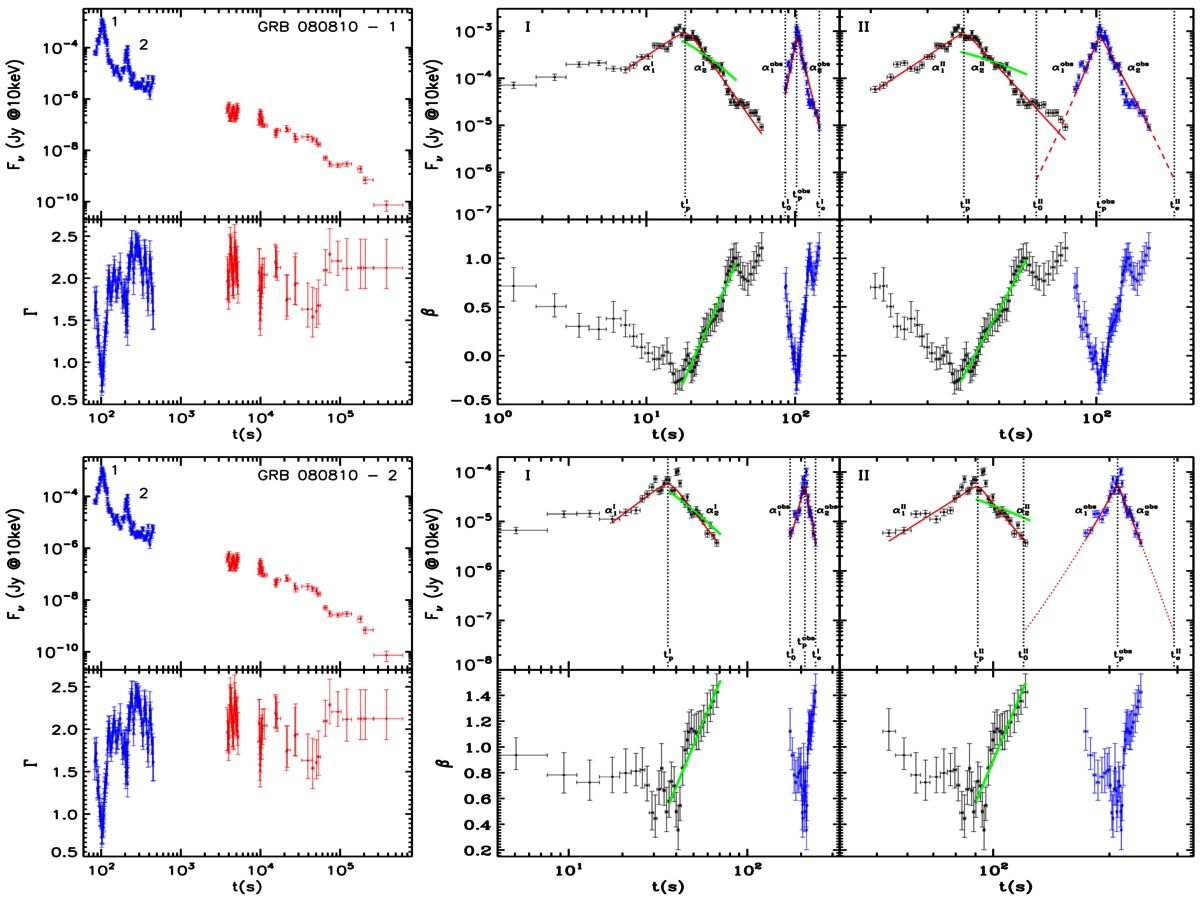}
\center{Fig. 1--- Continued}
\end{figure*}
\begin{figure*}\centering
\includegraphics[angle=0,scale=0.99,width=0.99\textwidth,height=0.75\textheight]{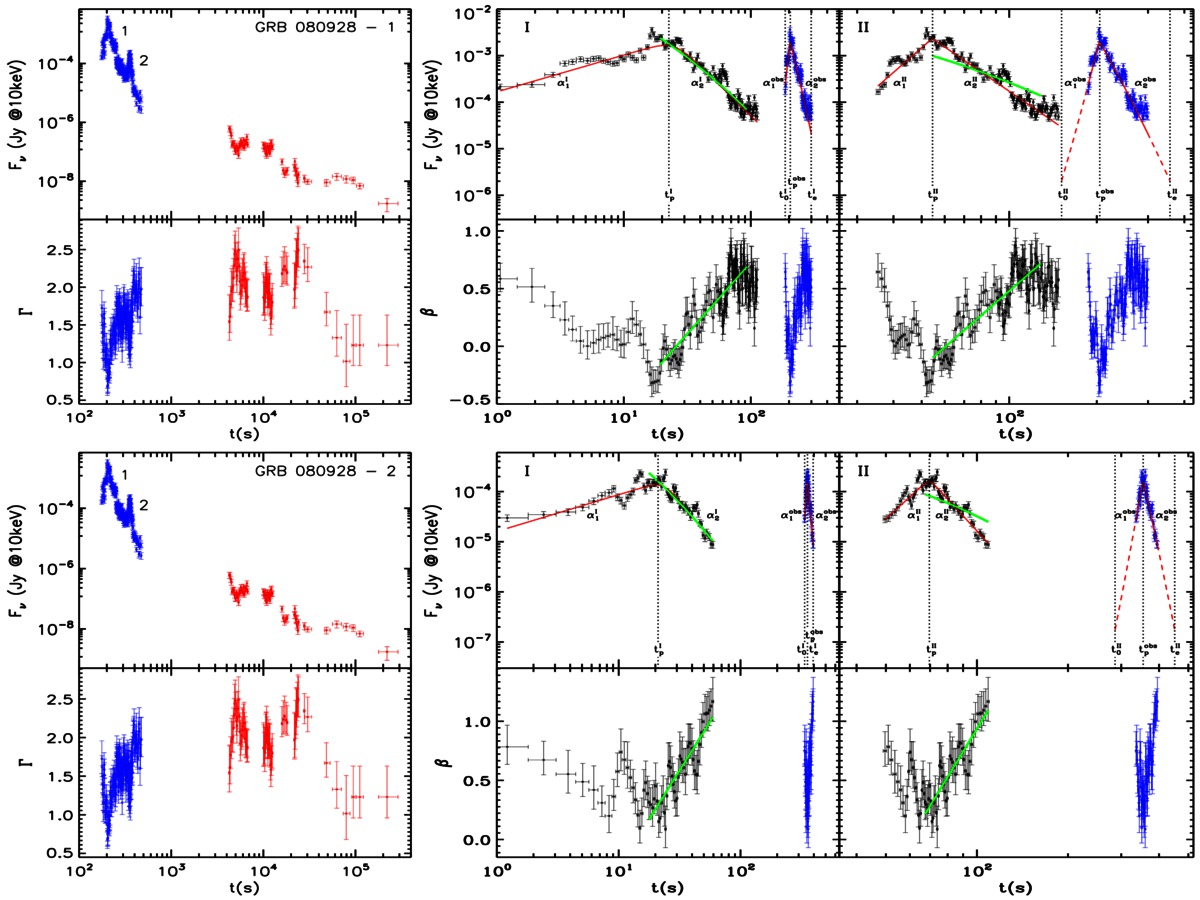}
\center{Fig. 1--- Continued}
\end{figure*}
\begin{figure*}\centering
\includegraphics[angle=0,scale=0.99,width=0.99\textwidth,height=0.75\textheight]{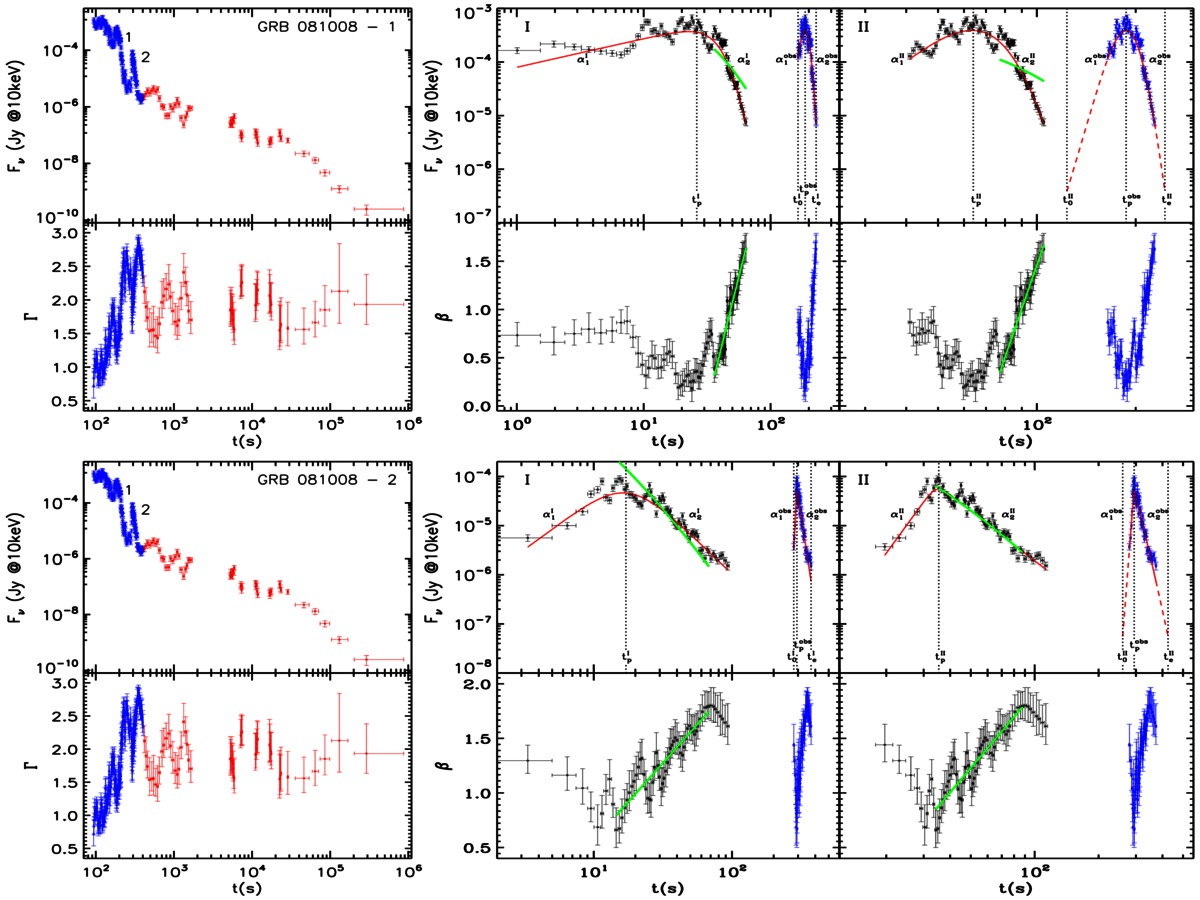}
\center{Fig. 1--- Continued}
\end{figure*}
\begin{figure*}\centering
\includegraphics[angle=0,scale=0.99,width=0.99\textwidth,height=0.75\textheight]{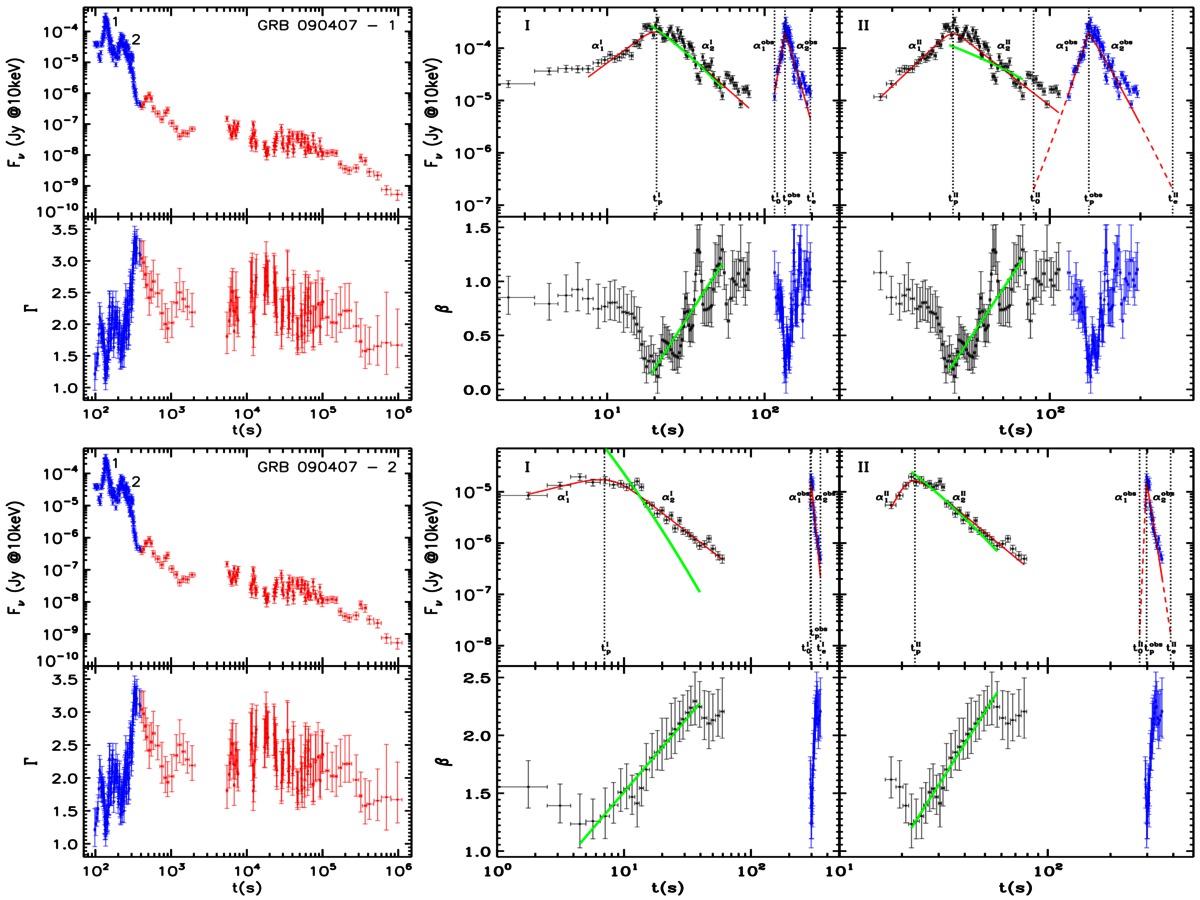}
\center{Fig. 1--- Continued}
\end{figure*}
\begin{figure*}\centering
\includegraphics[angle=0,scale=0.99,width=0.99\textwidth,height=0.75\textheight]{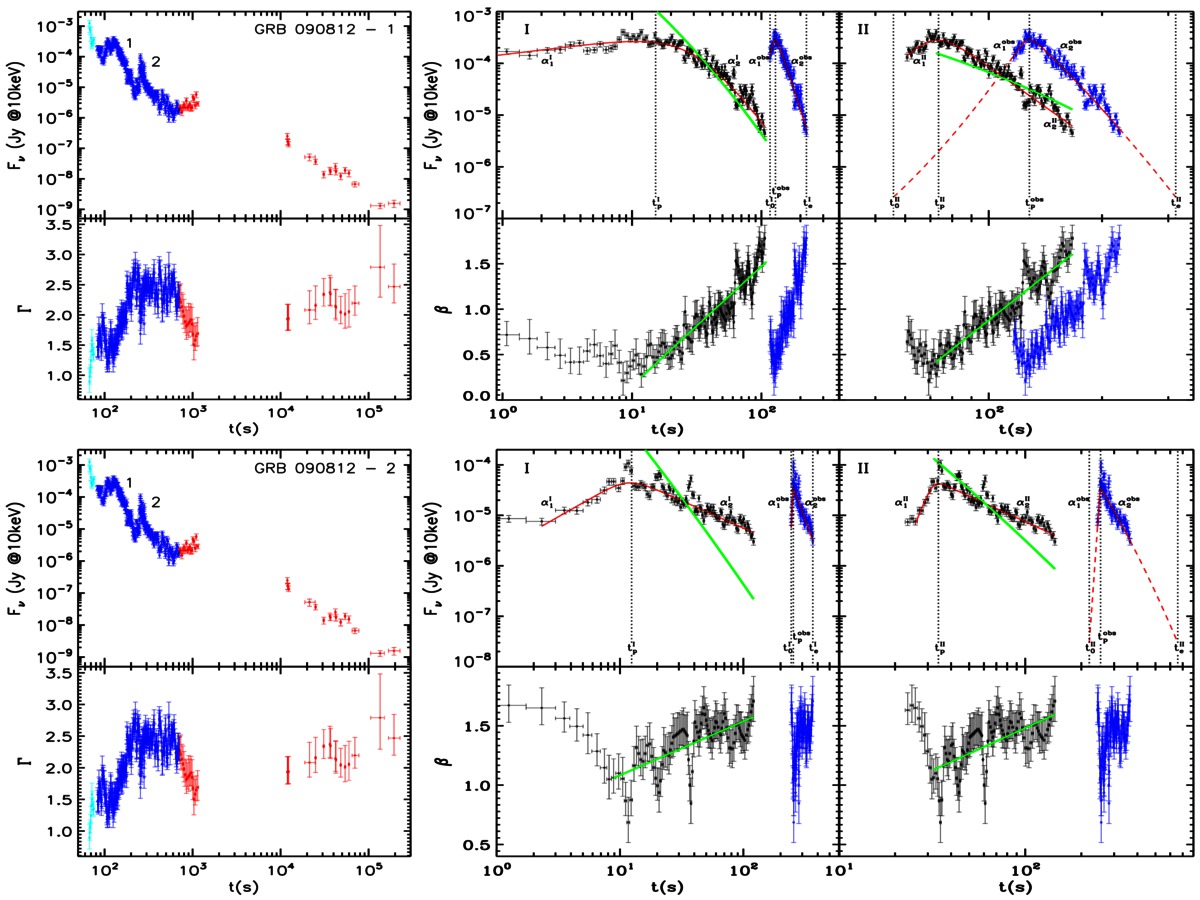}
\center{Fig. 1--- Continued}
\end{figure*}
\begin{figure*}\centering
\includegraphics[angle=0,scale=0.99,width=0.99\textwidth,height=0.75\textheight]{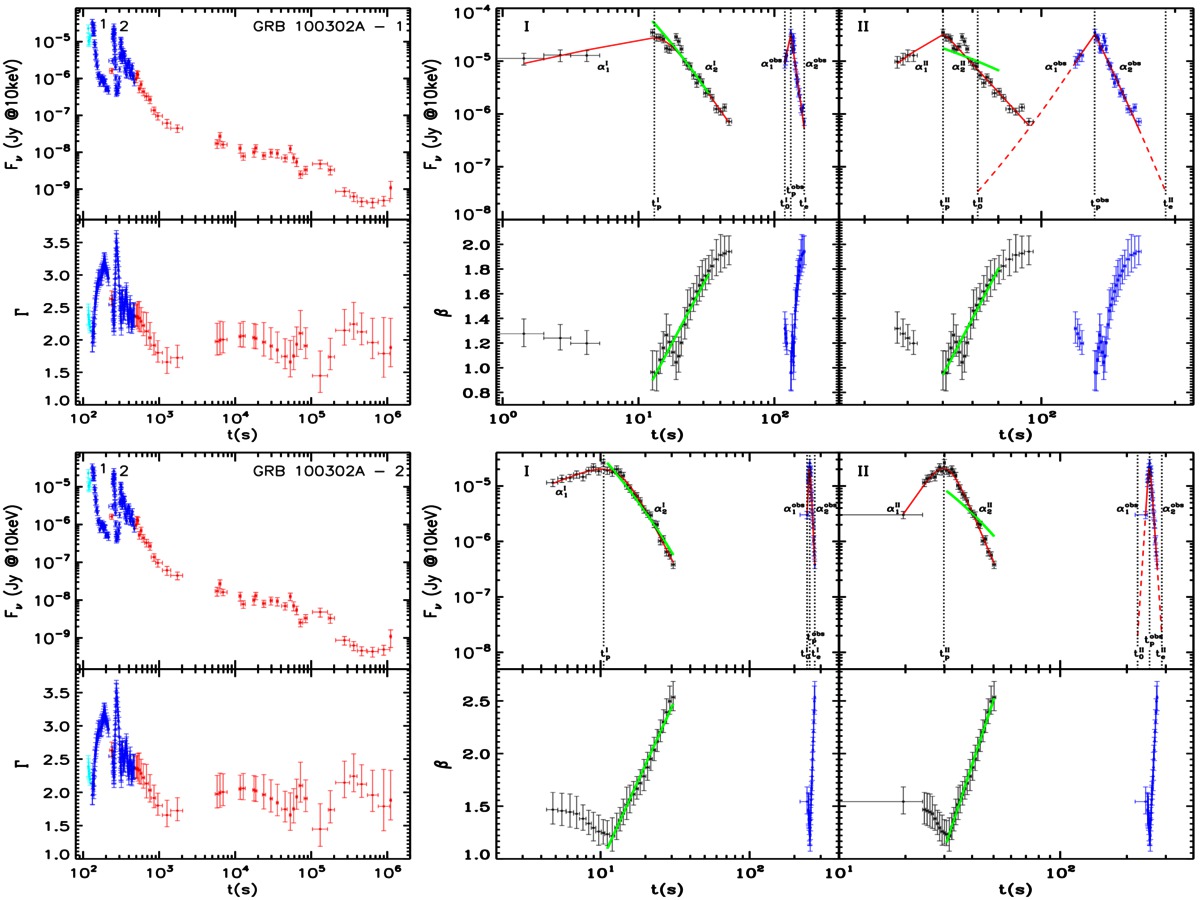}
\center{Fig. 1--- Continued}
\end{figure*}
\begin{figure*}\centering
\includegraphics[angle=0,scale=0.99,width=0.99\textwidth,height=0.75\textheight]{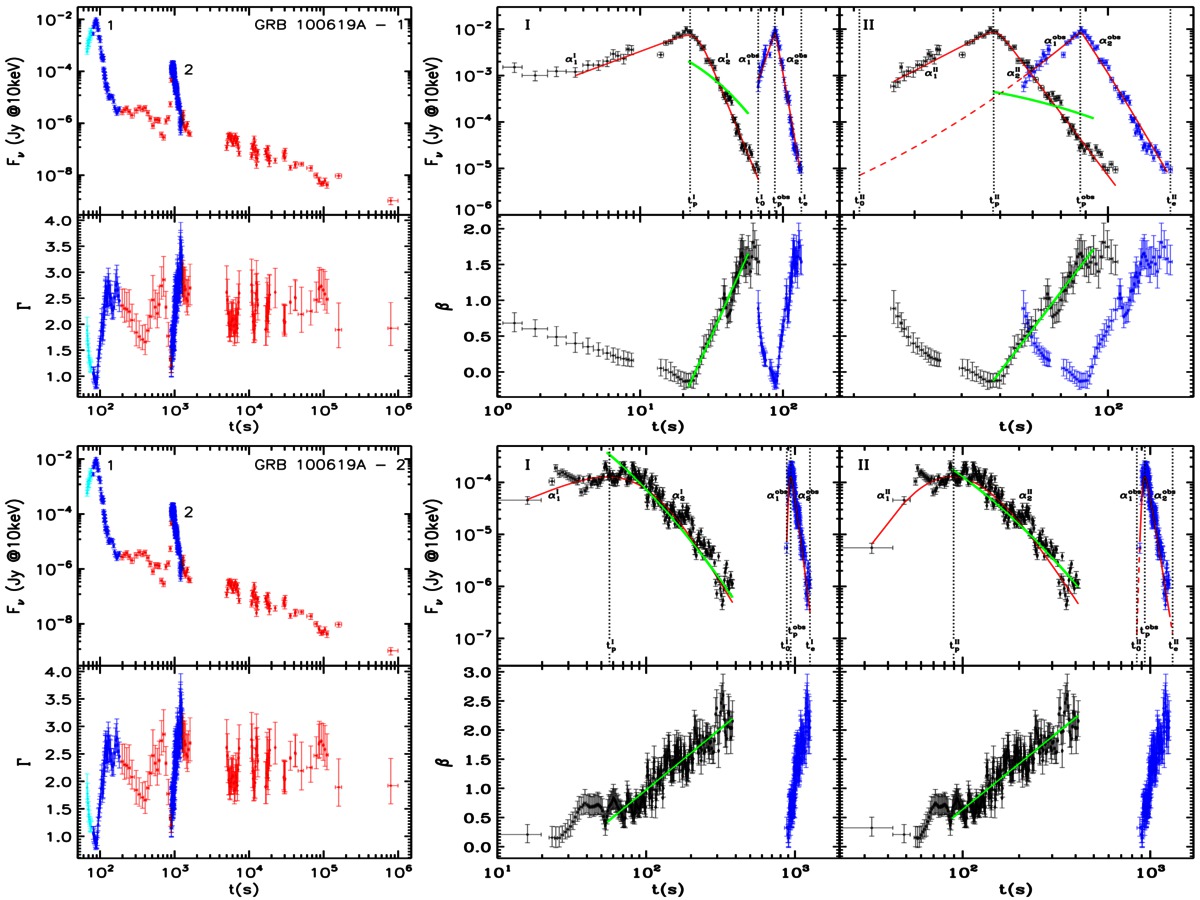}
\center{Fig. 1--- Continued}
\end{figure*}
\begin{figure*}\centering
\includegraphics[angle=0,scale=0.99,width=0.99\textwidth,height=0.75\textheight]{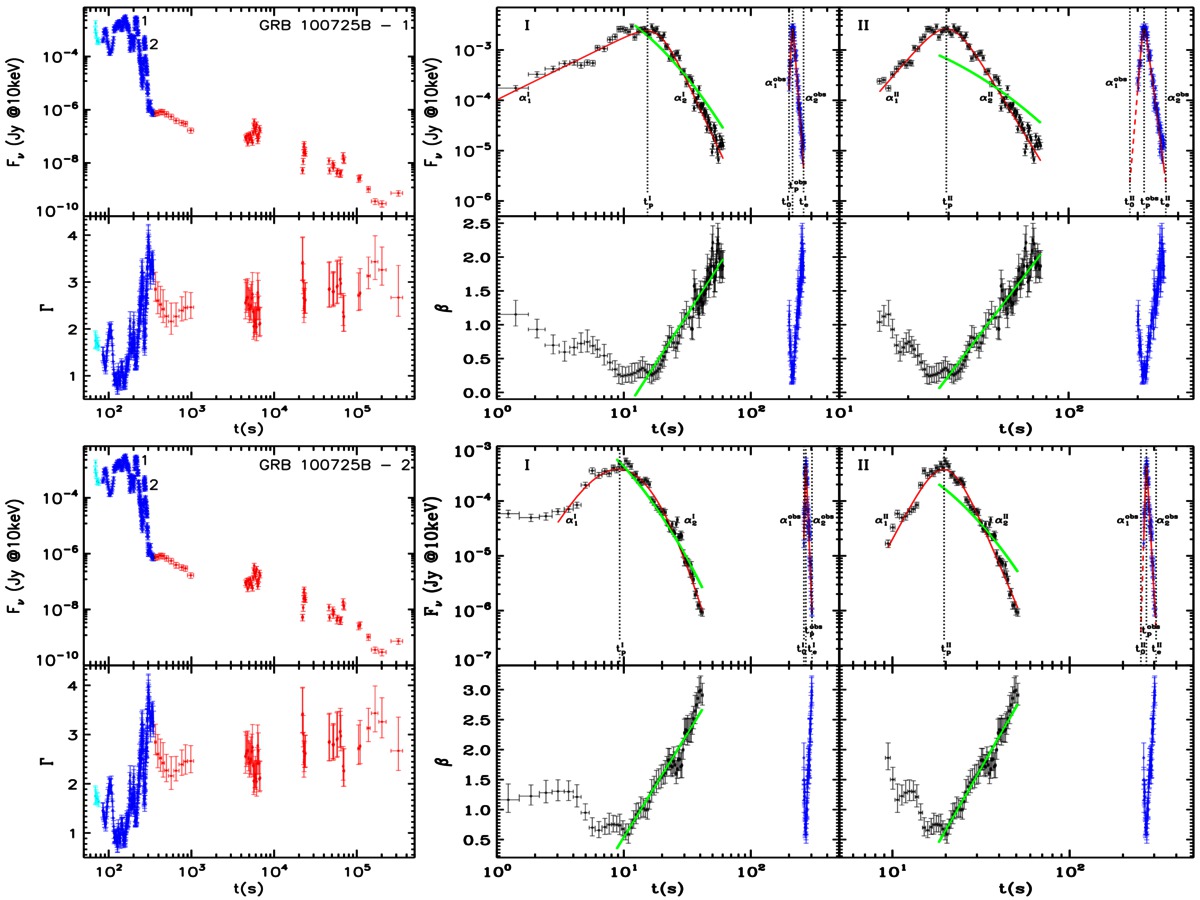}
\center{Fig. 1--- Continued}
\end{figure*}
\begin{figure*}\centering
\includegraphics[angle=0,scale=0.99,width=0.99\textwidth,height=0.75\textheight]{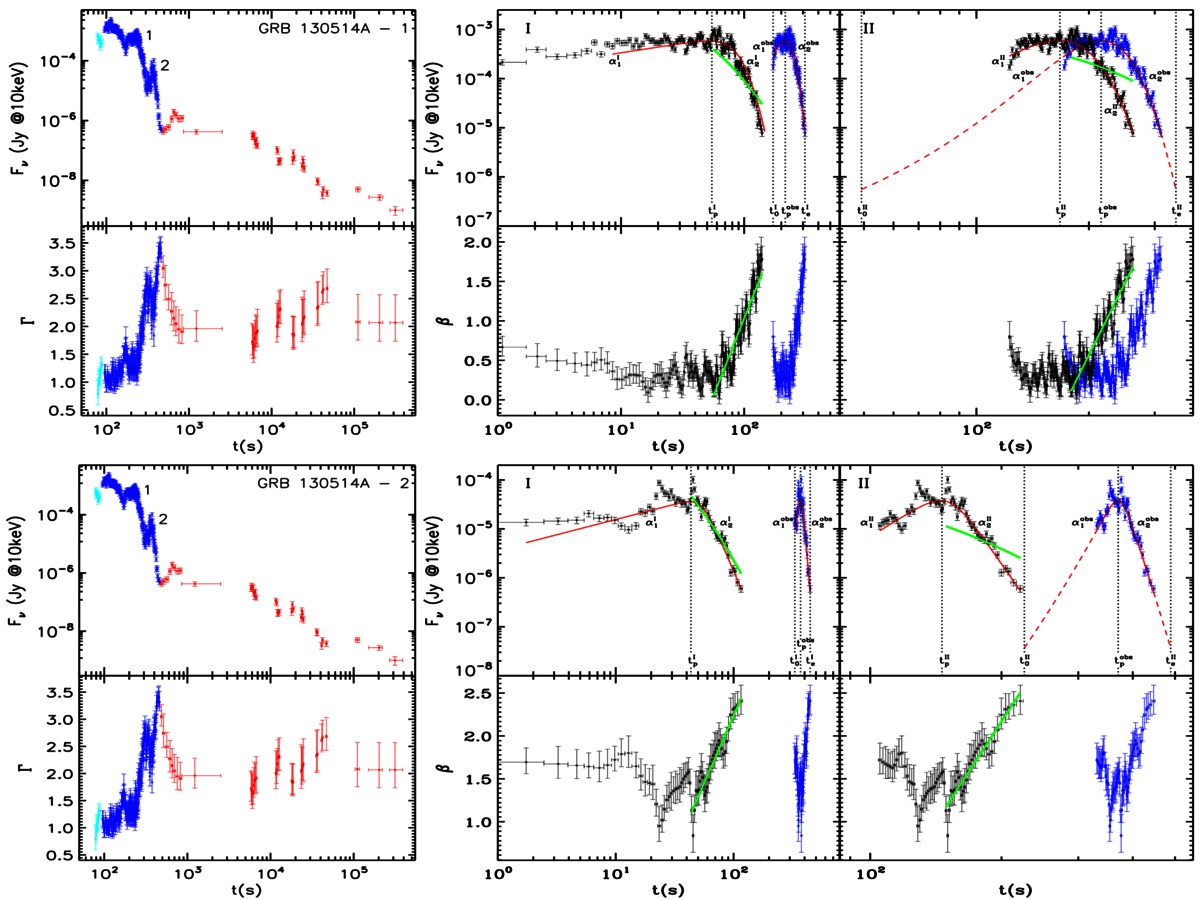}
\center{Fig. 1--- Continued}
\end{figure*}
\begin{figure*}\centering
\includegraphics[angle=0,scale=0.99,width=0.99\textwidth,height=0.75\textheight]{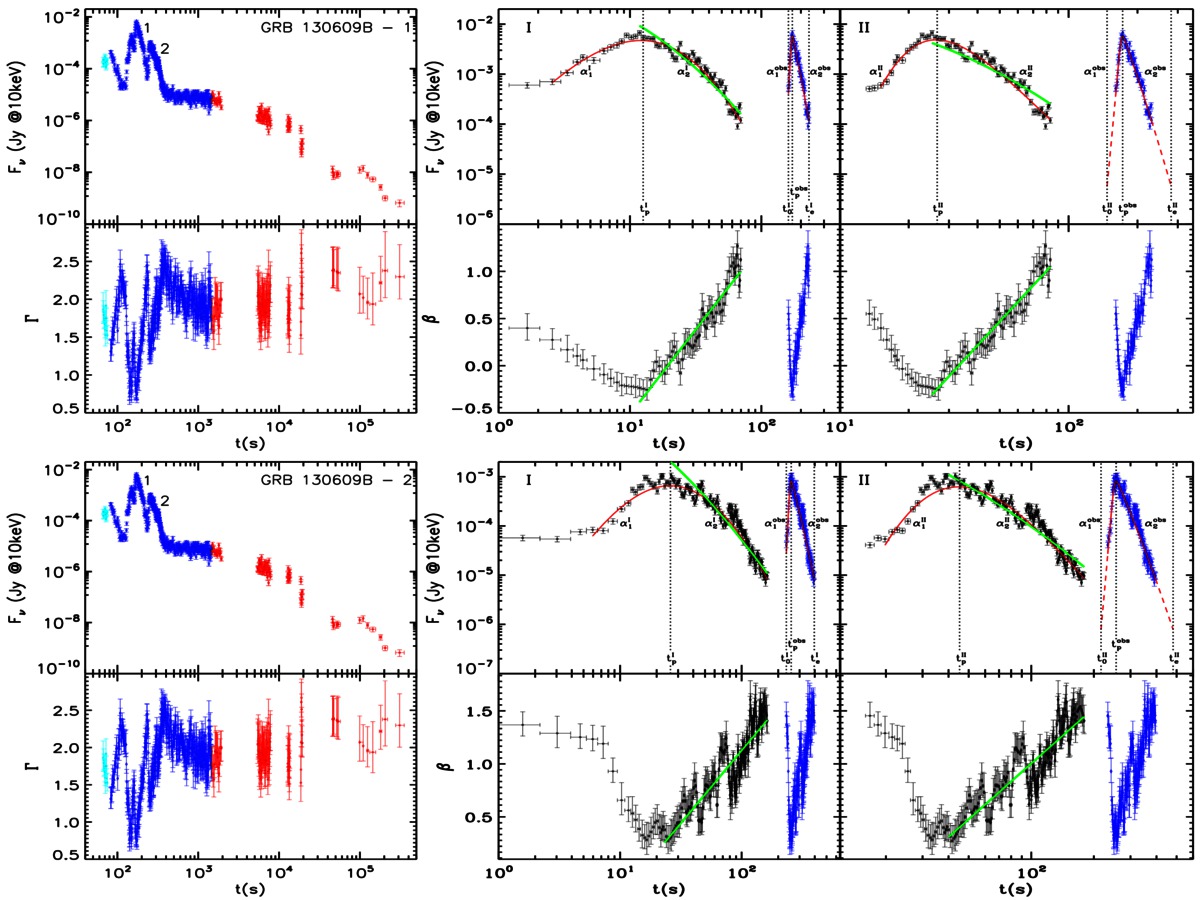}
\center{Fig. 1--- Continued}
\end{figure*}
\begin{figure*}\centering
\includegraphics[angle=0,scale=0.99,width=0.99\textwidth,height=0.75\textheight]{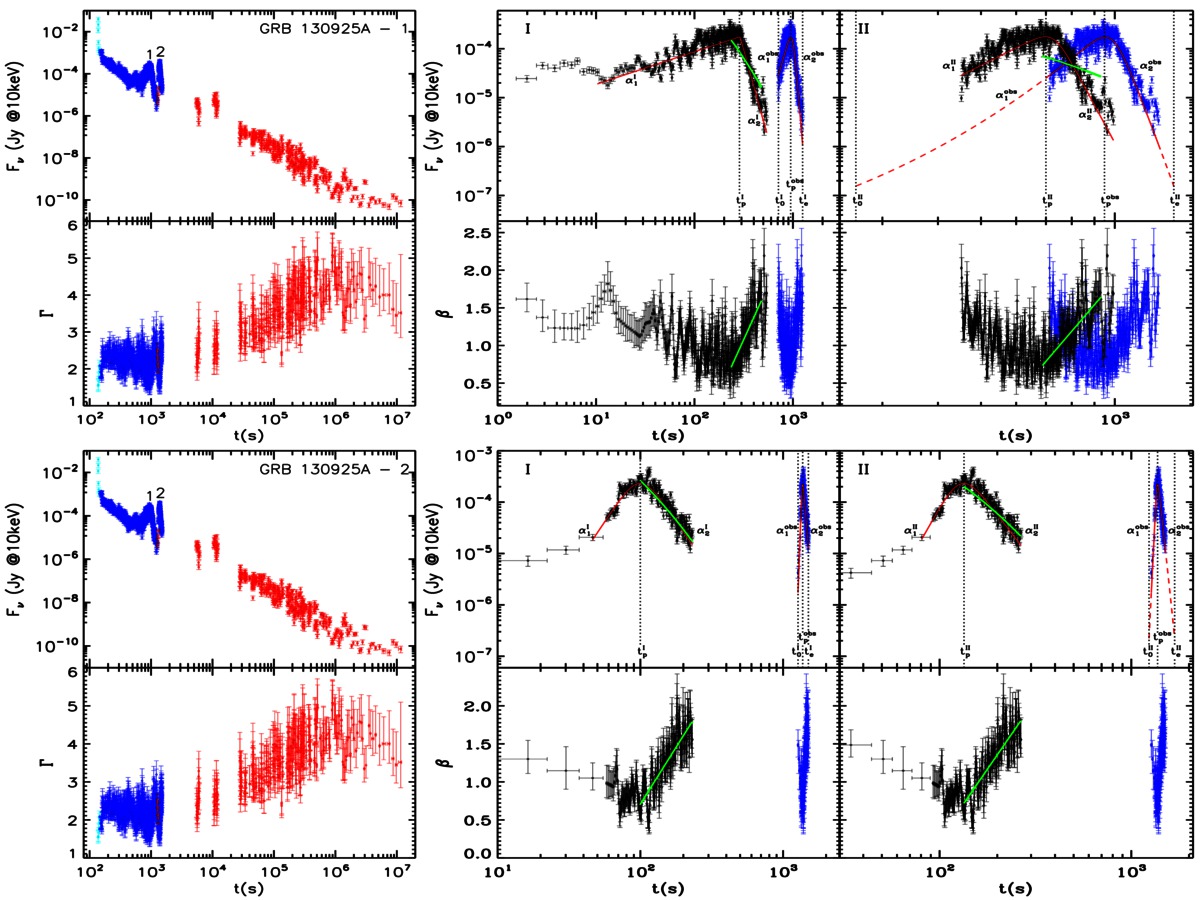}
\center{Fig. 1--- Continued}
\end{figure*}
\begin{figure*}\centering
\includegraphics[angle=0,scale=0.99,width=0.99\textwidth,height=0.75\textheight]{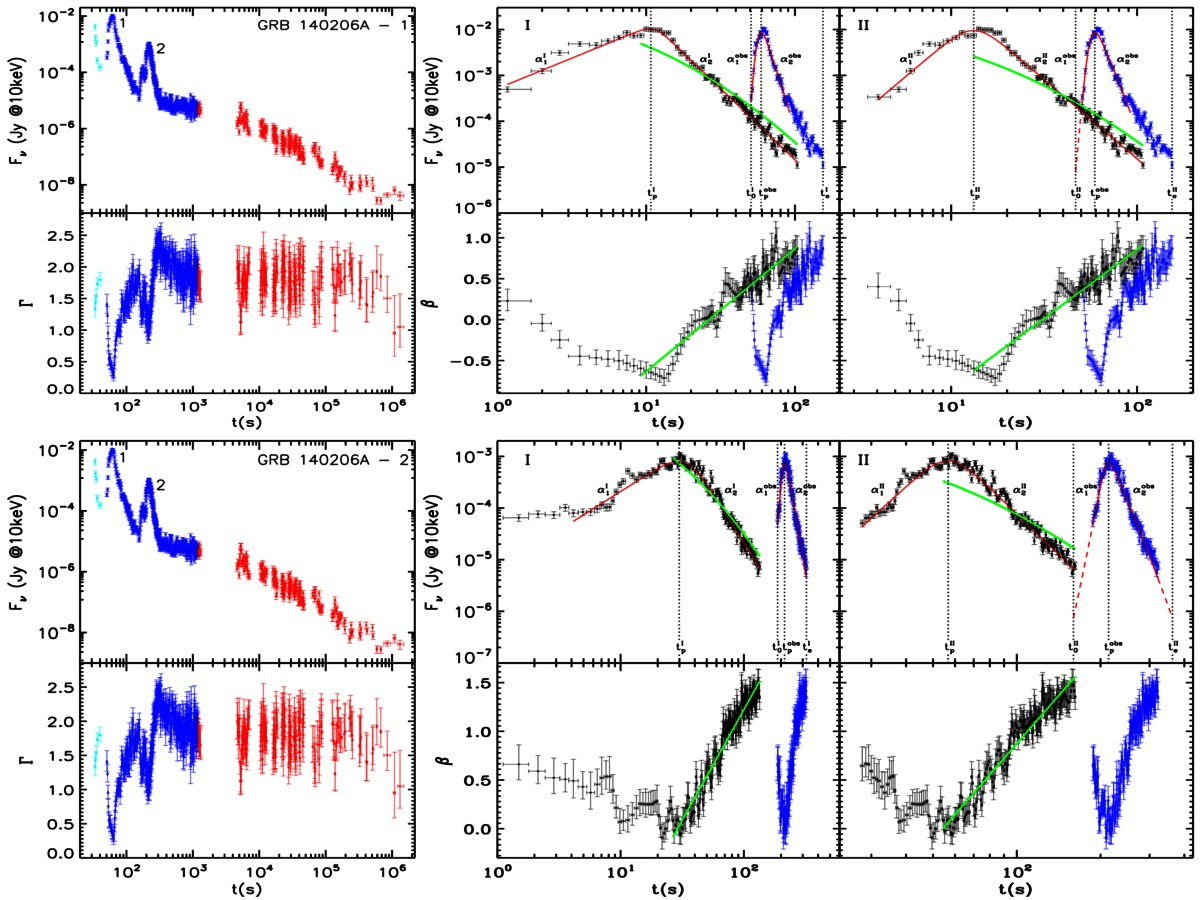}
\center{Fig. 1--- Continued}
\end{figure*}
\begin{figure*}\centering
\includegraphics[angle=0,scale=0.99,width=0.99\textwidth,height=0.75\textheight]{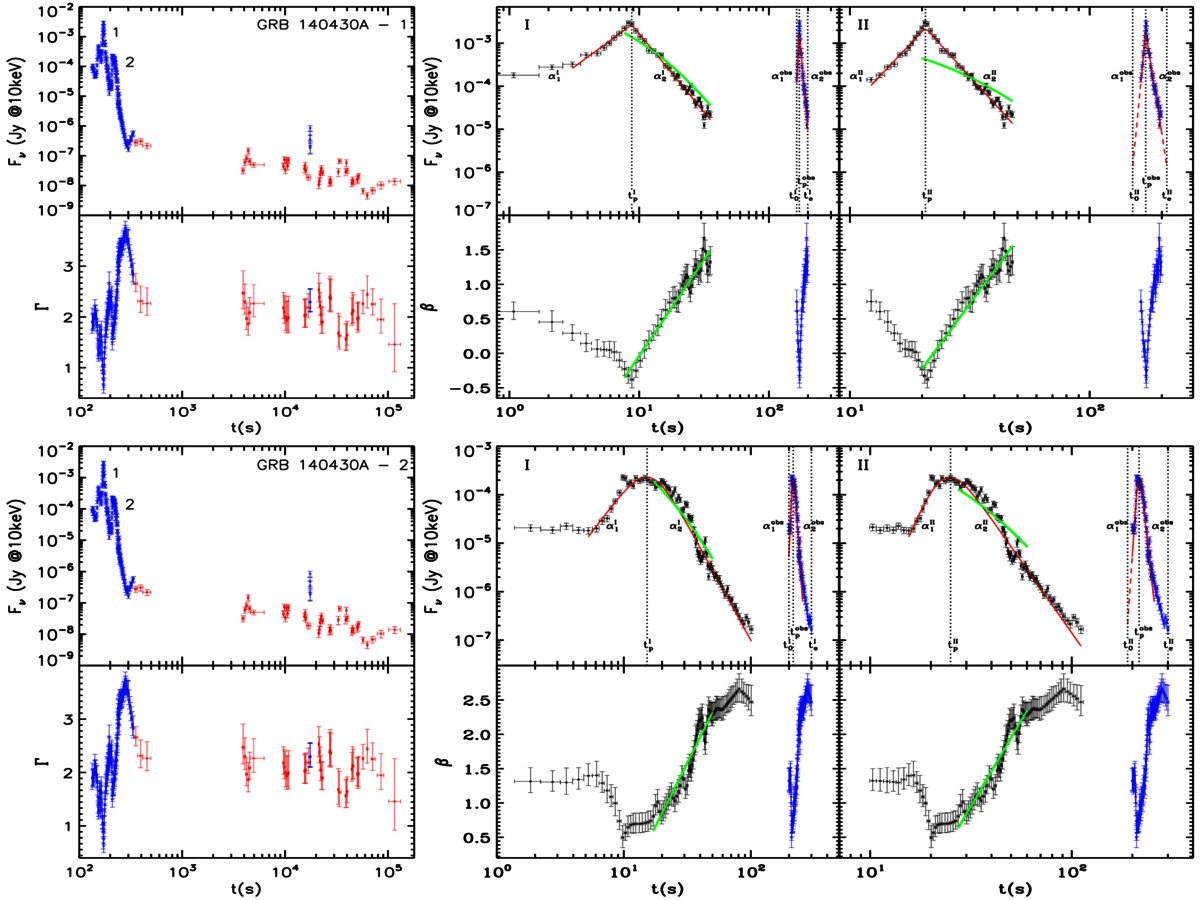}
\center{Fig. 1--- Continued}
\end{figure*}
\begin{figure*}\centering
\includegraphics[angle=0,scale=0.99,width=0.99\textwidth,height=0.75\textheight]{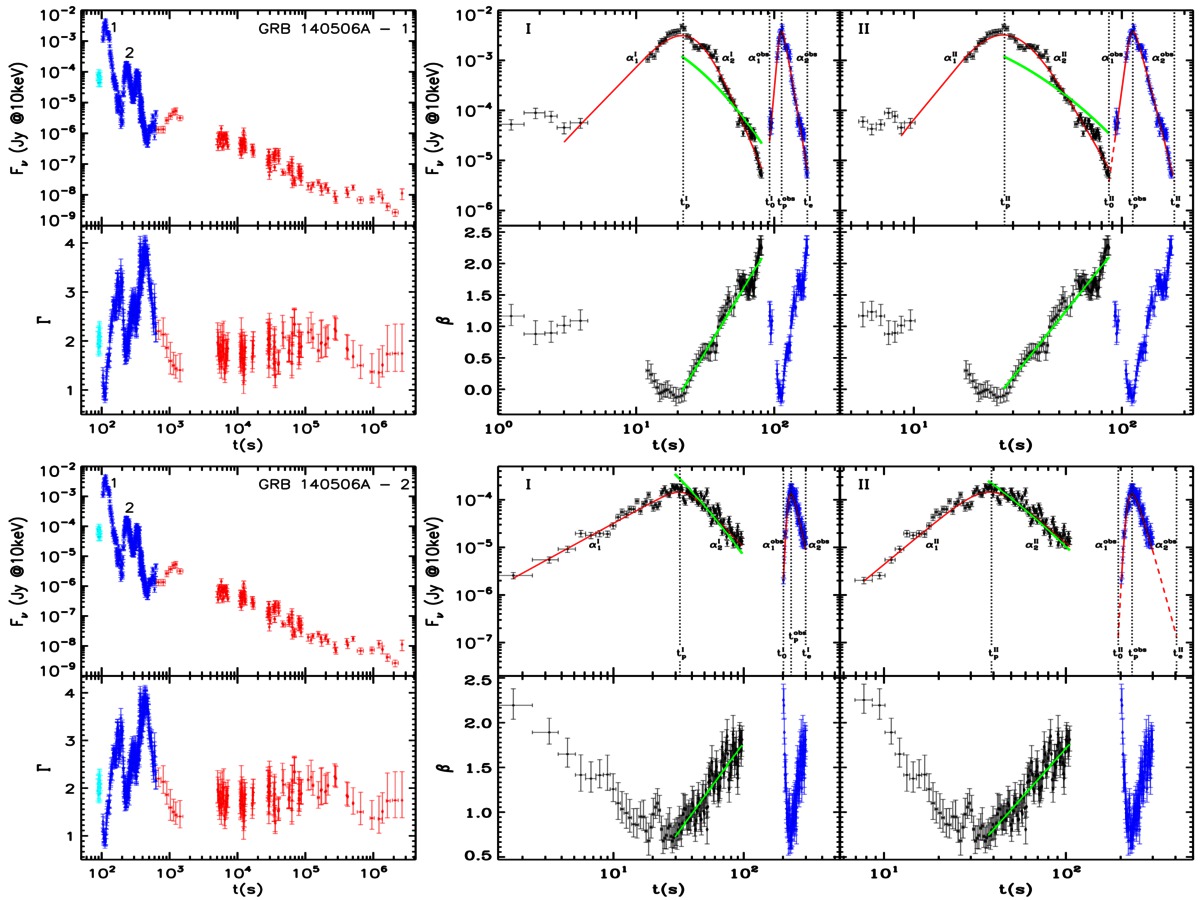}
\center{Fig. 1--- Continued}
\end{figure*}
\begin{figure*}\centering
\includegraphics[angle=0,scale=0.99,width=0.99\textwidth,height=0.75\textheight]{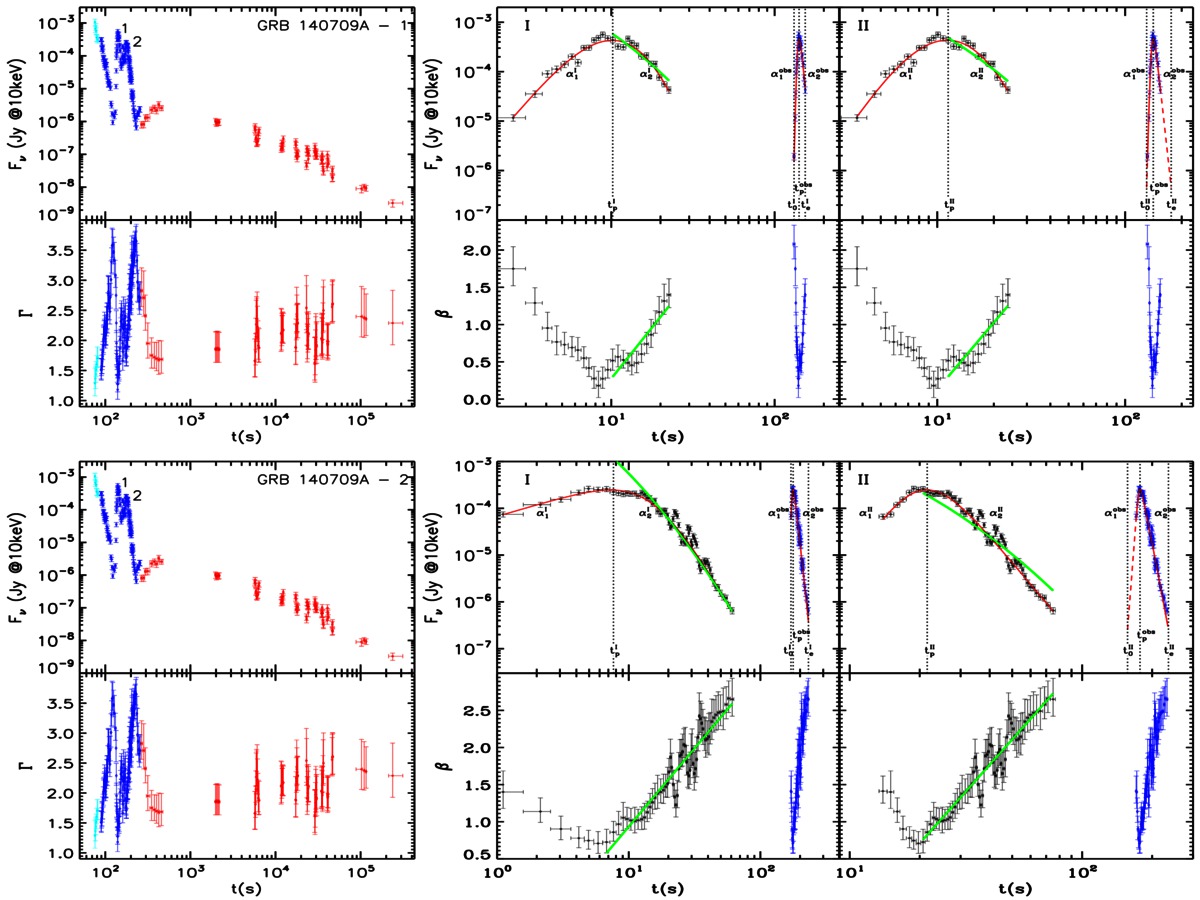}
\center{Fig. 1--- Continued}
\end{figure*}
\begin{figure*}\centering
\includegraphics[angle=0,scale=0.99,width=0.99\textwidth,height=0.65\textheight]{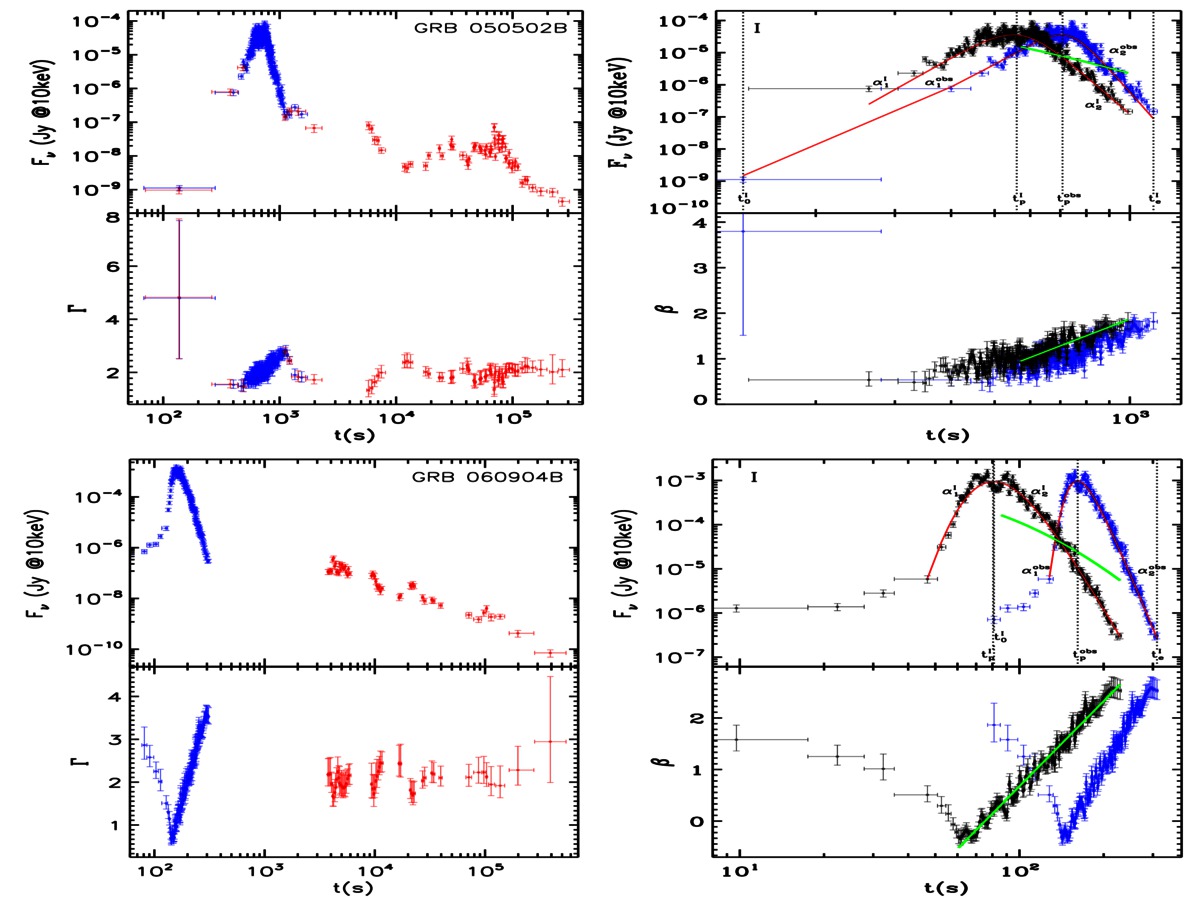}
\center{Fig. 1--- Continued}
\end{figure*}
\begin{figure*}\centering
\includegraphics[angle=0,scale=0.99,width=0.99\textwidth,height=0.75\textheight]{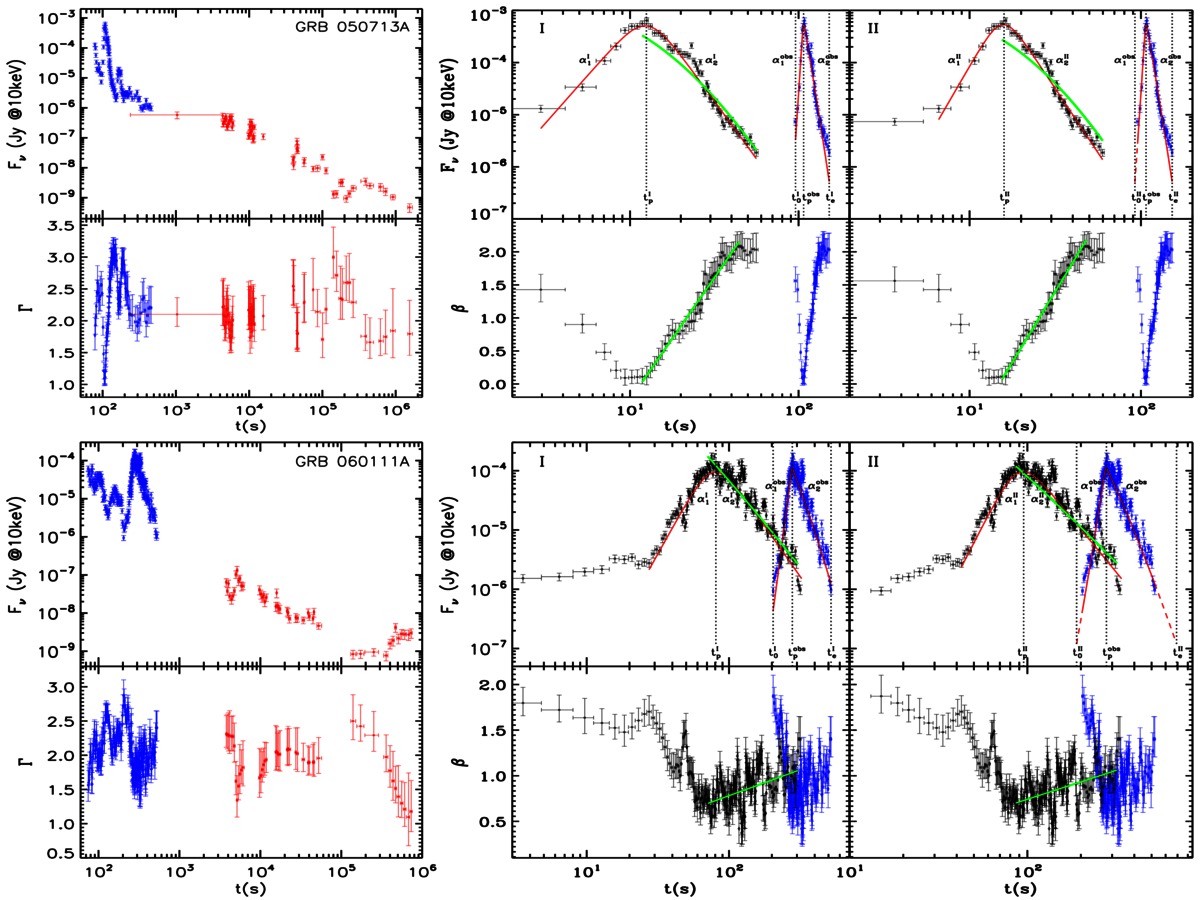}
\center{Fig. 1--- Continued}
\end{figure*}
\begin{figure*}\centering
\includegraphics[angle=0,scale=0.99,width=0.99\textwidth,height=0.75\textheight]{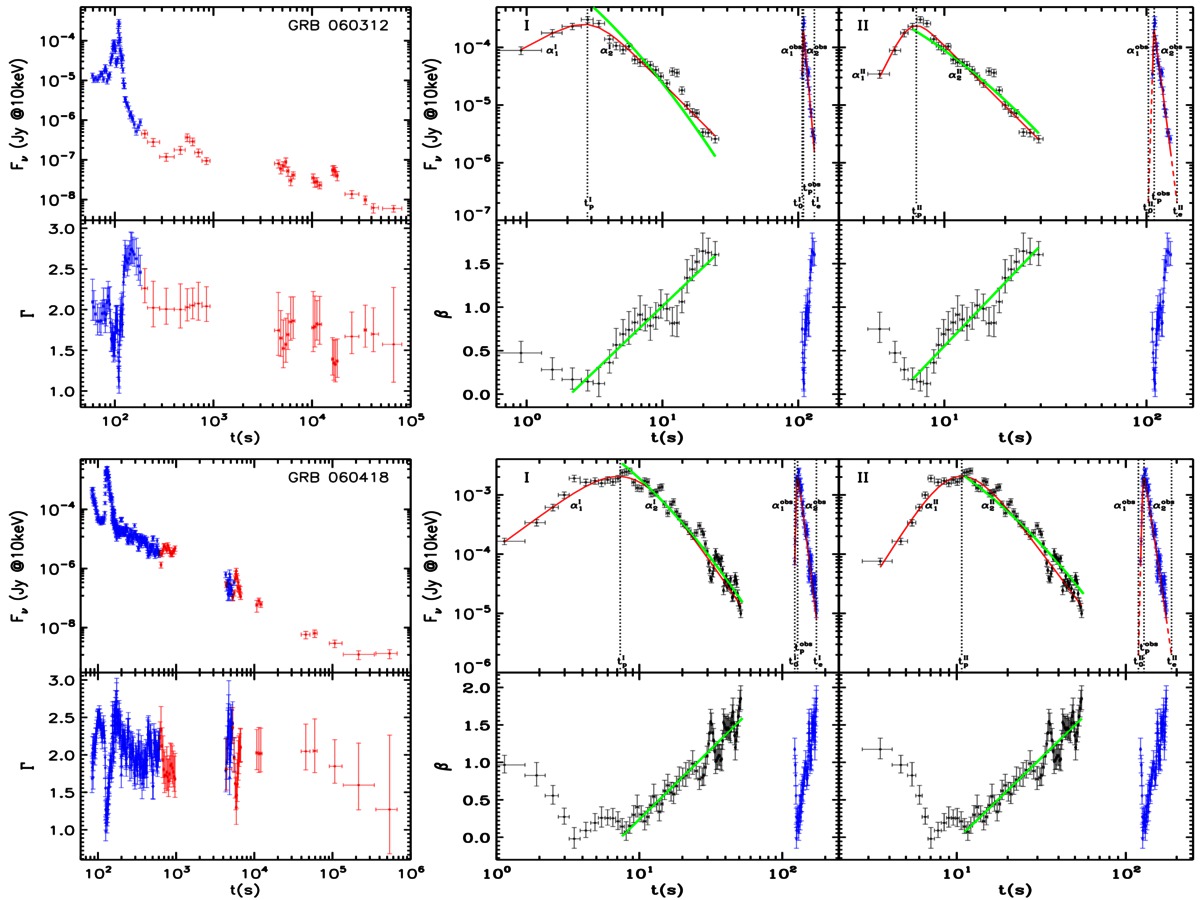}
\center{Fig. 1--- Continued}
\end{figure*}
\begin{figure*}\centering
\includegraphics[angle=0,scale=0.99,width=0.99\textwidth,height=0.75\textheight]{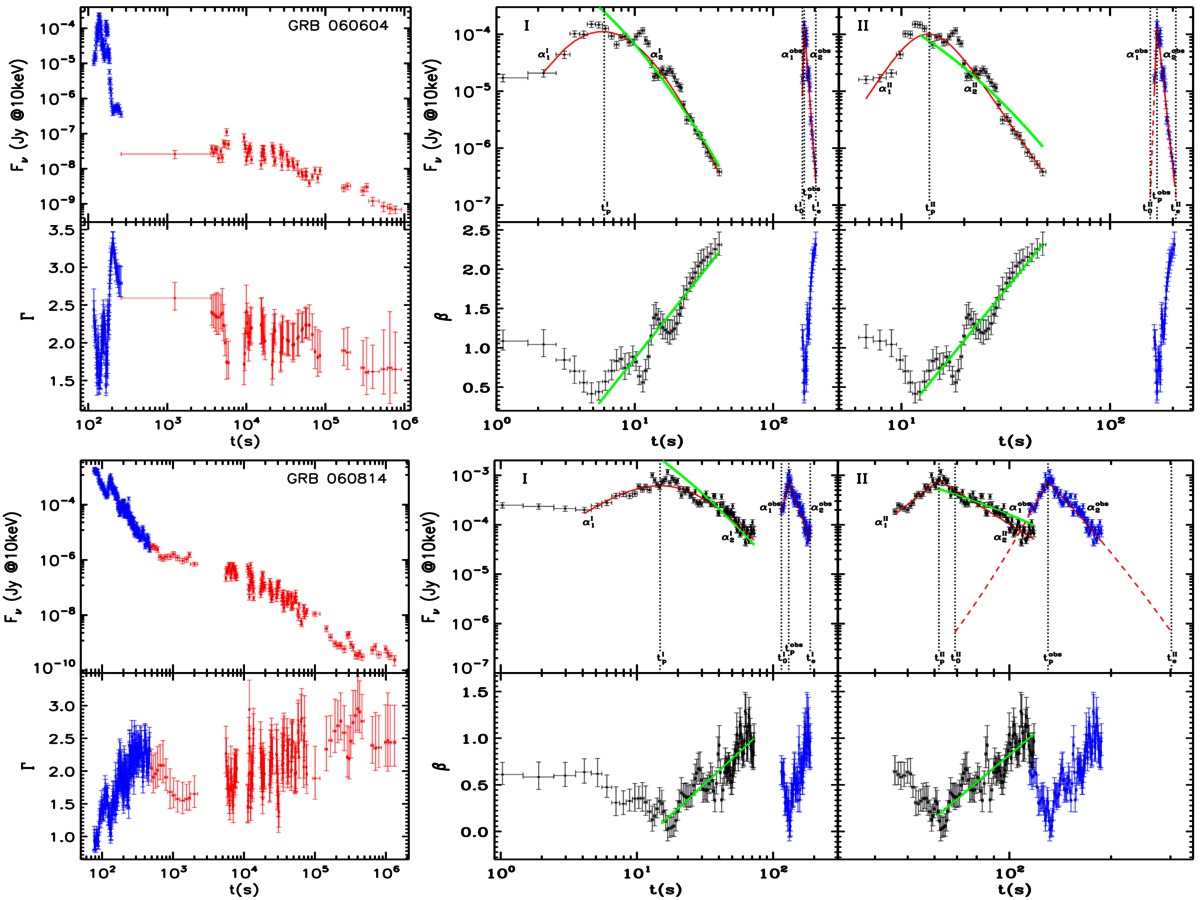} 
\center{Fig. 1--- Continued}
\end{figure*}
\begin{figure*}\centering
\includegraphics[angle=0,scale=0.99,width=0.99\textwidth,height=0.75\textheight]{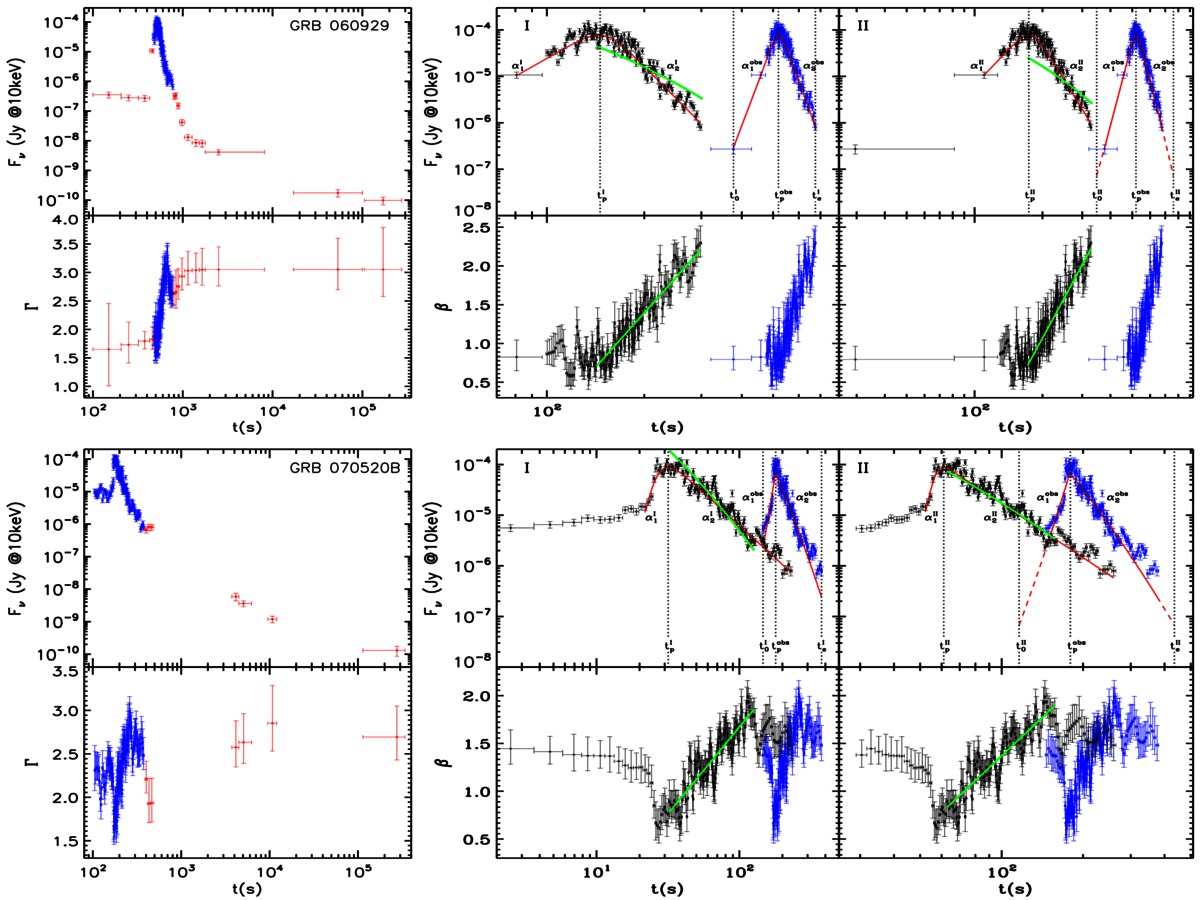}
\center{Fig. 1--- Continued}
\end{figure*}
\begin{figure*}\centering
\includegraphics[angle=0,scale=0.99,width=0.99\textwidth,height=0.75\textheight]{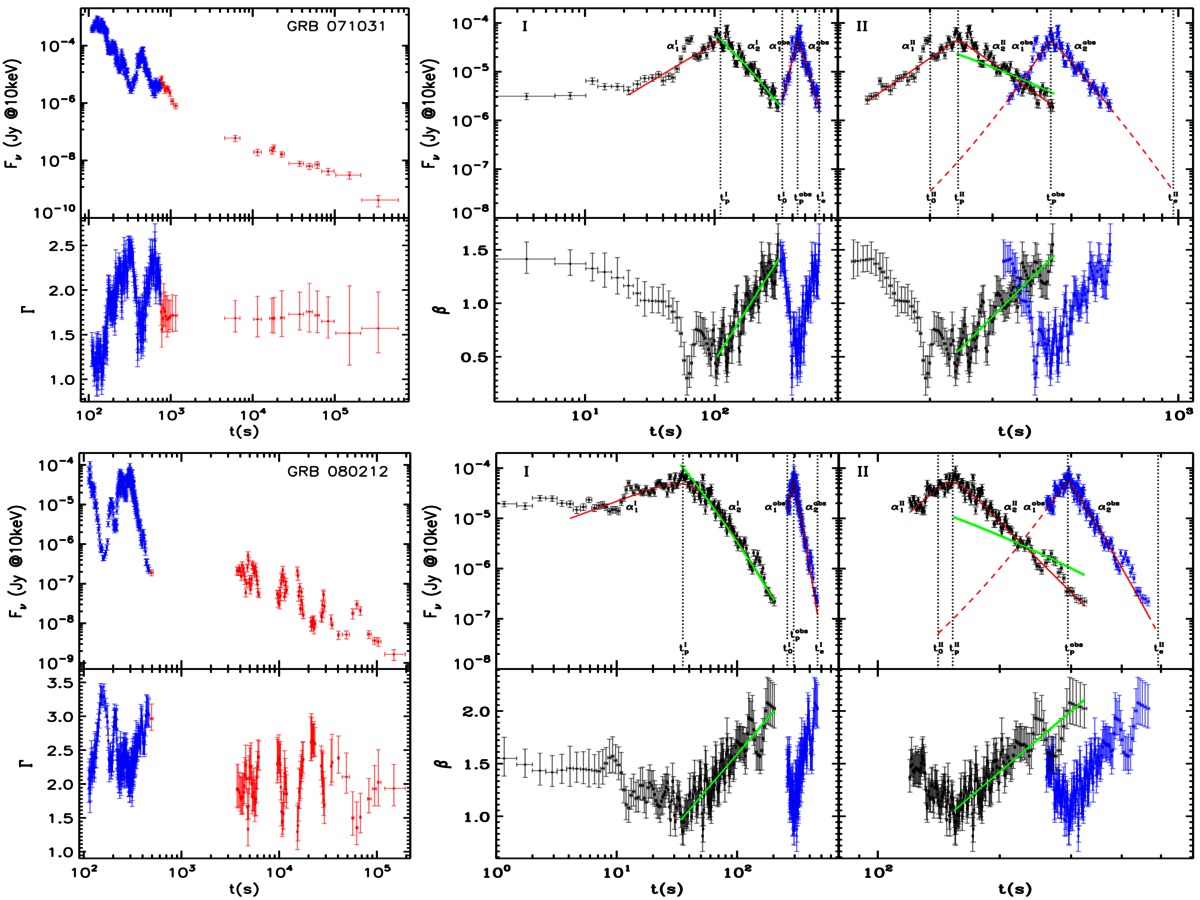}
\center{Fig. 1--- Continued}
\end{figure*}
\begin{figure*}\centering
\includegraphics[angle=0,scale=0.99,width=0.99\textwidth,height=0.75\textheight]{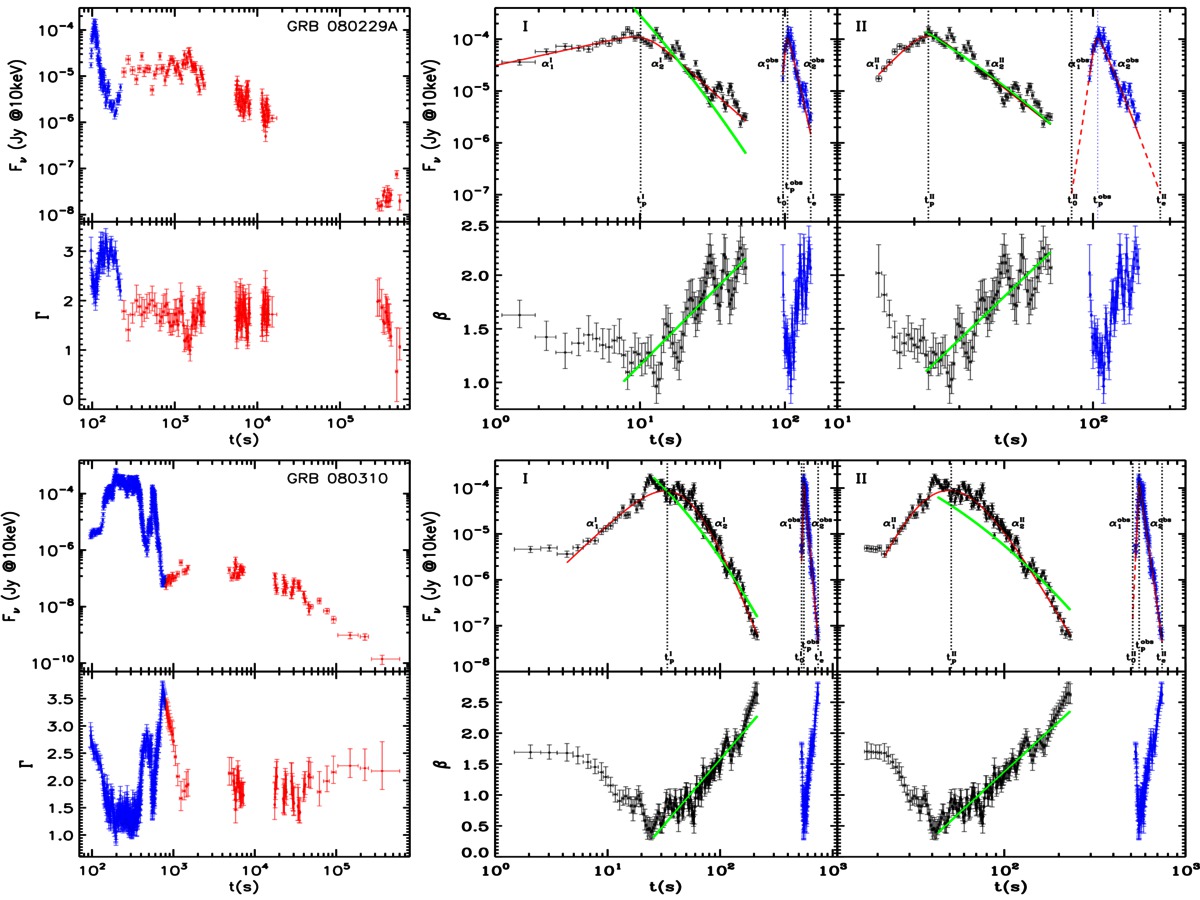}
\center{Fig. 1--- Continued}
\end{figure*}
\begin{figure*}\centering
\includegraphics[angle=0,scale=0.99,width=0.99\textwidth,height=0.75\textheight]{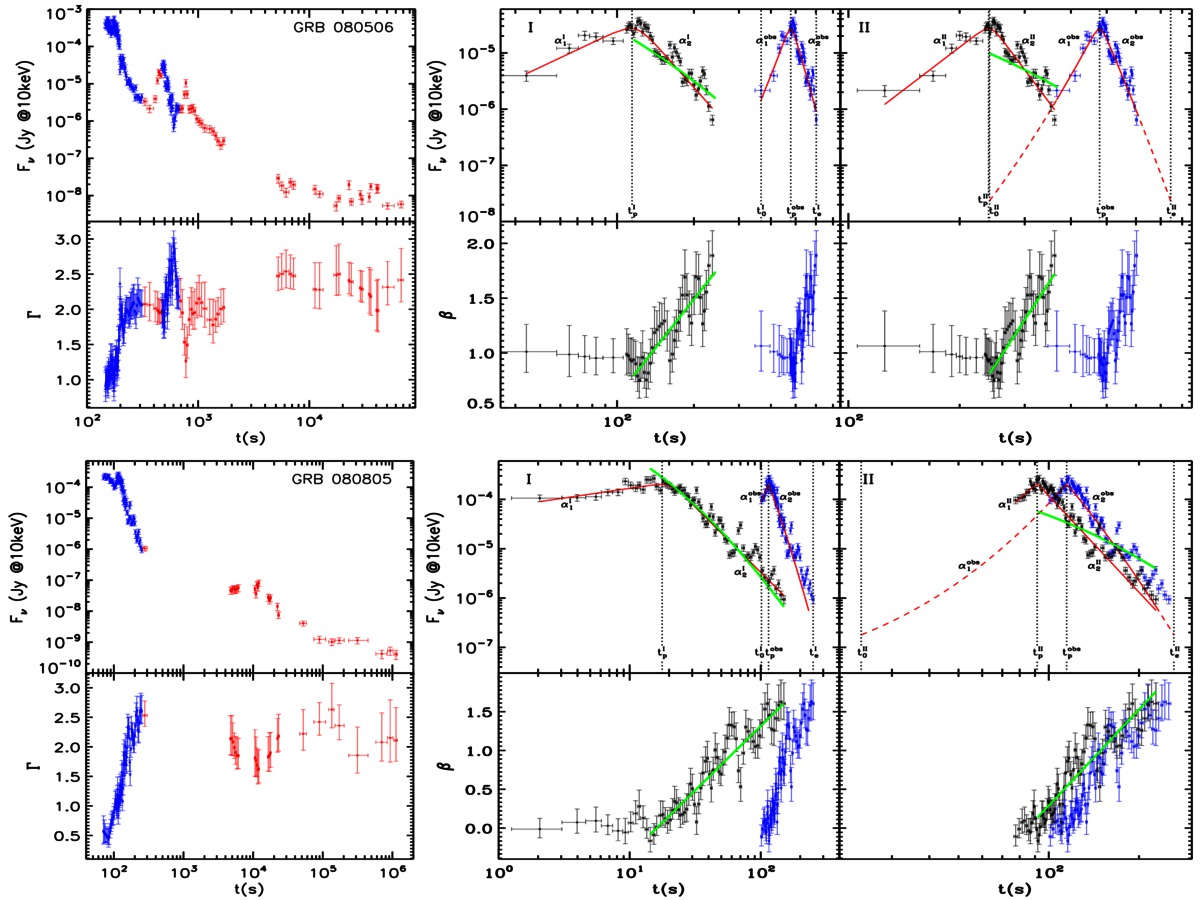}
\center{Fig. 1--- Continued}
\end{figure*}
\begin{figure*}\centering
\includegraphics[angle=0,scale=0.99,width=0.99\textwidth,height=0.75\textheight]{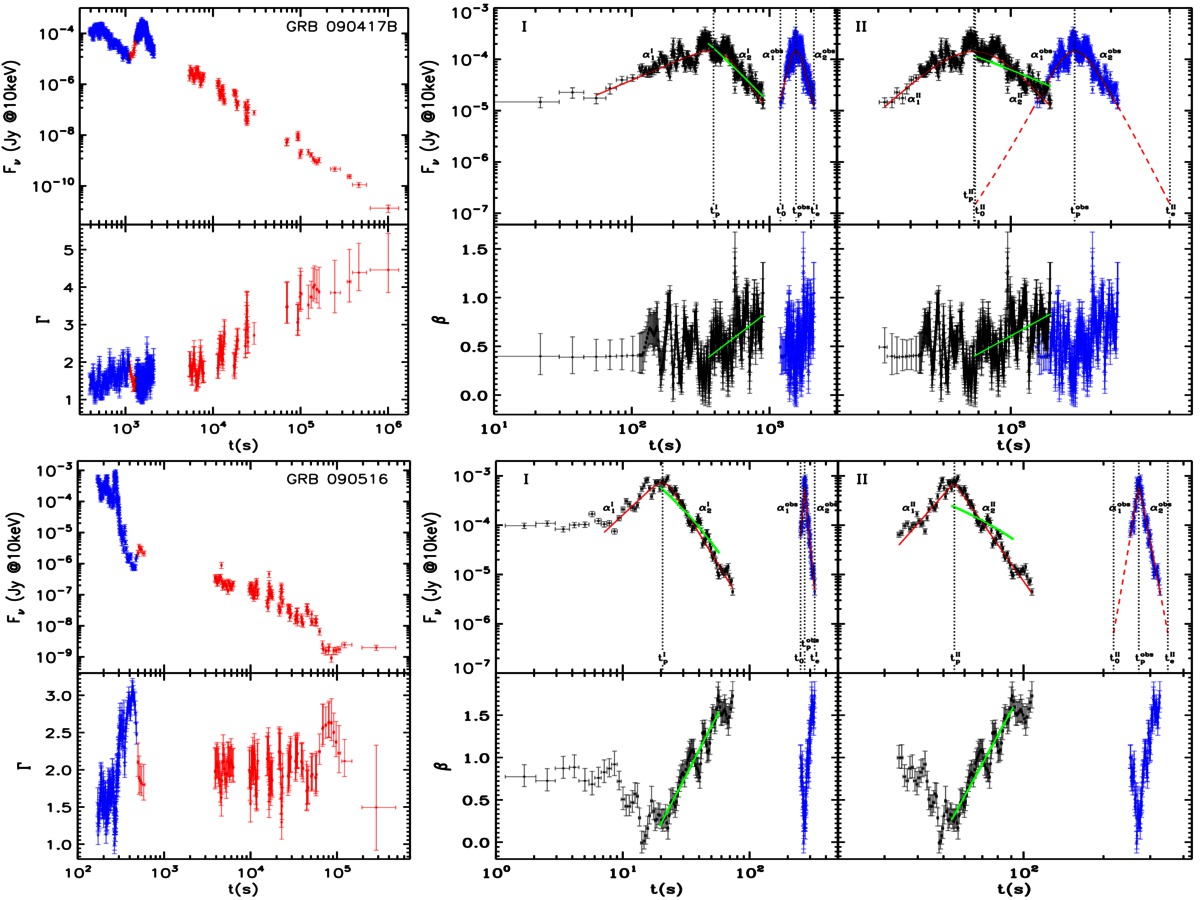}
\center{Fig. 1--- Continued}
\end{figure*}
\begin{figure*}\centering
\includegraphics[angle=0,scale=0.99,width=0.99\textwidth,height=0.75\textheight]{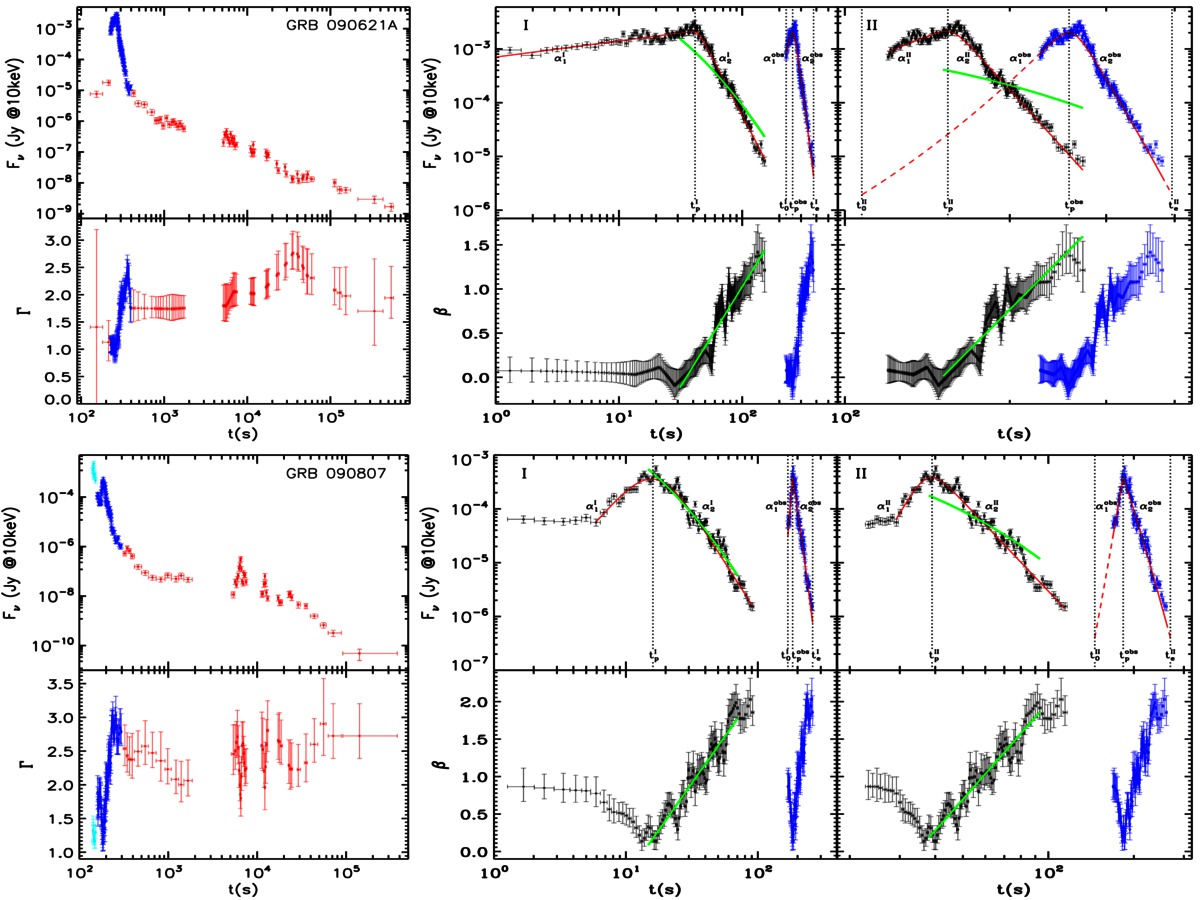}
\center{Fig. 1--- Continued}
\end{figure*}
\begin{figure*}\centering
\includegraphics[angle=0,scale=0.99,width=0.99\textwidth,height=0.75\textheight]{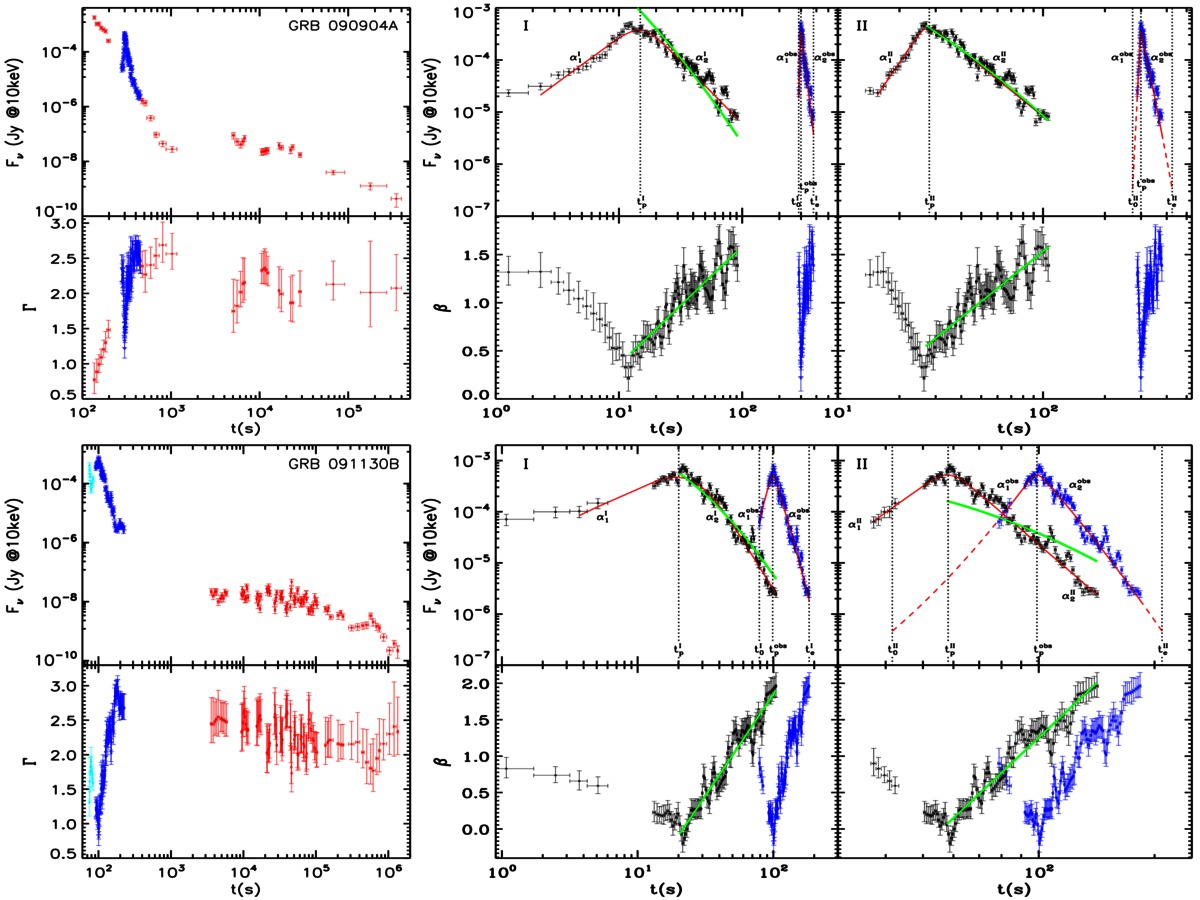}
\center{Fig. 1--- Continued}
\end{figure*}
\begin{figure*}\centering
\includegraphics[angle=0,scale=0.99,width=0.99\textwidth,height=0.75\textheight]{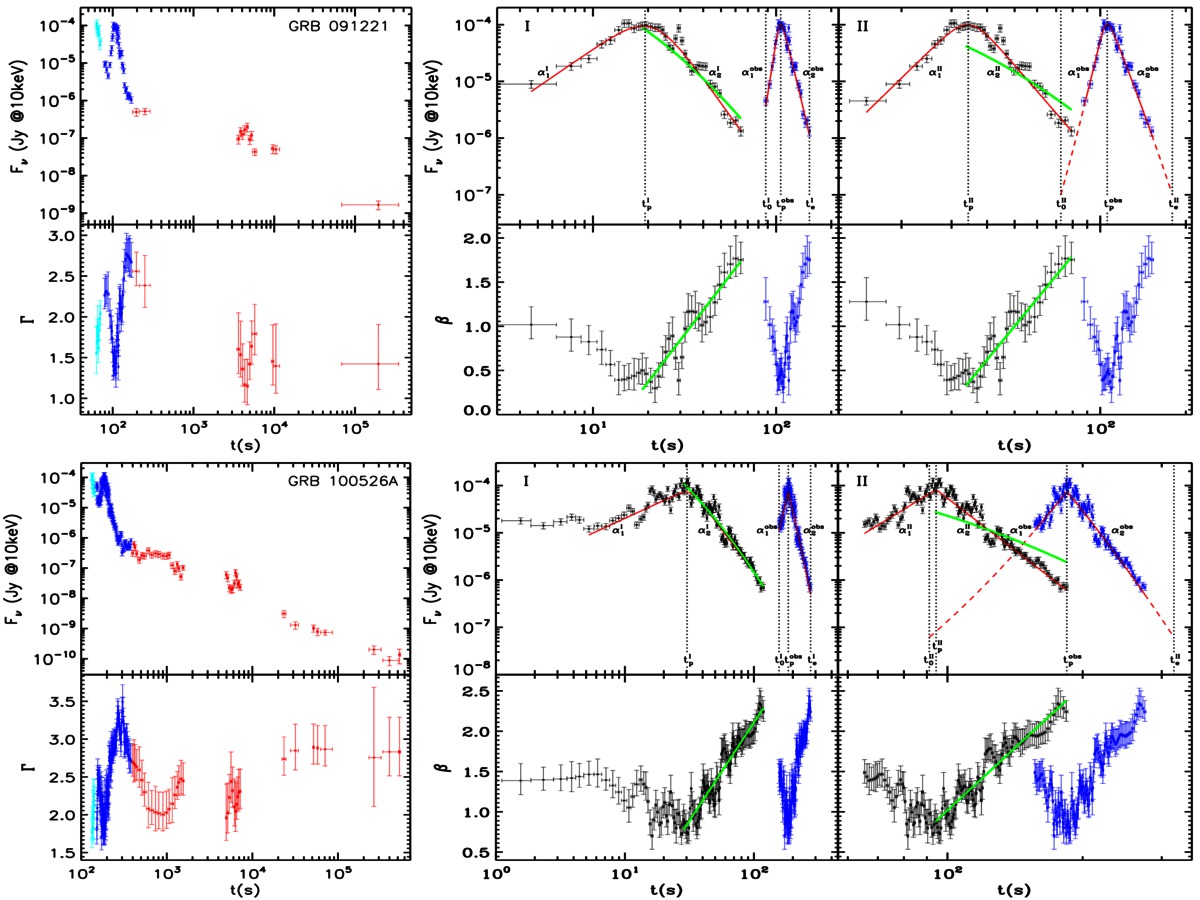}
\center{Fig. 1--- Continued}
\end{figure*}
\begin{figure*}\centering
\includegraphics[angle=0,scale=0.99,width=0.99\textwidth,height=0.75\textheight]{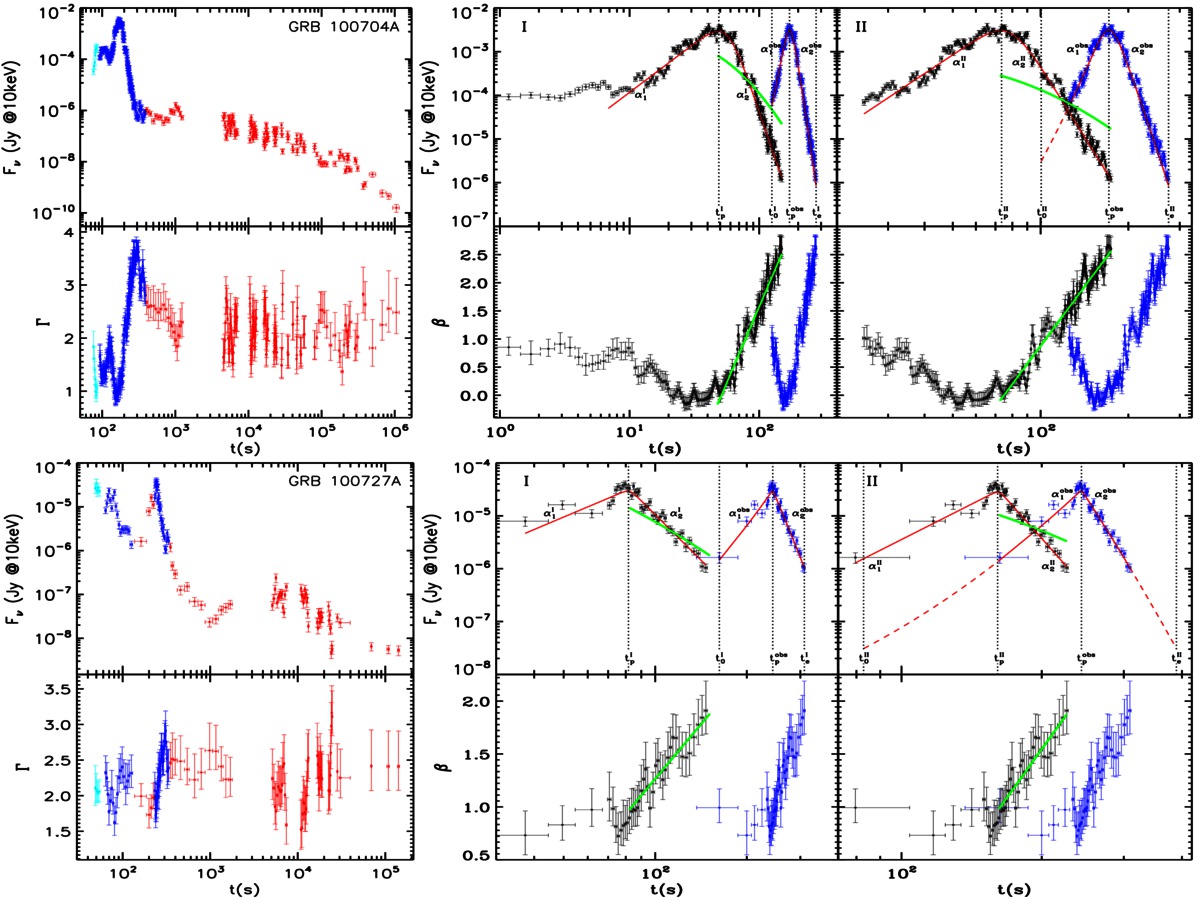}
\center{Fig. 1--- Continued}
\end{figure*}
\begin{figure*}\centering
\includegraphics[angle=0,scale=0.99,width=0.99\textwidth,height=0.75\textheight]{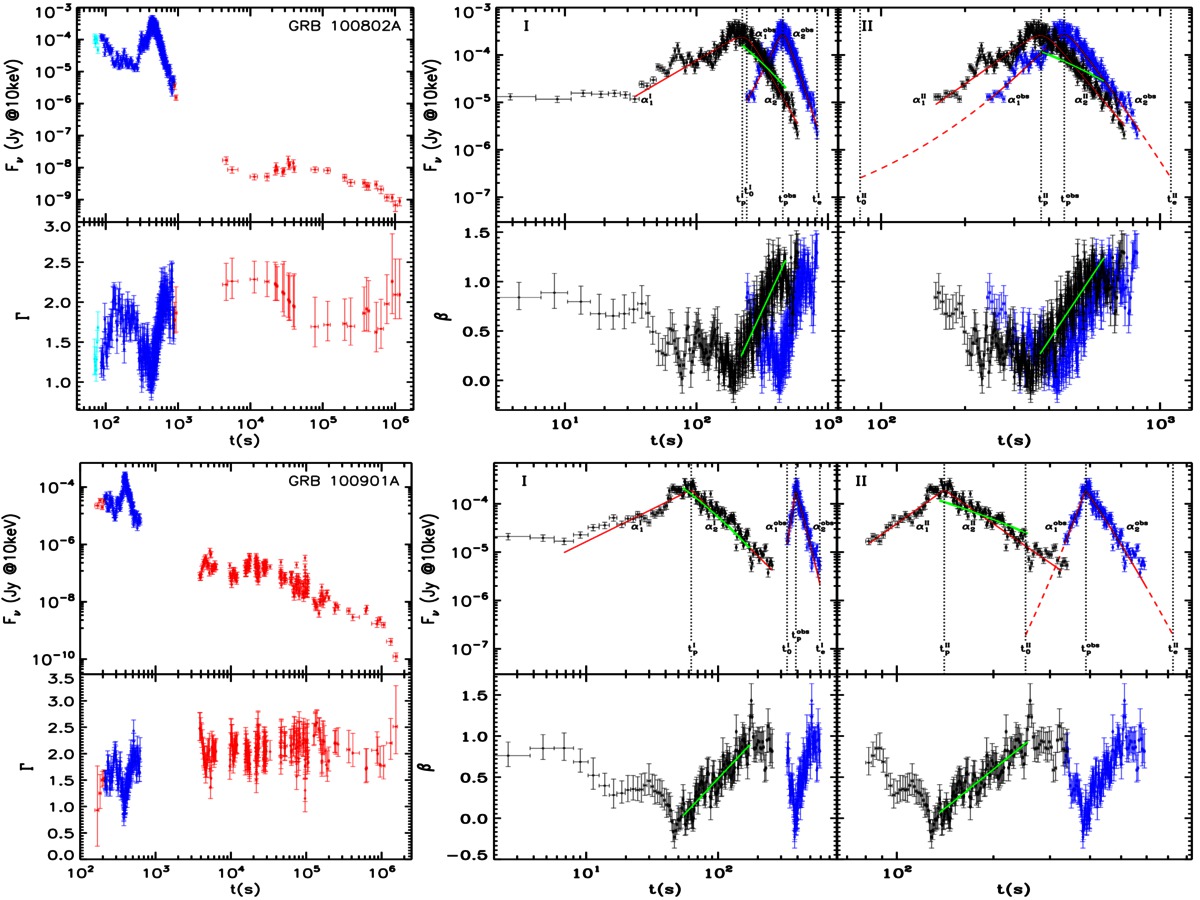}
\center{Fig. 1--- Continued}
\end{figure*}
\begin{figure*}\centering
\includegraphics[angle=0,scale=0.99,width=0.99\textwidth,height=0.75\textheight]{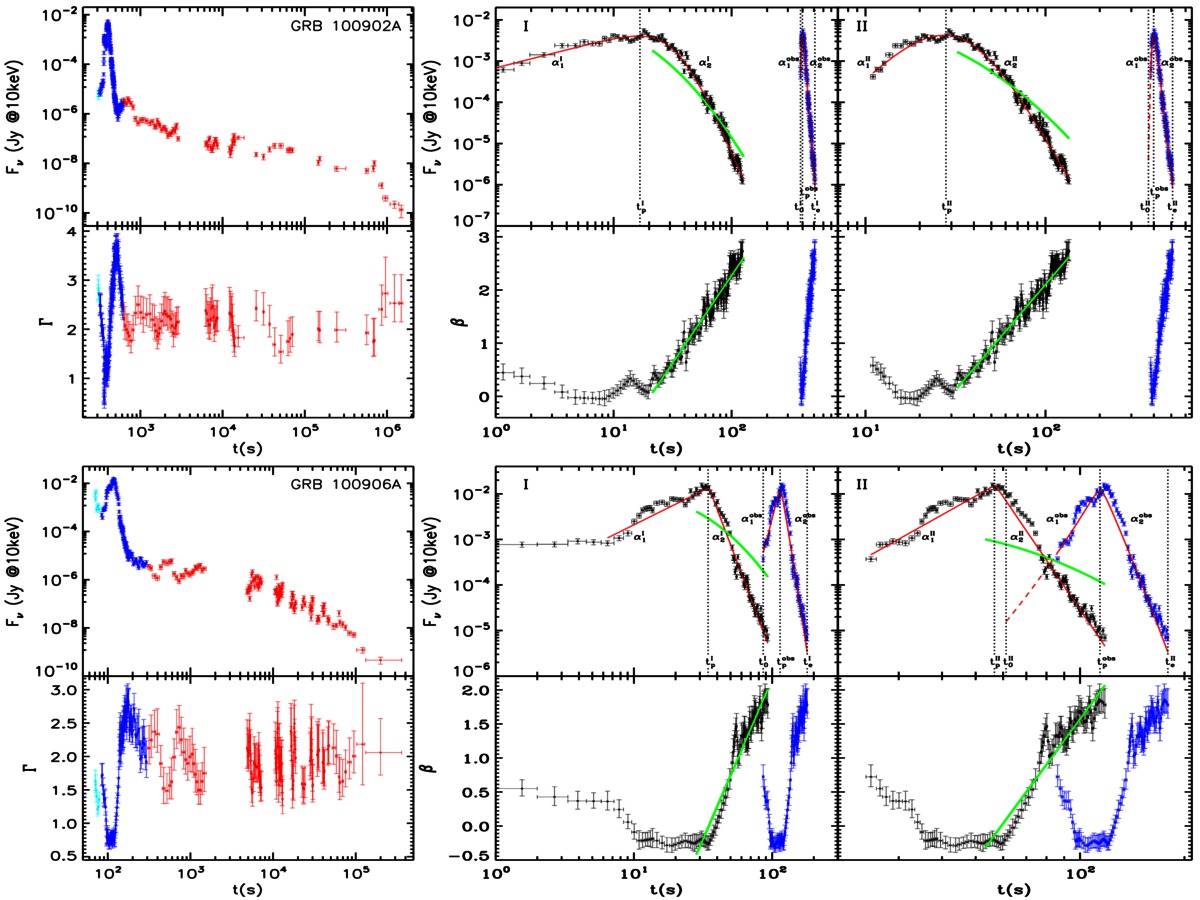}
\center{Fig. 1--- Continued}
\end{figure*}
\begin{figure*}\centering
\includegraphics[angle=0,scale=0.99,width=0.99\textwidth,height=0.75\textheight]{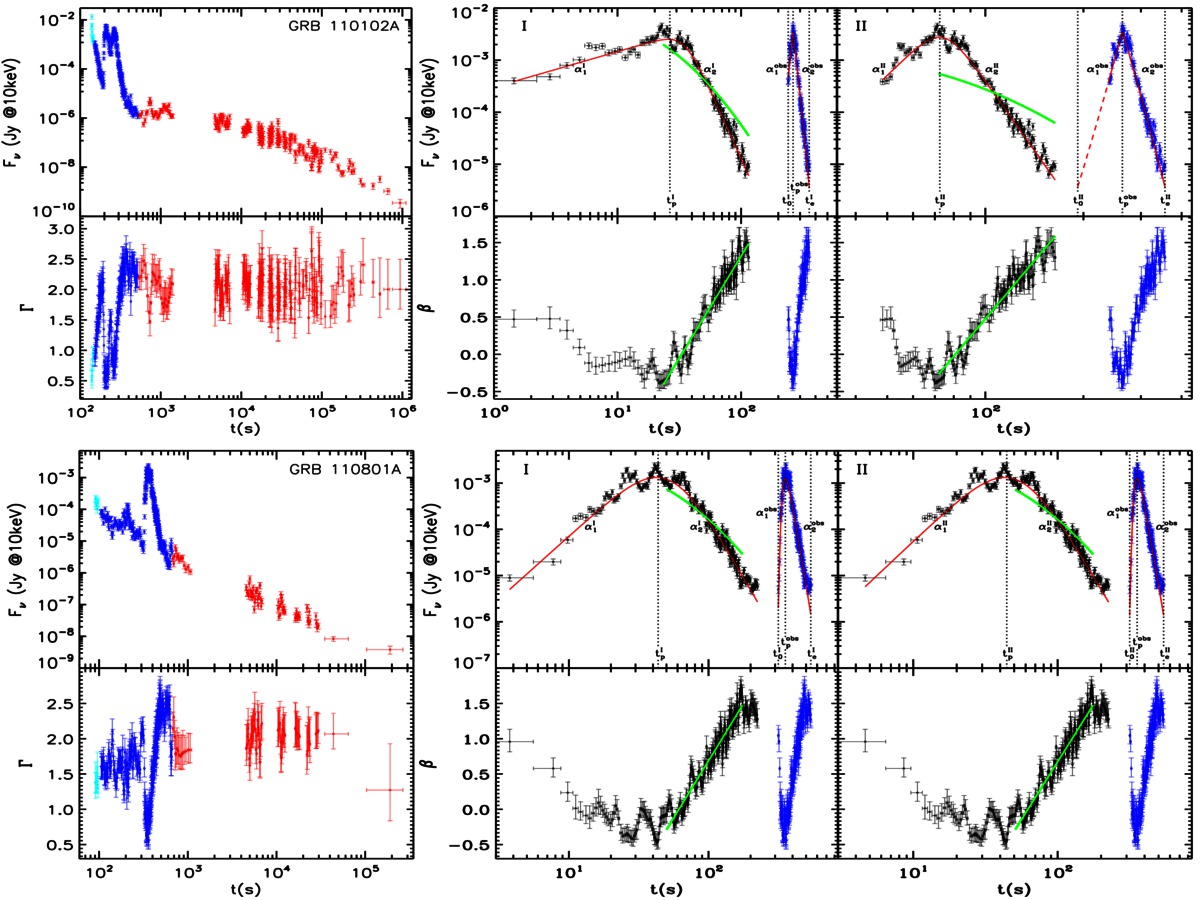}
\center{Fig. 1--- Continued}
\end{figure*}
\begin{figure*}\centering
\includegraphics[angle=0,scale=0.99,width=0.99\textwidth,height=0.75\textheight]{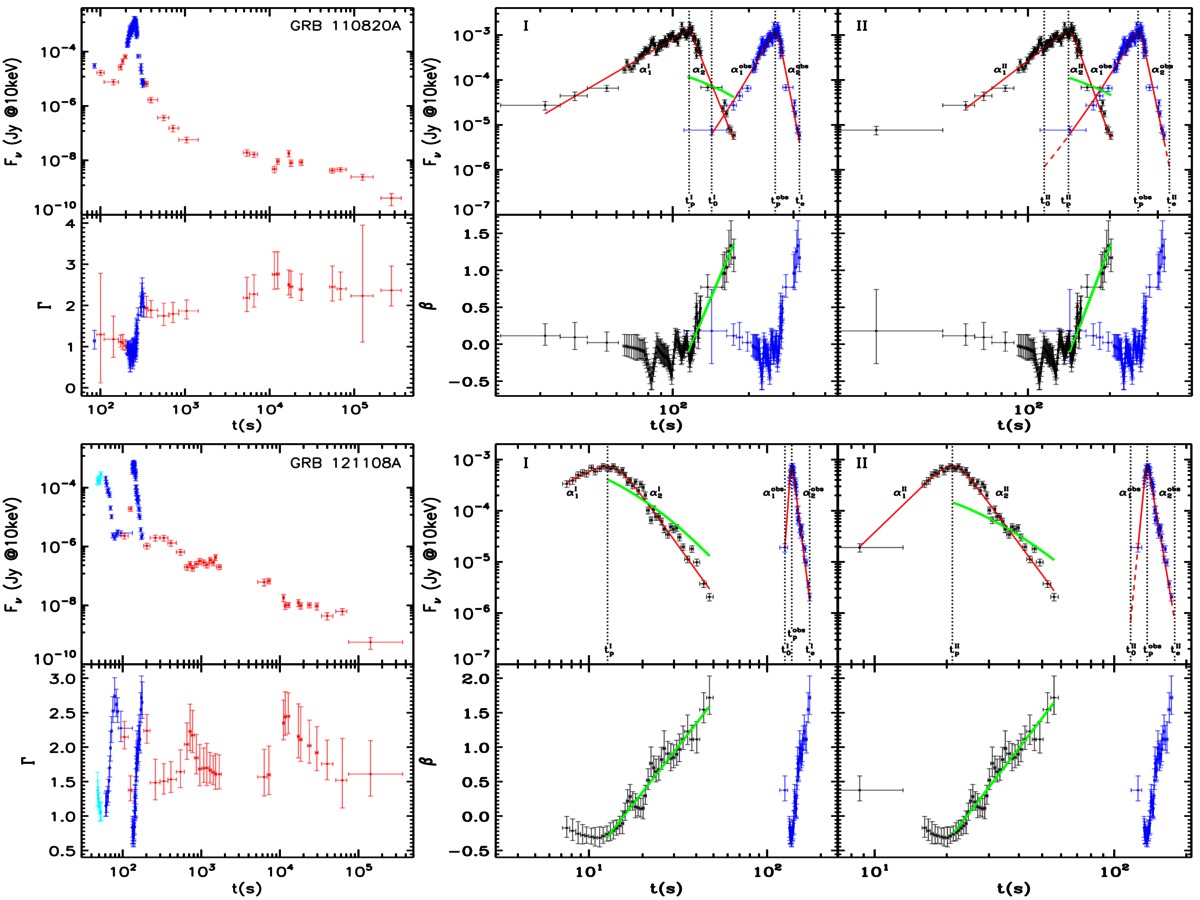}
\center{Fig. 1--- Continued}
\end{figure*}
\begin{figure*}\centering
\includegraphics[angle=0,scale=0.99,width=0.99\textwidth,height=0.75\textheight]{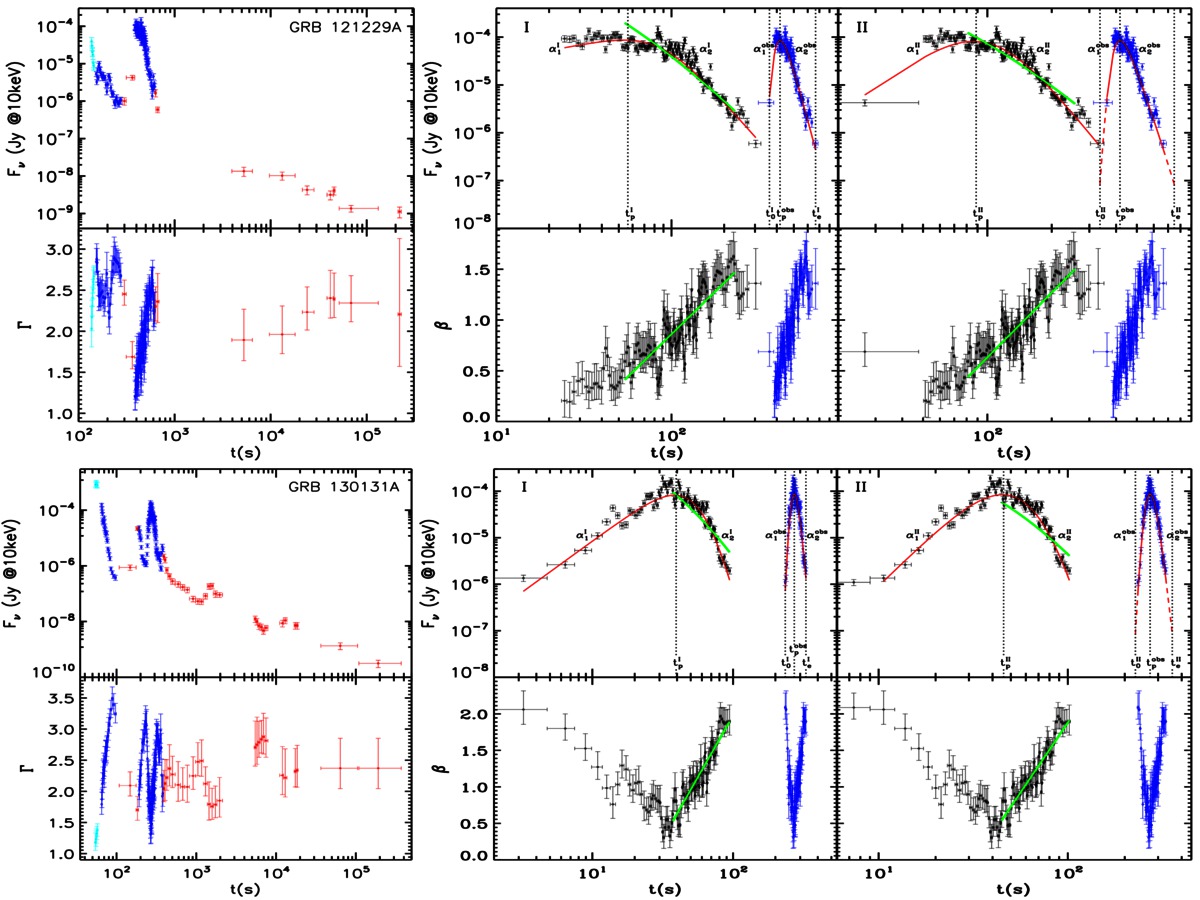}
\center{Fig. 1--- Continued}
\end{figure*}
\begin{figure*}\centering
\includegraphics[angle=0,scale=0.99,width=0.99\textwidth,height=0.75\textheight]{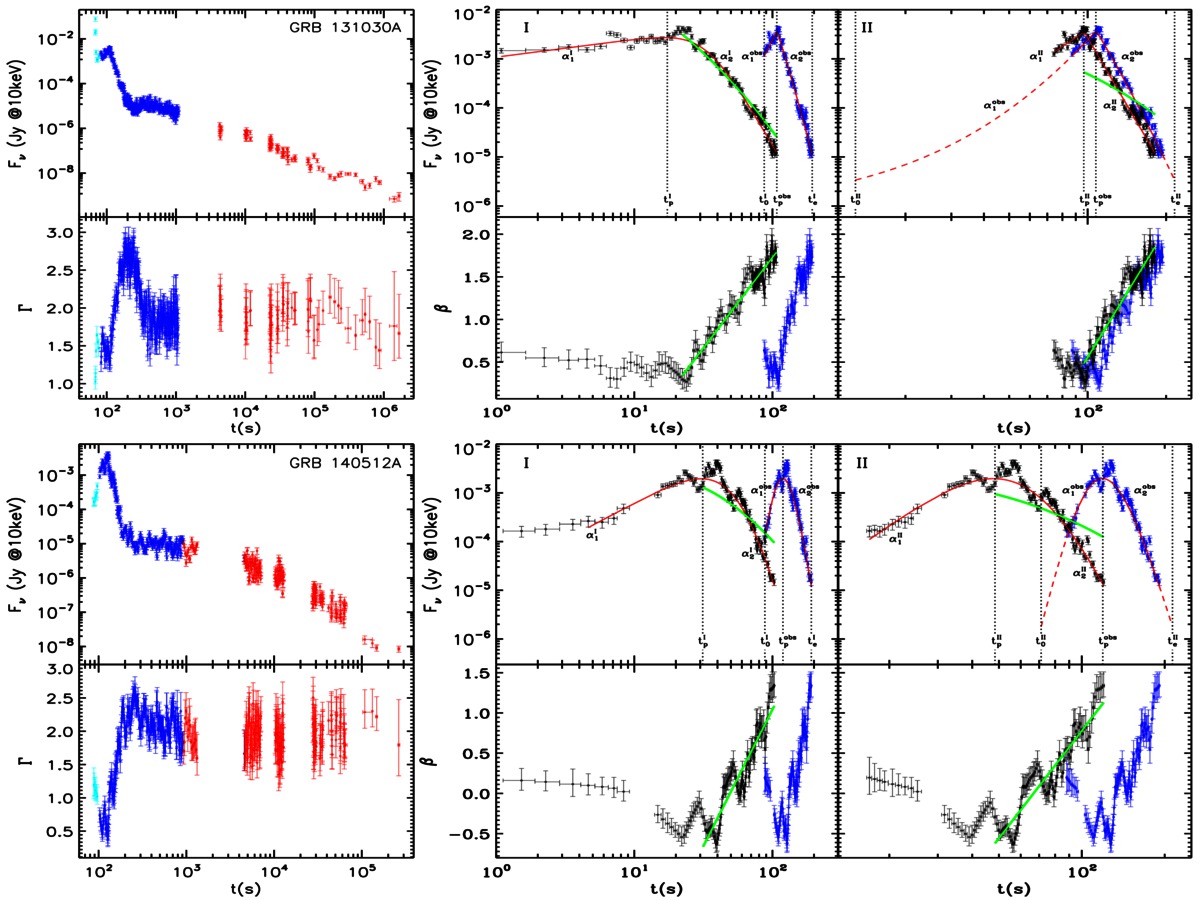}
\center{Fig. 1--- Continued}
\end{figure*}
\begin{figure*}\centering
\includegraphics[angle=0,scale=0.99,width=0.99\textwidth,height=0.75\textheight]{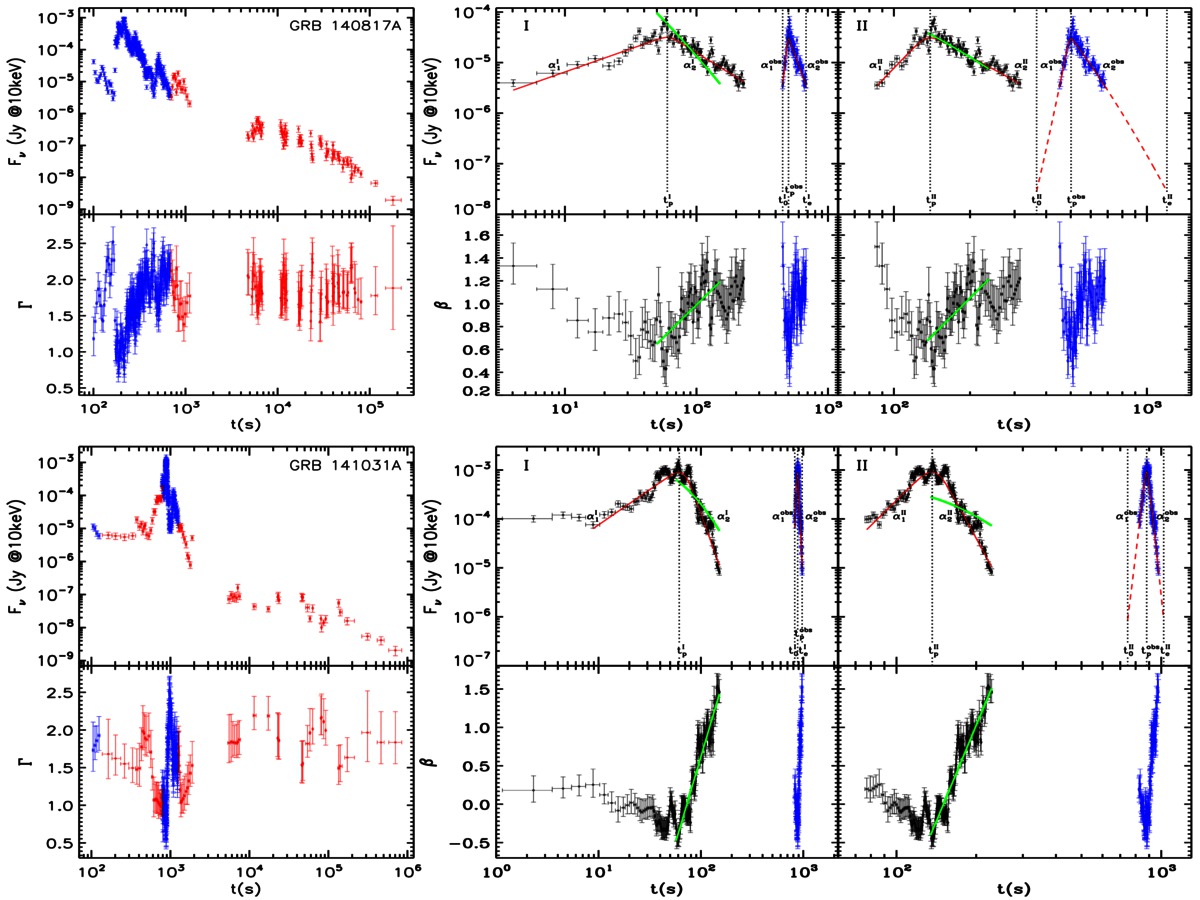}
\center{Fig. 1--- Continued}
\end{figure*}
\clearpage
\setlength{\voffset}{-18mm}
\begin{figure*}\centering
\includegraphics[angle=0,scale=0.85,width=0.95\textwidth,height=0.45\textheight]{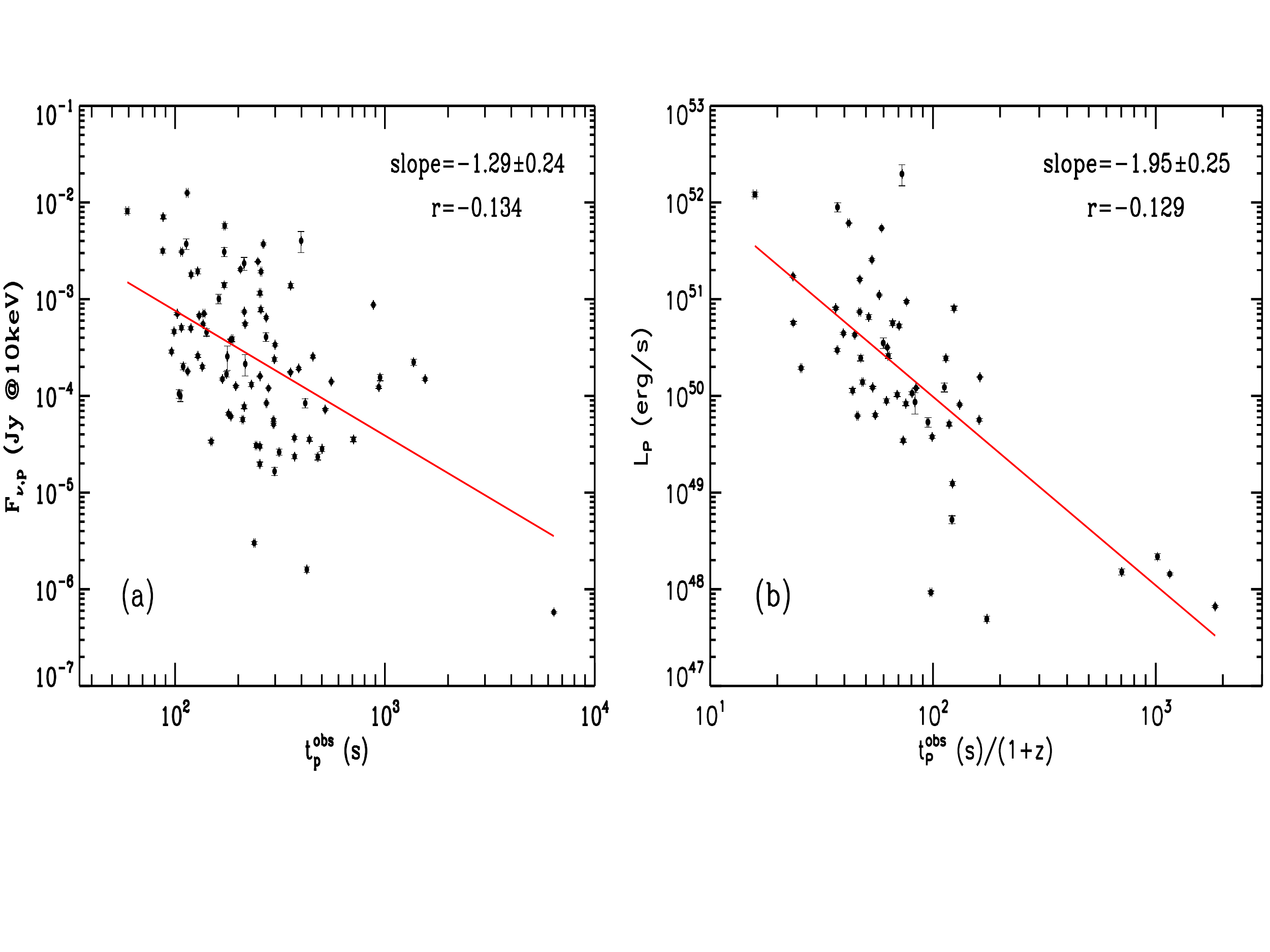}
\caption{Scatter plots of some pair-parameters listed in Table 1. (a) The peak flux density versus the observed peak time for the 85 flares in our sample. (b) The peak luminosity versus the rest-frame peak time for the 53 flares with measured redshifts. In both panels, the red solid lines indicate the best-fitting relations, with the Pearson correlation coefficients marked.}
\end{figure*}
\clearpage
\setlength{\voffset}{-18mm}
\begin{figure*}\centering
\includegraphics[angle=0,scale=0.85,width=0.95\textwidth,height=0.75\textheight]{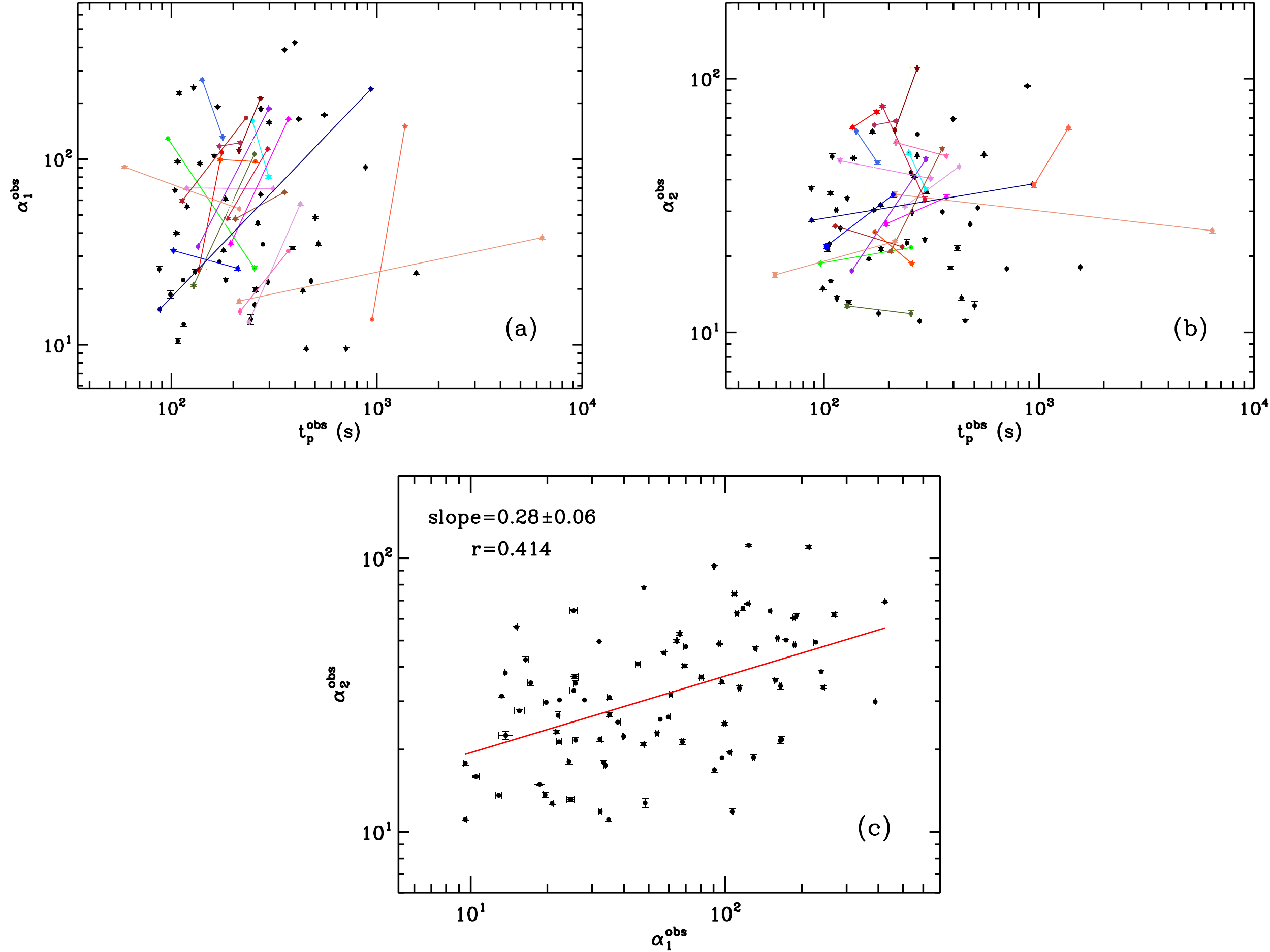}
\caption{Scatter plots of some pair-parameters listed in Table 1. (a) The observed rising slope of the light curves versus the observed peak time of the flare. The black points denote the GRBs with a single flare, and the colored points denote those GRBs with two flares that are connected. Different colors stand for different GRBs.  (b) The same as (a) but for the observed decaying slopes. (c) The observed rising slope versus the decaying slope of the flare light curves. The red solid line indicates the best-fitting relation, with the Pearson correlation coefficient marked.}
\end{figure*}
\clearpage
\setlength{\voffset}{-18mm}
\begin{figure*}\centering
\includegraphics[angle=0,scale=0.85,width=0.95\textwidth,height=0.90\textheight]{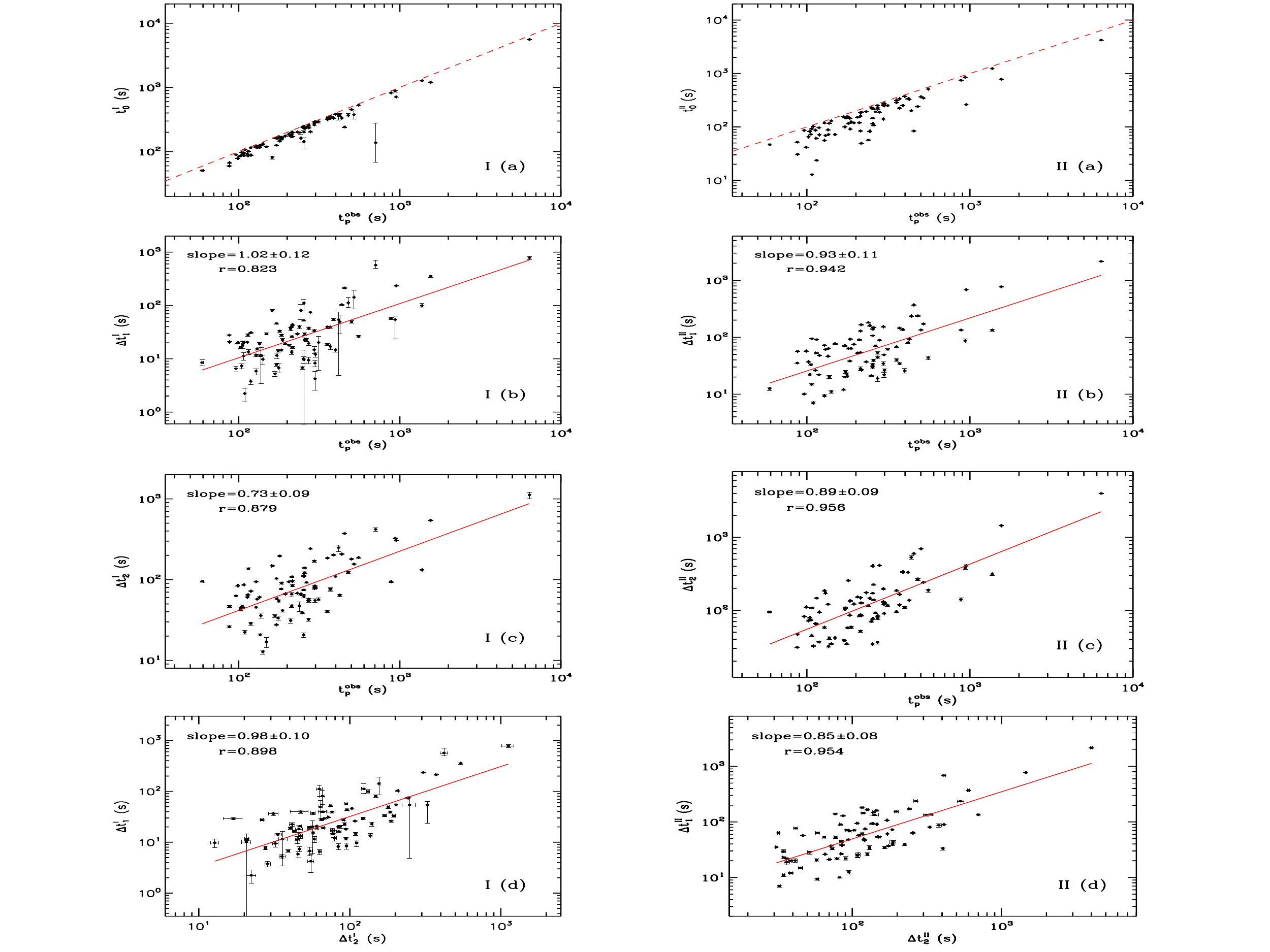}
\caption{Scatter plots of time-related pair-parameters listed in Tables 1 and 2. Symbols ``I" and ``II" stand for Method I and II, respectively. The red solid lines indicate the best-fitting relations in $\rm I$(b-d), $\rm II$(b-d). The Pearson correlation coefficients of each pair are also marked. In $\rm I$(a) and $\rm II$(a), the red dashed lines are the equality lines, which sets an upper boundary to the data points (i.e., the observed peak time is always greater than the starting time of the flare).} 
\end{figure*}
\clearpage
\setlength{\voffset}{-18mm}
\begin{figure*}\centering
\includegraphics[angle=0,scale=0.85,width=0.95\textwidth,height=0.90\textheight]{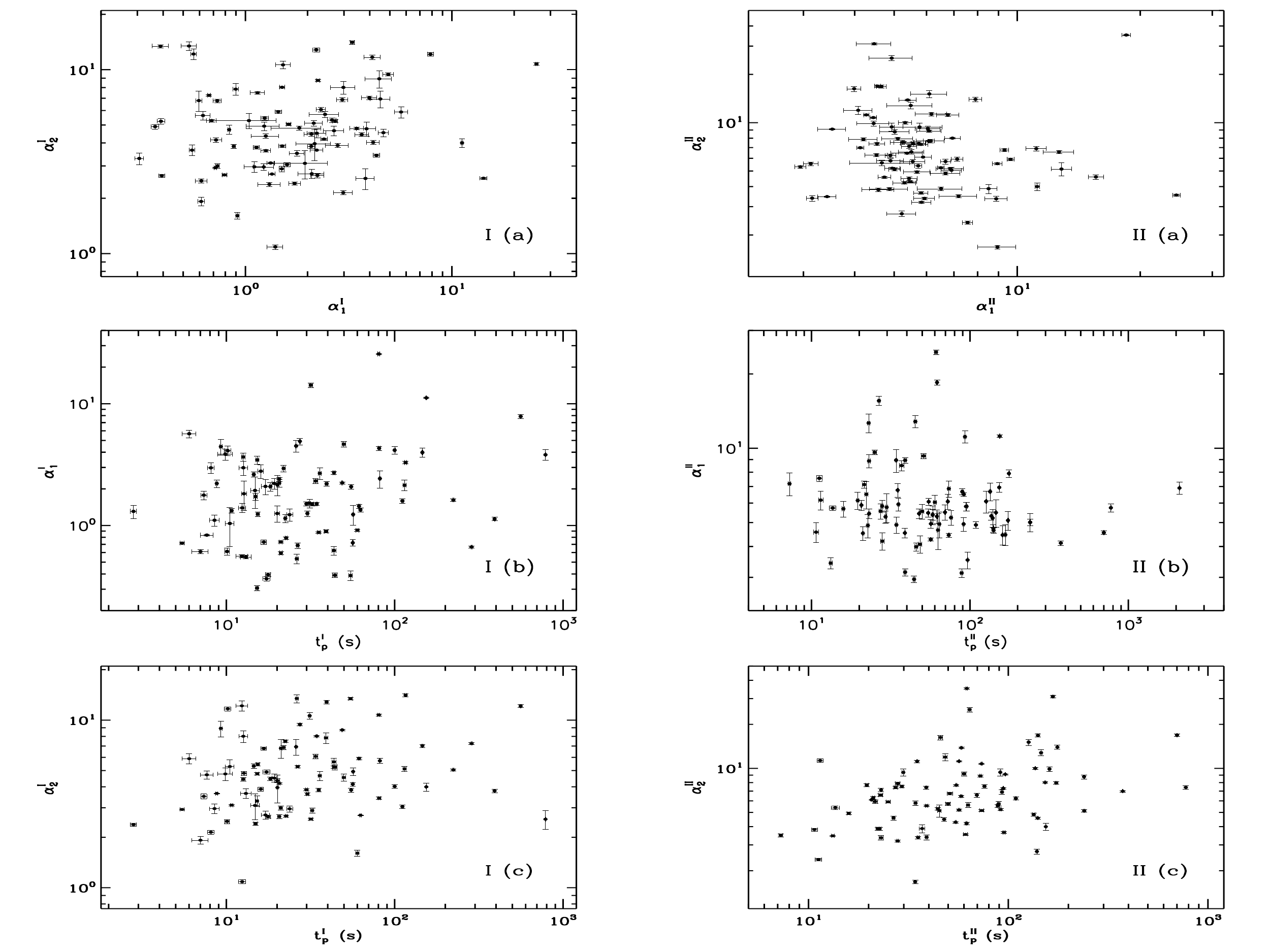} 
\caption{Scatter plots of $\alpha$-related pair-parameters listed in Tables 1 and 2. Symbols ``I" and ``II" stand for Method I and II, respectively. No clear correlation is seen.}
\end{figure*}
\clearpage
\setlength{\voffset}{-18mm}
\begin{figure*}\centering
\includegraphics[angle=0,scale=0.85,width=0.95\textwidth,height=0.90\textheight]{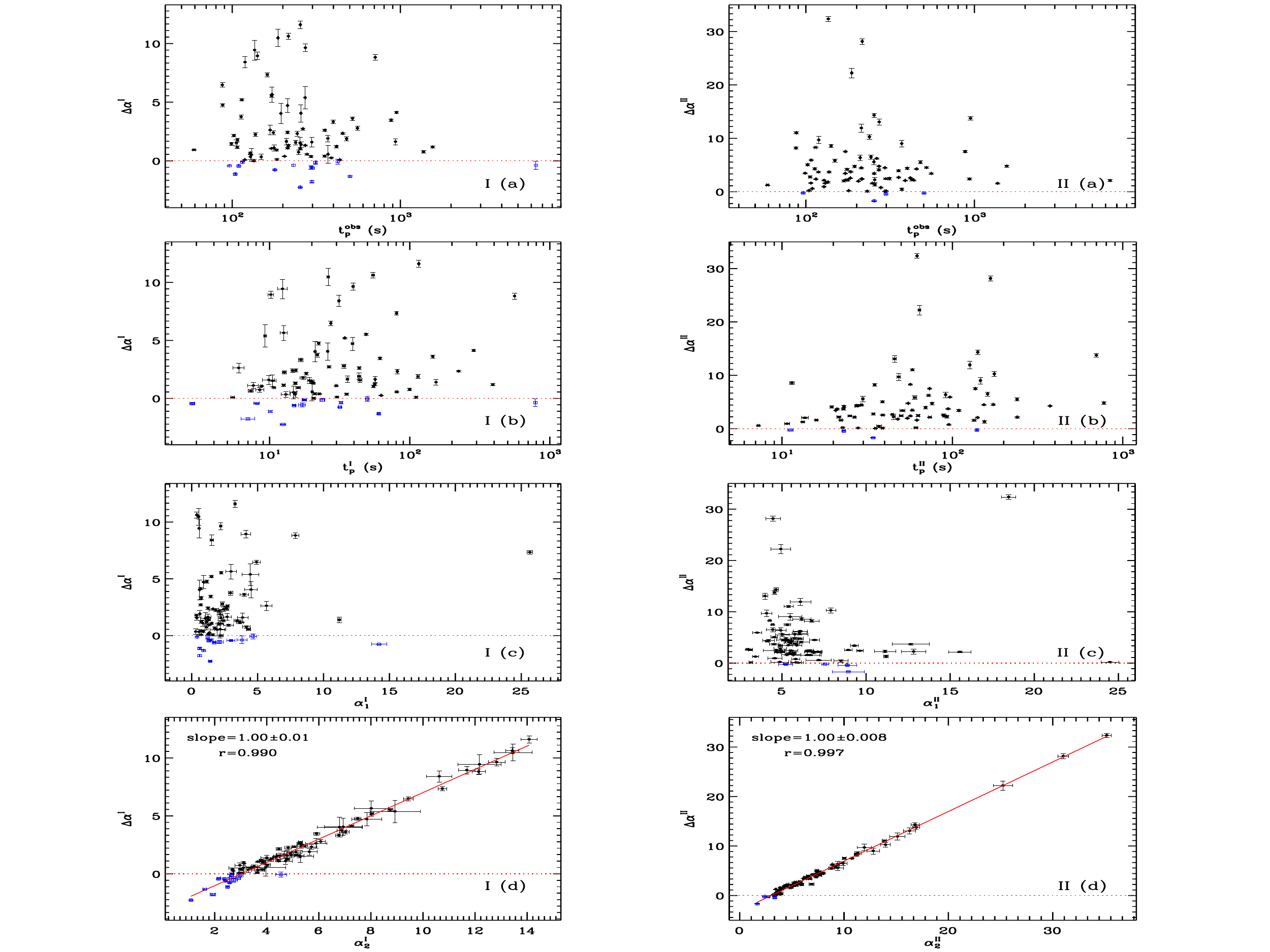}
\caption{Scatter plots of $\Delta\alpha$-related pair-parameters listed in Tables 1 and 2. Symbols ``I" and ``II" stand for Method I and II, respectively. The red dotted line in each panel denotes $\Delta \alpha=0$. The black dots have $\Delta \alpha>0$, which show evidence of acceleration. The blue dots have $\Delta \alpha \leq 0$, suggesting that the flare does not show evidence of acceleration (but could still be in the acceleration regime). The red solid lines  in panels I(d) and II(d) indicate the best-fit relation, with the Pearson correlation coefficients of each pair marked.}
 \end{figure*}\label{fig:delta-alpha}

\end{document}